\documentclass[iop, apj, numberedappendix, appendixfloats]{emulateapj}
\usepackage{graphics}
\usepackage{subfigure}
\usepackage{graphicx}
\usepackage{threeparttable}
\usepackage{multirow}
\usepackage{epsf}
\usepackage{bm}
\usepackage{color}
\usepackage{float}
\usepackage{times}
\usepackage{rotating}
\usepackage{CJK}
\usepackage[titletoc]{appendix}

\newcommand{\OIIIL}{[O~{\sc III}]\,\lambda 4959}
\newcommand{\OIIIB}{[O~{\sc III}]\,\lambda 5007}
\newcommand{\OI}{[O~{\sc I}]\,\lambda 6300}
\newcommand{\NIIL}{[N~{\sc II}]\,\lambda 6549}
\newcommand{\NIIB}{[N~{\sc II}]\,\lambda 6583}
\newcommand{\SIIL}{[S~{\sc II}]\,\lambda 6717}
\newcommand{\SIIB}{[S~{\sc II}]\,\lambda 6731}

\newcommand\nature{{Nature}}

\shortauthors{Zhang et al. 2021}
\shorttitle{A SYSTEMATIC SEARCH FOR DUAL AGNs IN MERGING GALAXIES}

\begin{document}

\title{ \textbf{A} \textbf{S}YS\textbf{T}EMATIC SEA\textbf{R}CH F\textbf{O}R \textbf{D}UAL \textbf{A}GNs IN ME\textbf{R}G\textbf{IN}G \textbf{G}ALAXIES \textbf{(ASTRO-DARING)}:\\ II: first results from long-slit spectroscopic observations }

   \author{Yang-Wei Zhang\altaffilmark{1,3},
   Yang Huang       \altaffilmark{2,5},
   Jin-Ming Bai     \altaffilmark{1,4,5},
   Xiao-Wei Liu     \altaffilmark{2,5},
   Jian-guo Wang     \altaffilmark{1},
   Xiao-bo Dong     \altaffilmark{1},
   }

\altaffiltext{1}{Yunnan Observatories, Chinese Academy of Sciences, Kunming, Yunnan 650011, People’s Republic of China; zhangyangwei@ynao.ac.cn; baijinming@ynao.ac.cn}
\altaffiltext{2}{South-Western Institute for Astronomy Research, Yunnan University, Kunming 650500, People’s Republic of China; yanghuang@ynu.edu.cn; x.liu@ynu.edu.cn}
\altaffiltext{3}{University of Chinese Academy of Sciences, Beijing 100049, People’s Republic of China}
\altaffiltext{4}{Key Laboratory for the Structure and Evolution of Celestial Objects, Chinese Academy of Sciences, Kunming 650011, People’s Republic of China}
\altaffiltext{5}{Corresponding authors}

\begin{abstract}

Building a large sample of kiloparsec (kpc)-scale dual active galactic nuclei (AGNs) amongst merging galaxies is of vital importance to understand the co-evolution between host galaxies and their central super massive black holes (SMBHs).
Doing so, with just such a sample, we have developed an innovative method of systematically searching and identifying dual AGNs of amongst kpc scale merging galaxies and selected 222 candidates at redshifts $\leqslant$ 0.25.
All the selected candidates have FIRST radio detection and at least one of two cores previously revealed as AGN spectroscopically.
We report the first results from A SysTematic seaRch fOr Dual Agns in meRgINg Galaxies (ASTRO-DARING), which consist of spatially resolved long-slit spectroscopic observations of 41 targets selected from our merging galaxies sample carried out between November 2014 and February 2017, using the Yunnan Faint Object Spectrograph and Camera (YFOSC) mounted on the 2.4 meter telescope in Lijiang of Yunnan Observatories.  
Of these 16 are likely dual AGNs and 15 are newly identified.
The efficiency of ASTRO-DARING is thus nearly 40 per cent.
With this method, we plan to build the first even sample of more than 50 dual AGNs constructed using a consistent approach. 
Further analysis of the dual AGN sample shall provide vital clues for understanding the co-evolution of galaxies and SMBHs.

\end{abstract}

\keywords{techniques: spectroscopic--
galaxies: active -- 
galaxies: nuclei --
galaxies: mergers --
galaxies: interactions -- 
galaxies: co-evolution 
}

\section{Introduction}

Galaxy merging plays an vital role for galaxy growth in the standard $\Lambda$CDM cosmology (e.g., \citealt{Kauffmann2000, Di_Matteo2005, Kormendy and Ho2013}). 
As essentially all massive galaxies are believed to host a central super massive black hole (SMBH; e.g., \citealt{Kormendy-Richstone1995, Richstone1998}), one expects to find SMBH pairs in the merging galaxies (e.g., \citealt{Begelman1980, Milosavljevic and David2001}). 
The merging would also trigger active galactic nuclei (AGN) activities when large amounts of gas fall into the central SMBHs via the gravitational interactions (e.g., \citealt{Hopkins2008, Kocevski2012, Treister2012}). 
In this scenario, dual AGNs are expected to form in merging galaxies (e.g., \citealt{Koss2012, Satyapal2017}).

For merging galaxies of projected separations smaller than 15 kpc (typical tidal radius of a galaxy), 
the gravitational interactions between the two merging galaxies become significant (e.g., \citealt{Begelman1980, Volonteri2003}). 
Identifying dual AGNs at this critical stage before final coalescence amongst gas-rich merging galaxies can provide vital information to investigate the co-evolution of SMBHs and their host galaxies (e.g., \citealt{Colpi and Dotti2011, Yu2011}), 
for example, the underlying physics of the tight scaling relations between black hole mass (M$_{\rm BH}$) and large-scale properties of the host galaxy found from observations, especially the well-known relation between central BH mass and host galaxy bulge velocity dispersion ($\sigma$; e.g., \citealt{Ferrarese2000, Gebhardt2000, Haring and Rix2004, Gultekin2009, Graham and Scott2013, Kormendy and Ho2013, McConnell2013}).
While considered fundamental and supported by bona fide observational evidence, it is still not clear whether the M$_{\rm BH}$-$\sigma$ relation is universal and followed by all types of galaxies including merging galaxies (e.g., \citealt{Komossa2007, Fu2011a, Savorgnan and Graham2015}).
The previous studies indicate that the dual AGN systems may not follow the typical M$_{\rm BH}$-$\sigma$ relation defined from isolated AGNs and thus can contribute scatter to this relation (e.g., \citealt{Blecha2011, Fu2011a}).
The physical mechanisms connecting the growth of BH mass and the host galaxy are also not well understood (e.g., \citealt{Ricci2017, Sheinis2017}). 
Other questions related include how the phenomenon of dual AGNs is triggered and how they evolve as in the merging progresses (e.g., \citealt{Treister2012, Satyapal2017}). 
Being the late stage products of galaxy merging, dual AGNs can clearly help address these fundamental questions (e.g., \citealt{Kosec2017, Solanes2019}).

To answer the above questions, a large sample of dual AGNs is required.
However, the current number of identified dual AGNs is limited, with a total number of no more than 50 (see the compilation in \citealt{Das2018} and Huang et al. 2021, in preparation; hereafter Paper I).
These confirmed dual AGNs are identified by a variety of methods, such as X-ray observations (e.g., 
\citealt{Komossa2003}; 
\citealt{Guainazzi2005}; 
\citealt{hudson2006};
 \citealt{liu2013};
\citealt{Comerford2011, comerford2015};
\citealt{Koss2016};
\citealt{Ellison2017}),
or radio observations (e.g., 
\citealt{rodriguez2006};
\citealt{Fu2011b,Fu2015};
\citealt{Muller-Sanchez2015};
\citealt{Rubinur2017}),
or optical spectroscopy (e.g., \citealt{Liu2010b, Liu2011, Shen2011, Huang2014}).

In the past decades, several systematic methods have been developed for identifying dual AGNs.
One of the most ambitious methods is finding dual AGNs from the double-peaked AGNs (DPAGNs).
As a potential mechanism, binary/dual AGN with projected separations raning from 100 pc to 10 kpc could contribute the DPAGNs (e.g., \citealt{Zhou2004, wang2009}).
Thanks to massive spectroscopic surveys, such as SDSS and DEEP2 Galaxy Survey, hundreds of DPAGNs have been spectroscopically selected (e.g., \citealt{Gerke2007, wang2009, Liu2010a, Smith2010, Rosario2011, Ge2012, Barrows2013, Shi2014, McGurk2015}).
However, spatially resolved longslit/integral-field spectroscopy (e.g., \citealt{Liu2011, Shen2011, McGurk2011, Comerford2012}) and high resolution imaging (e.g., \citealt{liu2013}, \citealt{comerford2015}, \citealt{Muller-Sanchez2015}) follow-up observations show that only 2 to 5 per cent of the DPAGNs are dual AGNs and most of DPAGNs are produced by gas kinematics related to a single AGN (e.g., rotating gas disks or biconical outflows from the narrow line region (NLR) of AGN; \citealt{Comerford2011, Fu2011a, Fu2012, Shen2011, Smith2011, Smith2012, gabanyi2014, Nevin2016}).
Most recently, the systematic search for dual AGNs based on radio imaging (\citealt{Fu2015}) and mid-infrared color (\citealt{Satyapal2014,Satyapal2017}) has been proposed but the number of identified dual AGNs is still very limited (see Section\,4 for more details).
 
It is a challenging task to build a homogeneous dual AGN sample for the purpose of studying the co-evolution of SMBHs and host galaxies.
A more robust and efficient new method is clearly desirable. 
Inspired by previous studies (e.g., \citealt{Hennawi2006, Hennawi2010, Ellison2011, Satyapal2014, Rubinur2019}), we have developed an innovative method to systematically find and identify dual AGNs amongst kpc scale merging galaxies (see Section\,2.1 and Paper I for more details). 
With this new method, a total of 222 targeted candidates (merging galaxies) have been selected.
To reveal their dual AGN nature, we have embarked on an observational campaign, a systematic search for dual AGNs in merging galaxies (ASTRO-DARING for short), using spatially resolved long-slit and aperture spectroscopy.

In this work (hereafter Paper II), we present the first results of ASTRO-DARING from November 2014 to February 2017 using the YFOSC mounted on the 2.4\,m telescope in Lijiang (LJT) of the Yunnan Observatories (YNAO). 
The observation and data reduction are described in Section~\ref{Obs and data}. 
The results are presented in Section~\ref{Result}. 
In Section~\ref{Discussions}, we discuss the implications and potential applications of our current results. 
Finally, a summary is given in Section~\ref{Conclusions}.
Cosmological constants $\rm H_{0}$ = 70 km\,s$^{-1}$ Mpc$^{-1}$, $\Omega_{\rm m}$ = 0.3, and $\Omega_{\rm \Lambda}=0.7$ are adopted throughout the paper and all wavelengths are vacuum.

\begin{table}[!htp]
\begin{center}
\caption{Instrumental setup}
\begin{tabular}{cccccccc}
\hline
 Grism       &   Spectral range  & Dispersion      & Resolution$_{1.8}$   &   Resolution$_{2.5}$  \\
	         &	 \AA         &   nm/pix   & (km $\mathrm{s}^{-1}$)  & 	(km $\mathrm{s}^{-1}$)     \\
\hline	
G8         &	 5100 $-$ 9600  &  0.15  &  $ 330 \pm 20$  &  $-$              \\
\hline
G14       &	 3600 $-$ 7500      &  0.17	 &  $ 500 \pm 40$  &  $-$               \\
\hline
G3         &	 3400 $-$ 9100  &  0.29  &  $730 \pm 60$  &  $970 \pm 70$   \\
\hline
\end{tabular}
\label{grism}
 \begin{flushleft}

Notes: Resolution$_{1.8}$: Median Resolution of the instrumental broadening for a slit width of 1.8$''$. 
Resolution$_{2.5}$: Median Resolution of the instrumental broadening for a slit width of 2.5$''$. 
The instrumental broadening was measured from the arc lamp spectra.

\label{slit} 
\end{flushleft}
\end{center}
\end{table}



\begin{table*}[!htp]
\scriptsize
\centering
\caption{Observational log}
\begin{threeparttable}
\begin{tabular}{cccccccccccc}
\hline
\hline
Name&   RA (J2000) & Dec (J2000) & Redshift& Sep0 & Sep1 &Grism&Slit width & Observing date & Seeing &PA\tnote{$a$}&Exposure time\\
   
    &   &  & &(\arcsec /kpc)& (\arcsec /kpc)&&(\arcsec)&UT&(\arcsec)&(\arcdeg)&(s)\\

\hline

\multicolumn{11}{c}{First phase (15 sources)}\\

\hline

\multirow{2}{*}{J0151$-$0245}& \multirow{2}{*}{01:51:07.74} & \multirow{2}{*}{$-$02:45:27.66} &\multirow{2}{*}{0.0479}&\multirow{2}{*}{5.3/5.0}&  \multirow{2}{*}{5.1/4.9}   & G14&1.8&20141211&2.2&\multirow{2}{*}{$98.3$ }&2500\\
&      &     &  & & &G14&1.8&20150207&2.6& &3000\\ 
\hline

\multirow{2}{*}{J0225$-$0824}& \multirow{2}{*}{02:25:11.63} & \multirow{2}{*}{$-$08:24:38.38} &\multirow{2}{*}{0.1099}&\multirow{2}{*}{5.6/11.3}&  \multirow{2}{*}{5.7/11.4}   &G14&1.8&20141210&2.4&\multirow{2}{*}{$146.6$ }&2246\tnote{$b$}\\
&      &    &   & & &G8&1.8&20141211&1.7&&2500$\times$2\\ 

\hline	

 \multirow{2}{*}{J0737$+$4651}& \multirow{2}{*}{07:37:44.10} & \multirow{2}{*}{$+$46:51:07.96} &\multirow{2}{*}{0.0951}&\multirow{2}{*}{1.6/2.9}&  \multirow{2}{*}{1.7/3.0}&  G8&1.8&20141212&1.6&\multirow{2}{*}{$25.2$ }&2500+2100\\
&      &    &   & & &G14&1.8&20150209&1.0&&3000\\

\hline	

\multirow{2}{*}{J0933$+$2114}& \multirow{2}{*}{09:33:47.76} & \multirow{2}{*}{$+$21:14:36.41} &\multirow{2}{*}{0.1722}&\multirow{2}{*}{4.1/11.9}&  \multirow{2}{*}{4.0/11.7}&  G8&1.8&20150210&0.9&\multirow{2}{*}{$52.9$ }&3000\\
&      &    &   & & &G8&1.8&20150211&1.4&&3000\\
\hline	
J1010$+$0612& 10:10:43.36 & $+$06:12:01.42 &0.0978&7.1/12.8&7.1/12.8 &G8&1.8&20150212&1.3&$61.0$&2410\tnote{$b$}\\	

\hline			

\multirow{2}{*}{J1017$+$3448}& \multirow{2}{*}{10:17:56.75} & \multirow{2}{*}{$+$34:48:50.36} &\multirow{2}{*}{0.1440}&\multirow{2}{*}{5.4/11.7}&   \multirow{2}{*}{5.4/11.7}&  G8&1.8&20150207&2.8&\multirow{2}{*}{$134.9$ }&3000\\
&      &    &   & & &G8&1.8&20150208&1.8&&3000\\
\hline
\multirow{2}{*}{J1105$+$1957}& \multirow{2}{*}{11:05:44.45} & \multirow{2}{*}{$+$19:57:46.29} &\multirow{2}{*}{0.1043}&\multirow{2}{*}{3.8/7.3}&  \multirow{2}{*}{3.7/7.1}&  G8&1.8&20150211&1.6& \multirow{2}{*}{$8.6$ } &3000\\
&      &   &    & & &G8&1.8&20150228&1.9&&3000\\
\hline

J1200$+$3147& 12:00:41.39  & $+$31:47:46.28 &0.1161&5.7/12.1& 5.7/12.1 &G8&1.8&20150212&1.6&$33.2$&3000\\
\hline		

J1201$-$0153& 12:01:49.74 & $-$01:53:27.55 &0.0907&7.0/11.8&  7.1/12.0   &G8&1.8&20150211&2.0& $143.2$&1292\tnote{$b$}\\
\hline	
J1535$+$3455& 15:35:02.26 & $+$34:55:38.43 &0.1307&5.3/12.3& 5.4/12.6 &G8&1.8&20150301&1.6&$93.1$&3000\\
\hline		
J1633$+$4718& 16:33:23.58 & $+$47:18:58.95 &0.1158&3.8/8.0& 4.0/8.4   &G8&1.8&20150301&1.4&$175.6$&3000\\
\hline			
\multirow{2}{*}{J2150$-$0052}& \multirow{2}{*}{21:50:24.70} &\multirow{2}{*}{$-$00:52:42.78} &\multirow{2}{*}{0.1108}&\multirow{2}{*}{5.4/11.0}&  \multirow{2}{*}{5.5/11.1}&  G8&1.8&20141115&1.8&\multirow{2}{*}{$53.1$ }&3000\\
&      &   &    & & &G14&1.8&20141115&1.8&&3000\\

\hline 	
\multirow{2}{*}{J2226$+$0143}& \multirow{2}{*}{22:26:21.65} & \multirow{2}{*}{$+$01:43:29.88} &\multirow{2}{*}{0.2231}&\multirow{2}{*}{2.6/9.3}&  \multirow{2}{*}{2.5/9.0}&  G8&1.8&20151115&1.6&\multirow{2}{*}{$79.8$ }&3000\\
&      &    &   & & &G14&1.8&20151115&1.6&&2387\tnote{$b$}\\
\hline 	

\multirow{2}{*}{J2233$+$0332}& \multirow{2}{*}{22:33:36.41} & \multirow{2}{*}{$+$03:32:34.70} &\multirow{2}{*}{0.1064}&\multirow{2}{*}{3.6/7.0}&  \multirow{2}{*}{3.7/7.2}&  G8&1.8&20141114&1.8&\multirow{2}{*}{$6.4$ }&3000\\
&      &   &    & & &G14&1.8&20141114&2.0&&3000\\
\hline  	

\multirow{2}{*}{J2258$-$0115}& \multirow{2}{*}{22:58:10.01} &  \multirow{2}{*}{$-$01:15:16.26} &\multirow{2}{*}{0.1170}&\multirow{2}{*}{3.4/7.1}&  \multirow{2}{*}{3.4/7.1}&  G14&1.8&20141111&2.0&\multirow{2}{*}{$10.1$ }&2700+3600\\
&      &    &   & & &G8&1.8&20141111&2.0&&2545\tnote{$b$}\\
\hline
\multicolumn{11}{c}{Second phase (26 sources)}\\ 

\hline
  J0101$-$0957&01:01:58.62 & $-$09:57:50.57  &0.1523&5.3/14.1& 5.4/14.3  &G3&1.8&20151105&2.0&$52.2$&2380\tnote{$b$}\\
  \hline  
   J0141$-$0105&01:41:56.81 & $-$01:05:32.03 &0.1392&6.0/14.7& 5.9/14.5  & G3&1.8&20151106&3.0&$5.6$&2700\\ 
\hline  
 \multirow{2}{*}{J0157$+$1155}& \multirow{2}{*}{01:57:23.82}  &\multirow{2}{*}{$+$11:55:47.61} &\multirow{2}{*}{0.0888}&\multirow{2}{*}{4.5/7.5}&    \multirow{2}{*}{4.5/7.5}&  G3&2.5&20161202&1.7&\multirow{2}{*}{$154.4$ }&3000\\ 
&      &   &    & & &G3&2.5&20161203&2.1&&2500\\ 
\hline  
 \multirow{2}{*}{J0204$-$0248}&\multirow{2}{*}{02:04:24.83} &\multirow{2}{*}{$-$02:48:41.27} &\multirow{2}{*}{0.0750}&\multirow{2}{*}{3.4/4.9}&   \multirow{2}{*}{3.5/5.0}&   G3&1.8&20151130&2.1&\multirow{2}{*}{$75.6$ }&2500\\ 
&      &   &    & & &G3&2.5&20161204&2.2&&1217\tnote{$b$}+2500\\ 

\hline
  J0206$-$0441& 02:06:28.41 & $-$04:41:10.54 &0.1364&5.5/13.2& 5.4/13.0  &G3&2.5&20161202&1.5&$82.1$&3600\\ 
\hline
J0217$-$0845 & 02:17:03.46 & $-$08:45:19.08 &0.1081&6.0/12.0&6.2/12.2   &G3&2.5&20161203&2.1&$59.0$&3000+3000\\ 
\hline
 \multirow{2}{*}{J0251$-$0837}& \multirow{2}{*}{02:25:11.63} &\multirow{2}{*}{$-$08:24:38.38} &\multirow{2}{*}{0.1323}&\multirow{2}{*}{4.9/11.4}&  \multirow{2}{*}{4.8/11.3}&  G3&2.5&20161202&1.7&\multirow{2}{*}{$7.8$ }&3500\\ 

&      &  &     & & &G3&2.5&20170105&1.8&&3000\\  
 
\hline
  J0750$+$3530& 07:50:57.26 & $+$35:30:37.67 &0.1762&3.4/10.3& 3.4/10.3 & G3&2.5&20170204&2.1&$141.8$&3000\\ 
\hline
 \multirow{2}{*}{J0752$+$3419}&\multirow{2}{*}{07:52:21.86} & \multirow{2}{*}{$+$34:19:35.58} &\multirow{2}{*}{0.1400}&\multirow{2}{*}{3.3/8.1}&  \multirow{2}{*}{3.1/7.6}&  G3&1.8&20151123&1.5&\multirow{2}{*}{$59.9$}&3800\\ 
                              &      &   &    & & &G3&1.8&20151220&1.7&&3000\\  
\hline
   J0756$+$2340&07:56:21.00 & $+$23:40:39.40 &0.0742&6.8/9.6& 6.8/9.6   &G3&2.5&20161204&3.0&$48.6$&3000\\ 
\hline
   J0758$+$2705&07:58:46.99& $+$27:05:15.61&0.0987&3.3/6.0& 3.4/6.2    &G3&1.8&20151124&1.3&$73.8$&3000\\ 
\hline
 \multirow{2}{*}{J0813$+$4941}&\multirow{2}{*}{08:13:47.49}  &\multirow{2}{*}{$+$49:41:09.83} &\multirow{2}{*}{0.0942}&\multirow{2}{*}{3.3/5.8}&  \multirow{2}{*}{3.4/5.9}&  G3&2.5&20170204&2.2&\multirow{2}{*}{$139.9$}&3000\\
                              &      &   &    & & &G3&2.5&20170205&2.0&&3000\\
\hline
  
    J0813$+$5529&08:13:26.77& $+$55:29:18.07 &0.0796&5.9/8.8& 5.9/8.8& G3&2.5&20161204&3.0&$132.1$&3000\\ 
\hline  
  
    J0832$+$0937& 08:32:02.71 & $+$09:37:59.17 &0.0750&4.3/6.2&4.2/6.0   & G3&2.5&20170206&2.7&$178.9$&2500\\  
\hline 
   
    J0833$+$1532&08:33:55.49 &$+$15:32:36.62 &0.1516&4.5/11.9   &4.5/11.9 &G3&2.5&20170205&2.0&$149.4$&3000\\ 
\hline 

 \multirow{2}{*}{J0848$+$3515}&\multirow{2}{*}{08:48:09.69} & \multirow{2}{*}{$+$35:15:32.12}  &\multirow{2}{*}{0.0570}&\multirow{2}{*}{5.6/6.2}&  \multirow{2}{*}{5.7/6.3}&  G3&2.5&20170205&2.5&\multirow{2}{*}{$66.8$}&3000\\ 

&      &   &    & & &G3&2.5&20170206&1.8&&2500\\ 
\hline

    J0907$+$5203&09:07:14.44  & $+$52:03:43.40 &0.0596&7.4/8.5& 7.4/8.5   & G3&2.5&20170206&2.1&$11.4$&2500\\ 
\hline 

    J1214$+$2931 & 12:14:18.25 & $+$29:31:46.70 & 0.0633 & 6.7/8.2 &  6.7/8.2  & G3&1.8&20160528 & 1.5 & $61.2$ &3200\\
\hline 
    J1645$+$2057&16:45:07.91 &$+$20:57:59.43 &0.1300&4.2/9.8& 4.2/9.8   &G3&1.8&20160519&2.0&$155.1$&3200\\ 
\hline 

    J2145$+$1144&21:45:30.39 &$+$11:44:03.66 &0.1122&3.8/7.8&3.7/7.6    &G3&1.8&20151106&2.2& $93.5$ &2200\\ 
\hline 

 \multirow{2}{*}{J2206$+$0003}& \multirow{2}{*}{22:06:35.08} & \multirow{2}{*}{$+$00:03:23.16} &\multirow{2}{*}{0.0461}&\multirow{2}{*}{4.7/4.3}  & \multirow{2}{*}{4.5/4.1} &G3&1.8&20151126&1.4&\multirow{2}{*}{$156.9$}&2000+2200\\ 

&      &     &  & & &G3&1.8&20151209&1.9&&2700\\ 
\hline

    J2210$+$0945& 22:10:58.06 &$+$09:45:00.92 &0.1170&4.7/9.9& 4.5/9.5    &G3&1.8&20151106&1.5&$19.1$&2500\\ 
\hline 

 \multirow{2}{*}{J2239$+$0012}&\multirow{2}{*}{ 22:39:32.21} &\multirow{2}{*}{$+$00:12:46.36}&\multirow{2}{*}{0.1615}&\multirow{2}{*}{4.1/11.4}&  \multirow{2}{*}{4.2/11.7}&  G3&1.8&20151125&1.4&\multirow{2}{*}{$178.8$}&3000\\ 

&      &    &   & & &G3&1.8&20161024&1.3&&2700\\ 

\hline

\multirow{2}{*}{J2252$+$0106}&\multirow{2}{*}{22:52:22.35} &\multirow{2}{*}{$+$01:06:59.98} &\multirow{2}{*}{0.0717}&\multirow{2}{*}{3.2/4.4}&  \multirow{2}{*}{3.1/4.2}&  G3&1.8&20161024&1.5&\multirow{2}{*}{$126.2$}&2700\\  

&      &   &    & & &G3&2.5&20161204&1.7&&3000\\ 

\hline
  
\multirow{2}{*}{J2314$+$0653}&\multirow{2}{*}{23:14:39.21} & \multirow{2}{*}{$+$06:53:12.97} & \multirow{2}{*}{0.0875}&\multirow{2}{*}{4.1/6.7}&  \multirow{2}{*}{4.2/6.9}&  G3&1.8&20151128&1.3&\multirow{2}{*}{$157.5$}&2300\\ 

&      &   &    & & &G3&1.8&20161024&1.8&&3000\\ 

\hline
  
\multirow{2}{*}{J2320$+$0741}&\multirow{2}{*}{23:20:41.53 }& \multirow{2}{*}{$+$07:41:48.24 } & \multirow{2}{*}{0.1316}&\multirow{2}{*}{4.4/10.2}&  \multirow{2}{*}{4.5/10.5}&  G3&1.8&20161024&1.3&\multirow{2}{*}{$106.1$}&3000\\ 

&  & &  & & &G3&2.5&20161204&1.6&&3000\\ 

\hline

\end{tabular}
\label{obs_log} 
\begin{flushleft}
 Notes: Sep0: The separation of two cores from the SDSS image.
 Sep1: The separation of two cores from the 2D spectrum.
\item[$a$]PA: Position angle of the slit on the sky, in degrees east of north.
\item[$b$]The exposure times of the observations were originally set to be longer (e.g., 3000s), but were cut short due to problems of the telescope tracking. 

\end{flushleft}
\end{threeparttable}
\end{table*} 



\begin{table*}[!htp]
\scriptsize
\centering
\caption{Fluxes of well detected emission lines}
\begin{threeparttable}
\begin{tabular}{ccccccccccc}
\hline
\hline

Name  & $f_{\rm \mathrm{H}\beta}$  & $f_{\rm \OIIIL}$ & $f_{\rm \OIIIB}$ & $f_{\rm \OI}$  & $f_{\rm \NIIL}$ & $f_{\rm \mathrm{H}\alpha}$  & $f_{\rm \NIIB}$ & $f_{\rm \SIIL}$ & $f_{\rm \SIIB}$ \\

\hline  
\multicolumn{10}{c}{First phase (6 Dual AGNs)}\\

\hline  

  J0151$-$0245WN  & $ 75 \pm 23 $ & $ 31 \pm 11 $ & $ 128 \pm 12 $ & $ 115 \pm 12 $ & $ 85 \pm 14 $ & $ 68 \pm 12 $ & $ 231 \pm 18 $ & $ 145 \pm 17 $ & $ 120 \pm 17 $ \\ 
  J0151$-$0245ES  & $ 76 \pm 22 $ & $ 52 \pm 15 $ & $ 153 \pm 13 $ & $ 109 \pm 15 $ & $ 213 \pm 18 $ & $ 247 \pm 18 $ & $ 445 \pm 11 $ & $ 272 \pm 14 $ & $ 164 \pm 14 $ \\ 
\hline  
 
J0933$+$2114EN & $ 131 \pm 84 $ & $ 437 \pm 59 $ & $ 1494 \pm 59 $ & $ 106 \pm 37 $ & $ 540 \pm 137 $ & $ 1663 \pm 140 $ & $ 1238 \pm 124 $ & $ 414 \pm 122 $ & $ 193 \pm 93 $ \\ 
J0933$+$2114WS & $ 399 \pm 33 $ & $ 382 \pm 31 $ & $ 1192 \pm 30 $ & $ 215 \pm 44 $ & $ 254 \pm 54 $ & $ 929 \pm 68 $ & $ 609 \pm 57 $ & $ 229 \pm 52 $ & $ 137 \pm 48 $ \\ 
\hline

J1017$+$3448NW & $ 313 \pm 106 $ & $ 286 \pm 100 $ & $ 976 \pm 104 $ & $ 128 \pm 52 $ & $ 365 \pm 81 $ & $ 1115 \pm 71 $ & $ 665 \pm 59 $ & $ 283 \pm 18 $ & $ 152 \pm 15 $ \\ 
J1017$+$3448SE & $ 118 \pm 52 $ & $ 309 \pm 40 $ & $ 1191 \pm 40 $ & $ 77 \pm 38 $ & $ 226 \pm 120 $ & $ 723 \pm 142 $ & $ 570 \pm 96 $ & $ 98 \pm 44 $ & $ 89 \pm 45 $ \\ 
\hline

J1105$+$1957EN & $ 197 \pm 14 $ & $ 58 \pm 15 $ & $ 195 \pm 17 $ & $ 57 \pm 8 $ & $ 376 \pm 14 $ & $ 1221 \pm 13 $ & $ 893 \pm 11 $ & $ 268 \pm 17 $ & $ 242 \pm 15 $ \\ 
J1105$+$1957WS & $ 110 \pm 11 $ & $ 51 \pm 19 $ & $ 230 \pm 21 $ & $ 33 \pm 6 $ & $ 51 \pm 8 $ & $ 247 \pm 13 $ & $ 176 \pm 13 $ & $ 83 \pm 10 $ & $ 58 \pm 8 $ \\ 
\hline

J1633$+$4718N & $ 205 \pm 23 $ & $ 93 \pm 19 $ & $ 320 \pm 20 $ & $ 96 \pm 32 $ & $ 104 \pm 32 $ & $ 1023 \pm 36 $ & $ 471 \pm 36 $ & $ 182 \pm 12 $ & $ 142 \pm 15 $ \\ 
J1633$+$4718S & $ 203 \pm 42 $ & $ 135 \pm 16 $ & $ 545 \pm 29 $ & $ 167 \pm 36 $ & $ 177 \pm 67 $ & $ 1163 \pm 176 $ & $ 576 \pm 76 $ & $ 128 \pm 17 $ & $ 114 \pm 17 $ \\ 
\hline

J2258$-$0115EN & $ 84 \pm 12 $ & $ 22 \pm 10 $ & $ 96 \pm 12 $ & $ 18 \pm 9 $ & $ 24 \pm 5 $ & $ 273 \pm 16 $ & $ 137 \pm 15 $ & $ 63 \pm 12 $ & $ 43 \pm 8 $ \\ 
J2258$-$0115WS & $ 76 \pm 39 $ & $ 127 \pm 13 $ & $ 425 \pm 23 $ & $ 61 \pm 8 $ & $ 64 \pm 10 $ & $ 128 \pm 16 $ & $ 109 \pm 11 $ & $ 21 \pm 4 $ & $ 20 \pm 5 $ \\ 
\hline

\multicolumn{10}{c}{Second phase (10 Dual AGNs)}\\
\hline

J0217$-$0845EN & $ 44 \pm 5 $ & $ 11 \pm 3 $ & $ 29 \pm 6 $ & $ 55 \pm 10 $ & $ 84 \pm 9 $ & $ 51 \pm 17 $ & $ 69 \pm 16 $ & $ 60 \pm 12 $ & $ 29 \pm 8 $ \\ 
J0217$-$0845WS & $ 47 \pm 7 $ & $ 24 \pm 4 $ & $ 48 \pm 7 $ & $ 62 \pm 11 $ & $ 57 \pm 11 $ & $ 99 \pm 13 $ & $ 124 \pm 19 $ & $ 82 \pm 17 $ & $ 83 \pm 18 $ \\

J0756$+$2340EN & $ 150 \pm 15 $ & $ 85 \pm 16 $ & $ 138 \pm 11 $ & $ 51 \pm 10 $ & $ 439 \pm 65 $ & $ 806 \pm 77 $ & $ 656 \pm 31 $ & $ 168 \pm 13 $ & $ 197 \pm 10 $ \\ 
J0756$+$2340WS & $ 80 \pm 15 $ & $ 93 \pm 16 $ & $ 195 \pm 16 $ & $ 66 \pm 12 $ & $ 369 \pm 37 $ & $ 329 \pm 39 $ & $ 397 \pm 15 $ & $ 231 \pm 29 $ & $ 168 \pm 19 $ \\   
\hline

J0813$+$4941WN & $ 92 \pm 30 $ & $ 229 \pm 29 $ & $ 866 \pm 44 $ & $ 105 \pm 15 $ & $ 189 \pm 47 $ & $ 200 \pm 60 $ & $ 278 \pm 46 $ & $ 94 \pm 27 $ & $ 78 \pm 29 $ \\ 
J0813$+$4941ES & $ 171 \pm 10 $ & $ 24 \pm 10 $ & $ 106 \pm 10 $ & $ 40 \pm 7 $ & $ 618 \pm 135 $ & $ 699 \pm 141 $ & $ 332 \pm 35 $ & $ 162 \pm 27 $ & $ 136 \pm 42 $ \\ 
\hline

J0833$+$1532WN & $ 30 \pm 6 $ & $ 45 \pm 10 $ & $ 174 \pm 16 $ & $ 25 \pm 5 $ & $ 33 \pm 13 $ & $ 140 \pm 25 $ & $ 222 \pm 10 $ & $ 69 \pm 12 $ & $ 45 \pm 12 $ \\ 
J0833$+$1532ES & $ 206 \pm 10 $ & $ 245 \pm 18 $ & $ 805 \pm 13 $ & $ 51 \pm 8 $ & $ 74 \pm 34 $ & $ 236 \pm 39 $ & $ 127 \pm 28 $ & $ 68 \pm 9 $ & $ 76 \pm 8 $ \\ 
\hline

J0848$+$3515EN & $ 277 \pm 22 $ & $ 475 \pm 23 $ & $ 1408 \pm 28 $ & $ 184 \pm 16 $ & $ 750 \pm 56 $ & $ 1025 \pm 88 $ & $ 610 \pm 62 $ & $ 156 \pm 24 $ & $ 235 \pm 19 $ \\ 
J0848$+$3515WS & $ 1773 \pm 50 $ & $ 2124 \pm 41 $ & $ 6856 \pm 70 $ & $ 1549 \pm 93 $ & $ 2137 \pm 163 $ & $ 2402 \pm 120 $ & $ 1529 \pm 90 $ & $ 2025 \pm 41 $ & $ 1383 \pm 42 $ \\ 
\hline

J0907$+$5203EN & $ 248 \pm 11 $ & $ 165 \pm 17 $ & $ 474 \pm 18 $ & $ 95 \pm 15 $ & $ 361 \pm 53 $ & $ 656 \pm 51 $ & $ 249 \pm 21 $ & $ 181 \pm 42 $ & $ 266 \pm 29 $ \\ 
J0907$+$5203WS & $ 116 \pm 12 $ & $ 209 \pm 10 $ & $ 633 \pm 10 $ & $ 141 \pm 14 $ & $ 109 \pm 37 $ & $ 455 \pm 31 $ & $ 193 \pm 14 $ & $ 180 \pm 12 $ & $ 180 \pm 17 $ \\ 
\hline

J1214$+$2931EN & $ 553 \pm 21 $ & $ 930 \pm 25 $ & $ 2743 \pm 22 $ & $ 287 \pm 15 $ & $ 463 \pm 16 $ & $ 1239 \pm 15 $ & $ 1467 \pm 13 $ & $ 598 \pm 44 $ & $ 375 \pm 35 $ \\ 
J1214$+$2931WS & $ 4452 \pm 117 $ & $ 10412 \pm 100 $ & $ 31133 \pm 105 $ & $ 1489 \pm 67 $ & $ 1182 \pm 97 $ & $ 9990 \pm 145 $ & $ 5413 \pm 218 $ & $ 2429 \pm 170 $ & $ 2544 \pm 204 $ \\ 
\hline

J1645$+$2057WN & $ 121 \pm 19 $ & $ 77 \pm 18 $ & $ 232 \pm 13 $ & $ 40 \pm 12 $ & $ 123 \pm 38 $ & $ 485 \pm 63 $ & $ 435 \pm 64 $ & $ 95 \pm 14 $ & $ 138 \pm 16 $ \\ 
J1645$+$2057ES & $ 293 \pm 27 $ & $ 76 \pm 19 $ & $ 339 \pm 16 $ & $ 37 \pm 11 $ & $ 327 \pm 63 $ & $ 765 \pm 56 $ & $ 547 \pm 15 $ & $ 137 \pm 12 $ & $ 100 \pm 10 $ \\ 
\hline

J2206$+$0003WN & $ 88 \pm 18 $ & $ 49 \pm 9 $ & $ 154 \pm 17 $ & $ 76 \pm 12 $ & $ 144 \pm 31 $ & $ 247 \pm 27 $ & $ 230 \pm 18 $ & $ 163 \pm 16 $ & $ 144 \pm 15 $ \\ 
J2206$+$0003ES & $ 60 \pm 13 $ & $ 67 \pm 6 $ & $ 251 \pm 12 $ & $ 174 \pm 14 $ & $ 335 \pm 52 $ & $ 436 \pm 79 $ & $ 804 \pm 30 $ & $ 437 \pm 21 $ & $ 401 \pm 17 $ \\ 
\hline

J2314$+$0653WN & $ 413 \pm 10 $ & $ 608 \pm 10 $ & $ 1736 \pm 10 $ & $ 216 \pm 12 $ & $ 401 \pm 27 $ & $ 1087 \pm 25 $ & $ 953 \pm 20 $ & $ 416 \pm 13 $ & $ 371 \pm 10 $ \\ 
J2314$+$0653ES & $ 430 \pm 19 $ & $ 149 \pm 13 $ & $ 376 \pm 11 $ & $ 84 \pm 17 $ & $ 361 \pm 20 $ & $ 1451 \pm 25 $ & $ 712 \pm 19 $ & $ 339 \pm 16 $ & $ 227 \pm 14 $ \\ 
\hline

\end{tabular}
\label{emision flux} 
\begin{flushleft}
Notes: Fluxes in units of 10$^{-17}$ erg cm$^{-2}$ s$^{-1}$. The Fluxes of $\rm \mathrm{H}\beta$ and $\rm \mathrm{H}\alpha$ emission lines are both for the narrow line component.
\end{flushleft}
\end{threeparttable}
\end{table*}


\begin{table*}[!htp]
\scriptsize
\centering
\caption{BPT classifications}
\begin{threeparttable}
\begin{tabular}{ccccccccccc}
\hline
\hline

Name &  log ([O\,{\sc iii}]/H$_{\beta}$)   & log ([N\,{\sc ii}]/${\mathrm{H}\alpha})$ & log ([S\,{\sc ii}]/${\mathrm{H}\alpha})$  & log ([O\,{\sc i}]/${\mathrm{H}\alpha})$  &  BPT$_{\rm [N II]}$ & BPT$_{\rm [S II]}$ & BPT$_{\rm [O I]}$ & SDSS$_{\rm C}$ \\

\hline  
\multicolumn{9}{c}{First phase (6 Dual AGNs)}\\
\hline 

J0151$-$0245WN  & $ 0.23 \pm 0.02 $ & $ 0.53 \pm 0.09 $ & $ 0.59 \pm 0.09 $ & $ 0.23 \pm 0.08 $ &  AGN  &  LINER  &  LINER  & $ -- $ \\
J0151$-$0245ES  & $ 0.30 \pm 0.02 $ & $ 0.25 \pm 0.03 $ & $ 0.25 \pm 0.04 $ & $ -0.35 \pm 0.04 $ &  AGN  &  LINER  &  LINER  & $ -- $ \\ 
\hline

J0933$+$2114EN & BLA & BLA & BLA & BLA & BLA & BLA & BLA &  Type I AGN  \\         
J0933$+$2114WS  & BLA & BLA & BLA & BLA & BLA & BLA & BLA & $ -- $ \\   
  \hline  

J1017$+$3448WN  & $ 0.49 \pm 0.08 $ & $ -0.22 \pm 0.02 $ & $ -0.41 \pm 0.02 $ & $ -0.94 \pm 0.09 $ & AGN & Seyfert & Seyfert & $ -- $ \\ 
J1017$+$3448ES  & BLA & BLA & BLA & BLA & BLA & BLA & BLA & Type I AGN \\   
  \hline 
 
J1105+1957EN & $ -0.00 \pm 0.02 $ & $ -0.14 \pm 0.01 $ & $ -0.38 \pm 0.01 $ & $ -1.33 \pm 0.01 $ & Comp & H\,{\sc ii} & H\,{\sc ii} & $--$ \\         
J1105+1957WS & $ 0.32 \pm 0.06 $ & $ -0.15 \pm 0.01 $ & $ -0.24 \pm 0.01 $ & $ -0.87 \pm 0.04 $ & AGN & Seyfert & Seyfert & Seyfert \\
\hline

J1633$+$4718N  & $ 0.19 \pm 0.03 $ & $ -0.34 \pm 0.02 $ & $ -0.50 \pm 0.01 $ & $ -1.03 \pm 0.07 $ & Comp & H\,{\sc ii} & Seyfert & $ -- $ \\          
J1633$+$4718S  & BLA & BLA & BLA & BLA & BLA & BLA & BLA & Type I AGN \\          
 \hline

J2258$-$0115EN & $ 0.06 \pm 0.04 $ & $ -0.30 \pm 0.01 $ & $ -0.41 \pm 0.02 $ & $ -1.18 \pm 0.06 $ & Comp & H\,{\sc ii} & H\,{\sc ii} & $ -- $ \\ 
J2258$-$0115WS   & BLA & BLA & BLA & BLA & BLA & BLA & BLA & Type I AGN \\ 
\hline

\multicolumn{9}{c}{Second phase (10 Dual AGNs)}\\ 
 
\hline  
 
J0217$-$0845EN & $ -0.18 \pm 0.04 $ & $ 0.13 \pm 0.18 $ & $ 0.24 \pm 0.16 $ & $ 0.03 \pm 0.15 $ & AGN & LINER & LINER & $ -- $ \\ 
J0217$-$0845WS & $ 0.01 \pm 0.01 $ & $ 0.10 \pm 0.07 $ & $ 0.22 \pm 0.09 $ & $ -0.20 \pm 0.08 $  & AGN & LINER & LINER & LINER \\ 
 \hline

J0756$+$2340EN & $ -0.03 \pm 0.04 $ & $ -0.09 \pm 0.05 $ & $ -0.34 \pm 0.05 $ & $ -1.20 \pm 0.06 $ & Comp & H\,{\sc ii} & H\,{\sc ii} & $ -- $\\  
J0756$+$2340WS & $ 0.38 \pm 0.03 $ & $ 0.08 \pm 0.05 $ & $ 0.08 \pm 0.06 $ & $ -0.70 \pm 0.07 $ & AGN & LINER & LINER & LINER \\  
\hline

 J0813$+$4941WN  & BLA & BLA & BLA & BLA & BLA & BLA & BLA & Type I AGN \\
 J0813$+$4941ES   & $ -0.21 \pm 0.03 $ & $ -0.32 \pm 0.05 $ & $ -0.37 \pm 0.06 $ & $ -1.24 \pm 0.06 $ & Comp & H\,{\sc ii} & H\,{\sc ii} & $ -- $ \\ 
\hline

 J0833$+$1532WN & $ 0.75 \pm 0.09 $ & $ 0.20 \pm 0.08 $ & $ -0.09 \pm 0.10 $ & $ -0.73 \pm 0.08 $ & AGN & Seyfert & Seyfert & $ -- $ \\ 
 J0833$+$1532ES  & BLA & BLA & BLA & BLA & BLA & BLA & BLA & Type I AGN \\
 \hline

 J0848$+$3515EN  & $ 0.71 \pm 0.04 $ & $ -0.23 \pm 0.06 $ & $ -0.30 \pm 0.04 $ & $ -0.75 \pm 0.04 $ & AGN & Seyfert & Seyfert &  Seyfert \\    
 J0848$+$3515WS  & BLA & BLA & BLA & BLA & BLA & BLA & BLA & Type I AGN \\    
\hline

J0907$+$5203EN & $ 0.28 \pm 0.02 $ & $ -0.42 \pm 0.05 $ & $ -0.17 \pm 0.06 $ & $ -0.84 \pm 0.04 $ & Comp & LINER & LINER & Ambiguous AGN \\  
J0907$+$5203WS & $ 0.73 \pm 0.05 $ & $ -0.37 \pm 0.04 $ & $ -0.10 \pm 0.03 $ & $ -0.51 \pm 0.05 $ & AGN  & Seyfert & Seyfert & Seyfert \\ 
\hline

J1214$+$2931EN  & $ 0.69 \pm 0.02 $ & $ 0.07 \pm 0.01 $ & $ -0.10 \pm 0.03 $ & $ -0.64 \pm 0.01 $ & AGN & Seyfert & Seyfert & Seyfert \\
J1214$+$2931WS  & BLA & BLA & BLA & BLA & BLA & BLA & BLA & Type I AGN \\  
 \hline

J1645$+$2057WN & $ 0.28 \pm 0.04 $ & $ -0.05 \pm 0.09 $ & $ -0.32 \pm 0.06 $ & $ -1.07 \pm 0.06 $ & AGN & Seyfert & Seyfert & Seyfert \\ 
J1645$+$2057ES & $ 0.06 \pm 0.05 $ & $ -0.15 \pm 0.03 $ & $ -0.51 \pm 0.04 $ & $ -1.32 \pm 0.11 $ & Comp & H\,{\sc ii} & H\,{\sc ii} & Comp \\
\hline

J2206$+$0003WN & $ 0.24 \pm 0.05 $ & $ -0.03 \pm 0.06 $ & $ 0.10 \pm 0.05 $ & $ -0.51 \pm 0.05 $ & AGN & LINER & LINER & LINER \\
J2206$+$0003ES & $ 0.62 \pm 0.03 $ & $ 0.27 \pm 0.08 $ & $ 0.28 \pm 0.08 $ & $ -0.40 \pm 0.09  $ & AGN & LINER & LINER & LINER \\     
\hline

J2314$+$0653WN & $ 0.62 \pm 0.01 $ & $ -0.06 \pm 0.01 $ & $ -0.14 \pm 0.01 $ & $ -0.62 \pm 0.03 $ & AGN & Seyfert & Seyfert & Seyfert \\        
J2314$+$0653ES & $ -0.06 \pm 0.02 $ & $ -0.31 \pm 0.01 $ & $ -0.41 \pm 0.01 $ & $ -1.23 \pm 0.04 $ & Comp & H\,{\sc ii} & H\,{\sc ii} & $ -- $ \\

\hline 
      
\end{tabular}
\label{BPT classify} 
\begin{flushleft}

   Notes: BPT$_{\rm [N\,{\sc II}]}$, BPT$_{\rm [S\,{\sc II}]}$, BPT$_{\rm [O\,{\sc I}]}$: Result of BPT classifications in the log ([O\,{\sc iii}]/${\mathrm{H}\beta})$ -- log ([N\,{\sc ii}]/${\mathrm{H}\alpha})$, log ([O\,{\sc iii}]/${\mathrm{H}\beta})$ -- log ([S\,{\sc ii}]/${\mathrm{H}\alpha})$
   and log ([O\,{\sc iii}]/${\mathrm{H}\beta})$ -- log ([O\,{\sc i}]/${\mathrm{H}\alpha})$ planes, respectively.  
   Seyfert: Seyfert galaxy.
   BLA: Broad line AGN (Type I AGN).
   Comp: AGN/Star-forming composite galaxy.
   H\,{\sc ii}: Star--forming region.
  LINER: Low ionization nuclear emission line region.
  Ambiguous AGN: Galaxy of ambiguous classification, i.e. those classified as one type of AGN in one or two of the BPT diagram(s) but as another type of AGN in other remaining diagram(s).
  SDSS$_{\rm C}$: Classification based on BPT diagrams redrawed with the SDSS data.
\end{flushleft}
\end{threeparttable}
\end{table*} 


\begin{figure*}[hp]
  \centering
  \includegraphics[width=5.20cm,height=5.0cm]{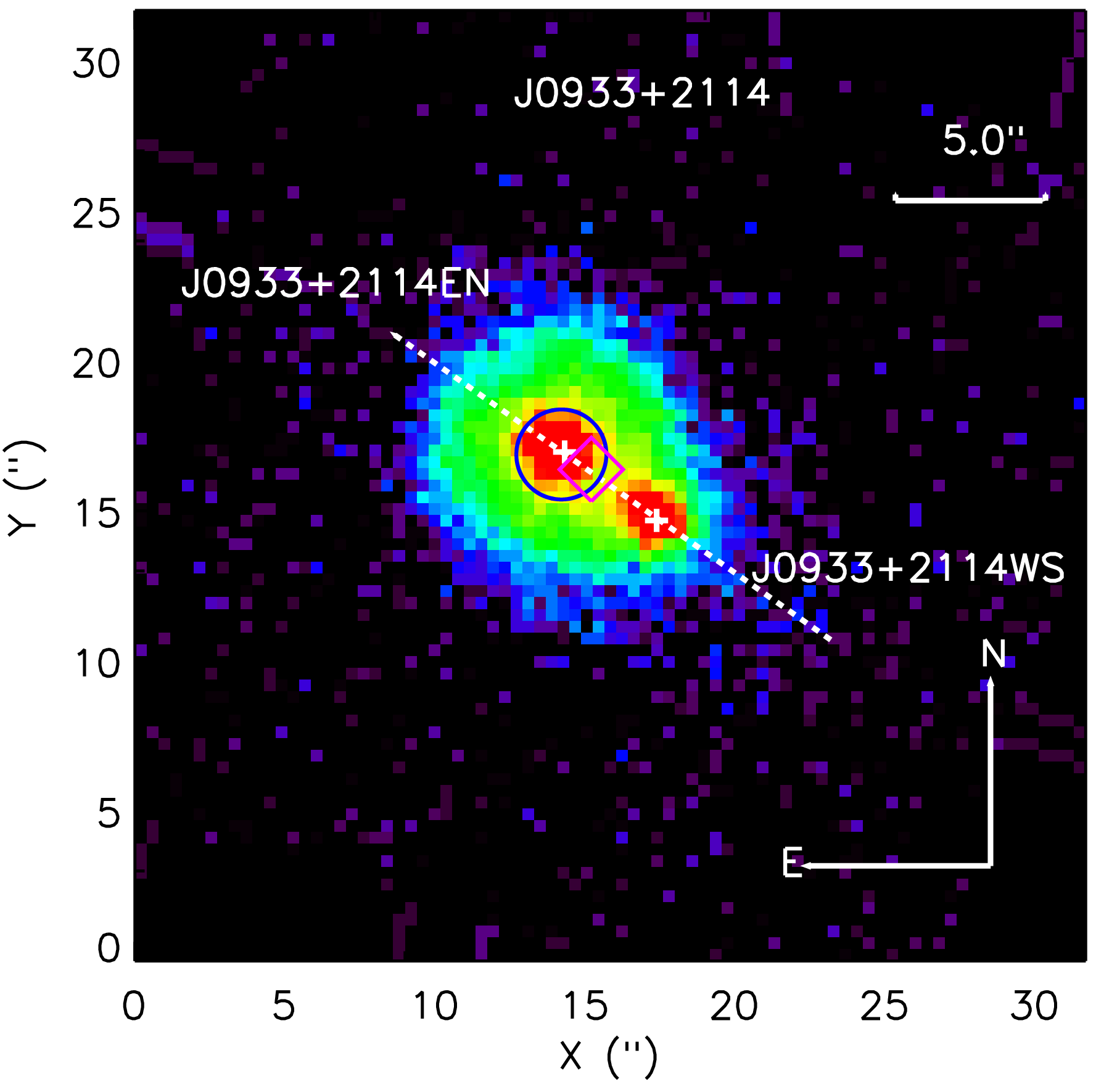}
  \includegraphics[width=12.2cm,height=5.0cm]{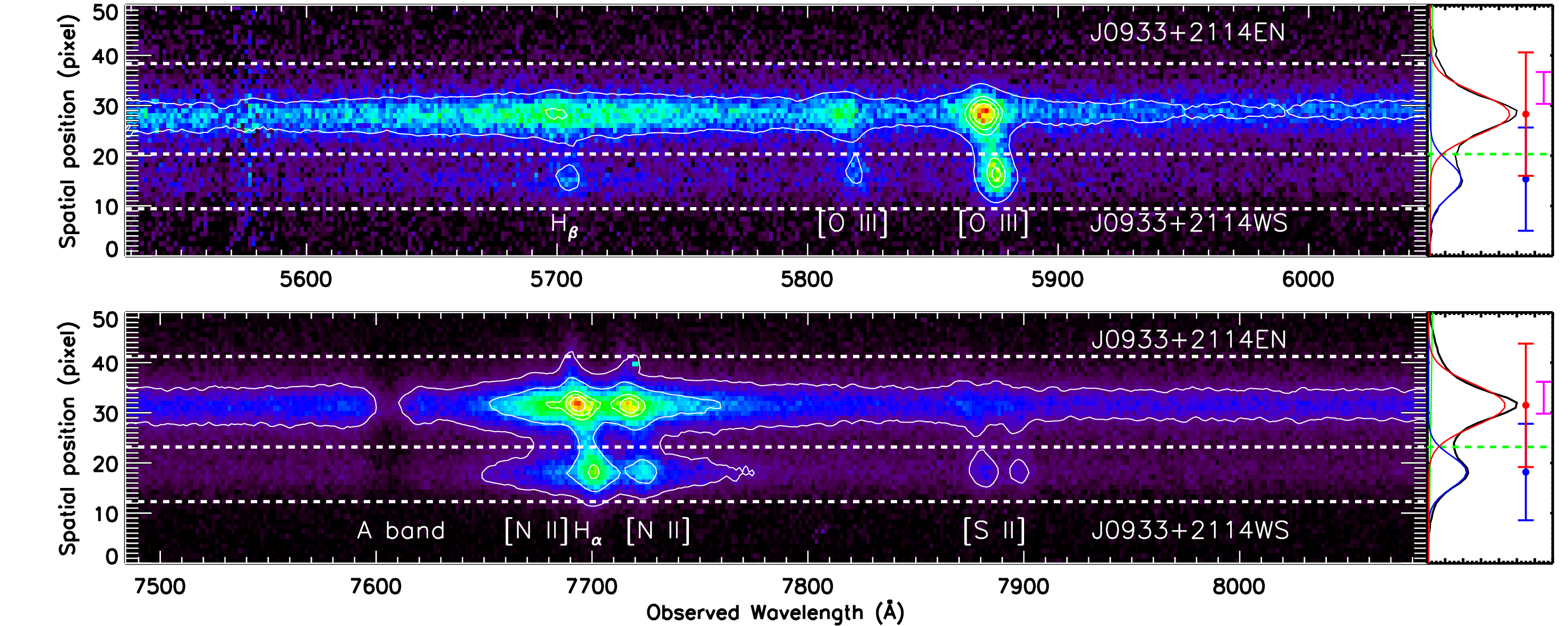}
\caption{ Left: Pseudo color image of J0933+2114 in SDSS $g$ band. The slit of our long-slit observation was positioned to cross the centers of J0933+2114EN and J0933+2114WS, as marked by the white dashed line. The position of the radio source given in the FIRST catalogue is marked by the magenta diamond. The fiber position of the SDSS spectrum is marked by the blue circle.
The white pluses indicate positions of the two optical cores.
Right: Segments of the two-dimensional long-slit spectrum of J0933+2114. 
The spectrum exhibits two sets of spatially resolved AGN spectra, corresponding to J0933+2114EN and J0933+2114WS, respectively. 
The white dashed lines set the boundaries used to extract the 1D spectra of the two cores. 
More details about extracting 1D spectra are given in Section 2.3.2.}
\label{slitimages of J0933+2114}
\end{figure*}

\section{Observation \& Data Reduction}
\label{Obs and data}

\subsection{Sample Selection}
\label{Sample Selection}


As mentioned above, ASTRO-DARING proposes an efficient method to search and identify dual AGNs amongst merging galaxies.
To do so, we first select galaxies that are potentially undergoing a merging process from the SDSS photometric catalog from DR9 (\citealt{Ahn2013}): galaxies exhibiting two optical cores\footnote{Both cores are classified as galaxy in SDSS DR9 (\citealt{Ahn2013}).} separated by less than 8$\arcsec$ and at least one optical core with SDSS fiber spectra from DR12/14 (\citealt{Alam2015, Abolfathi2018}).
Secondly, the candidates are selected with at least one radio detection from the FIRST survey (\citealt{Helfand2015}) in order to exclude physically unrelated pairs as merging galaxies\footnote{The FIRST survey has already released the final catalogue (\citealt{Helfand2015}) and presented definitive high-resolution 20 cm maps covering in total 10\,575 deg$^2$ sky area at 1.4 {\rm GHz}, using the Very Large Array (VLA) with a detection sensitivity of 1\,mJy.}. 
Typically, enhanced radio radiation are expected from merge-driven star formations or starbursts (e.g., \citealt{Sanders1988, Bell2006, Jogee2009, Robaina2009}).
Based on this, the requirement of radio detection can help remove the physically unrelated pairs from the merging galaxies to a certain extent.
Finally, only systems with at least one core previously identified as an AGN with redshift $z < 0.25$ by analyzing the SDSS DR12/14 spectra (\citealt{Paris2017, Paris2018}) are selected as dual AGN candidates.
In this manner, a total of 222 dual AGN candidates ($z < 0.25$) are selected amongst the merging galaxies sample from SDSS catalog. 
We note that the current sample has a rough limiting magnitude of $i$ band model magnitude $\sim$ 19.5-20 due to the redshift cut ($z < 0.25$), and the sample will miss dual AGN systems with radio radiation below the FIRST detection threshold (i.e., 1 mJy).
More details about sample selection are described in Paper I.

\subsection{Observations}

To identify the candidates as real dual AGNs, we have carried out the ASTRO-DARING campaign to collect spatially resolved long-slit spectroscopic observations of the candidates. 
The 2D spatially resolved long-slit spectra allow one to explore the activity nature of the two cores in the merging system (e.g., \citealt{Shen2011}) and thus help us diagnose whether the observed target is a real dual AGN or not.
The long-slit spectra are required to cover the emission lines of H$\beta$, [O {\sc iii}] $\lambda$5007, [O {\sc I}] $\lambda$6300, H$\alpha$, [N {\sc ii}] $\lambda\lambda$6549,6583 and [S {\sc ii}] $\lambda\lambda$6717,6731 in optical band with resolving power $R$ $>$ 300 (corresponding to a velocity resolution $>$ 1000 km\,s$^{-1}$), thus enabling us to identify AGN by either measuring the Full Width at Half Maximum (FWHM) of the Balmer emission lines or using the classical Baldwin-Phillips-Terlevich (BPT, \citealt{Baldwin1981}) diagrams with emission line ratios.

All spectroscopic observations reported here were obtained with the YFOSC mounted on the LJT of the YNAO as part of the ASTRO-DARING campaign.
YFOSC is a multi-mode instrument for both imaging and low/medium resolution spectroscopy, working at the Cassegrain focus (\citealt{Fan2015}). 
The CCD sensor is a 2k$\times$4k back-illuminated deep depletion chip, with a pixel size of 13.5\, $\mu$m that projects to 0.283 arcsec per pixel on the sky, covering a field of view of about 9.6$\times$9.6 square arcmin. 

Several grisms of different spectral resolutions and wavelength range are available for YFOSC.
The spectrograph can be used in long-slit mode with several slits of different widths available.
We note the overall sensitivity of the grisms\footnote{The wavelength of the adopted grisms are present in Table\,1.} (including telescope and detectors) are about 10 to 20 per cent and even $> 5$ per cent at the blue beginning and red end.
For each candidate, the two optical cores are spatially resolved in the SDSS images and the slit is set to pass through the two cores of the merging galaxies. 
For this purpose, the spectrograph is rotated with a specific position angle (PA) for each galaxy pair and the values of PA for the individual systems are presented in Table\,\ref{obs_log}.
In this way, we obtain 2D spatially resolved spectra for the two cores in the merging systems, under good seeing condition.

The ASTRO-DARING project was started in 2014 November.
All observations presented here were carried out with the YFOSC in long-slit mode using several slits of fixed widths, depending on the observing condition, especially the seeing and transparency.
Depending on the grisms used, our observations can be divided into two phases.
The first phase is from November 2014 to November 2015.
During this period, we mainly use two grisms: a blue one G14 covering 3600 to 7500 \AA \, and a red one G8 covering 5100 to 9600 \AA \,.  
A slit of width 1.8$\arcsec$ matching the typical seeing was used during this phase.
With the above combinations, the typical resolution are about 330 $\pm$ 20 km $\mathrm{s}^{-1}$, 500 $\pm$ 40 km $\mathrm{s}^{-1}$ for G8, G14 Grisms respectively (Table\,\ref{slit}).
In total, 15 targets were observed during this period.

Due to the increasing seeing and decreasing transparency (possibly casused by the El Ni$\rm \widetilde{n}$o effects\footnote{El Ni$\rm \widetilde{n}$o effects results from the rise of sea surface temperatures in the Indian Ocean. It may not bring drought but cause heavy rainfall at certain periods.}), we changed to use G3 that has a higher efficiency than G14 and G8 but a lower resolution in the second phase (from November 2015 to February 2017). The spectra obtained cover the whole optical wavelength range (3400 to 9100 \AA \,)\footnote{We note that the effect of the second-order contamination at the  wavelength smaller than 4550\,\AA\,is largely smaller than 5-10 per cent. Moreover, the current analysis is not affected by this issue since all of our concerned emission lines are longer than 4550\,\AA.}.
Depending on the seeing condition, a slit of width 1.8\arcsec or 2.5\arcsec was adopted.
The resulting spectral resolution was relatively low with FWHMs of $730 \pm 60$ km $\mathrm{s}^{-1}$, $970 \pm 70$ km $\mathrm{s}^{-1}$, respectively, for the 1.8\arcsec\,, 2.5\arcsec\, slits of G3 Grism.
In this phase, a total of 26 targets were observed.

Totally, we obtained long-slit spectra for 41 candidates between November 2014 and February 2017 using the YFOSC. 
For more details, see the observational log in Table\,\ref{obs_log}.


\subsection{Data Reduction}
\subsubsection{Wavelength \& Flux calibration}

The data were reduced with IRAF\footnote{http://ast.noao.edu/data/software} and IDL routines. 
The 2D spectra were bias subtracted, flat-fielded calibrated, cosmic--rays fully removed (using the IRAF task \texttt{crmedian}), wavelength calibrated, and flux calibrated on 2D with IRAF.
The wavelength and flux calibrations were applied to all 2D spectra. 
The wavelength distortions in the spatial direction were carefully corrected using 2D wavelength map constructed by arc frame performed on IRAF.
The ESO spectroscopic standard stars observed in the same night was applied to the flux calibrations.
The sky background is subtracted in 2D manner by the \texttt{background} implemented in IRAF.
The above steps were all carried out with IRAF for the 2D spectra. 
After those, the IDL routine is adopted to extract the final 1D spectra. More details about this step are presented below.

\subsubsection{Extraction of 1D spectra}
\label{extraction}

To explore the activity natures of the two cores in the merging system, 1D spectra of the two cores in the system are extracted from the spatially resolved 2D spectrum.
When extracting the 1D spectra, flux loss and contamination are unavoidable as the spectra of the two cores are close to each other. 
We therefore developed a 1D extraction method by achieving a tradeoff between reducing flux loss and contamination.

An example of the 1D extraction is shown in Fig.\,\ref{slitimages of J0933+2114}. 
The flux distribution along the spatial direction is contributed by the two galaxies/AGNs and the sky background residuals.
We assumed that the contribution from each of the two nuclei can be represented by a Gaussian, while that of the residual sky background can be represented by a first-order polynomial accounting for any systematic pattern left in the subtracted background. 
To extract the spectrum of each galaxy/AGN, the flux distribution is fit by two Gaussians plus a first-order polynomial and the 1D spectra of each two galaxy/AGN are obtained as shown in the right panel of Fig.\,\ref{slitimages of J0933+2114}.
The inner/outer boundaries in Fig.\,\ref{slitimages of J0933+2114} is set when the 1D spectra of two cores are extracted from the 2D spectrum.
The outer boundary of each of the two galaxies/AGNs is defined as the position of two-sigma width from the center of the fitted Gaussian. 
This was chosen in order to enclose as much of the flux from the target but as possible with minimum noise from the background (see the two white dashed lines on the two-sides of the galaxies/AGNs in the right panel of Fig.\,\ref{slitimages of J0933+2114}).      
For the central overlapping region of the two galaxies/AGNs, the position of equal flux of the two fitted Gaussians is selected as the inner boundary to minimize the flux loss and contamination of each target (see the middle white dashed line in the right panel of Fig.\,\ref{slitimages of J0933+2114}).
Given the above extraction procedures, the true flux of one target is assumed by the total flux under the fitted Gaussian and then the flux loss is given by the total area beyond the inner and outer boundaries of the fitted Gaussian. The flux contamination is defined as the total flux tail between the extracted inner and outer boundaries, extended from the nearby target (true flux also represented by another fitted Gaussian).
We note that the extraction limits (the inner/outer boundaries) are for purposes of summing across the spatial dimension.

\begin{figure}[tp]
\centering

\subfigure{\includegraphics[width=8.0cm,height=7.0cm]{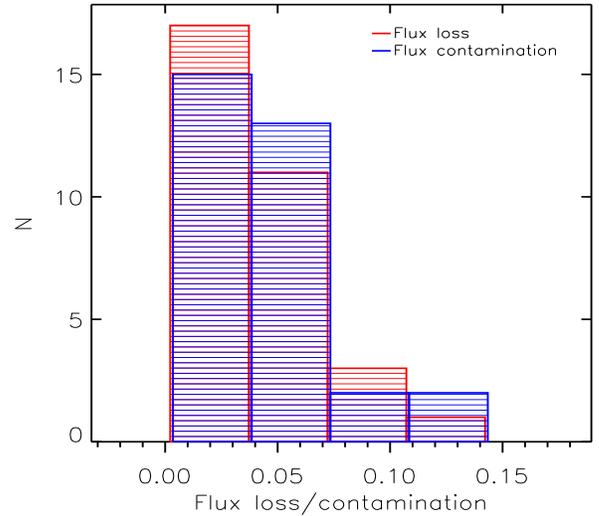}}

\caption{Distribution ratios of flux loss and contamination of the extracted 1D spectra. The flux loss ratio is given by the ratio between the flux loss and the true flux of specified target and the flux contamination ratio is defined as the flux contamination to the actual extracted flux between the inner and outer boundaries of specified target (see Fig. 1).} 

\label{flux_loss_cont}
\end{figure}

\begin{figure*}[htp]
\centering
  \includegraphics[scale=0.51]{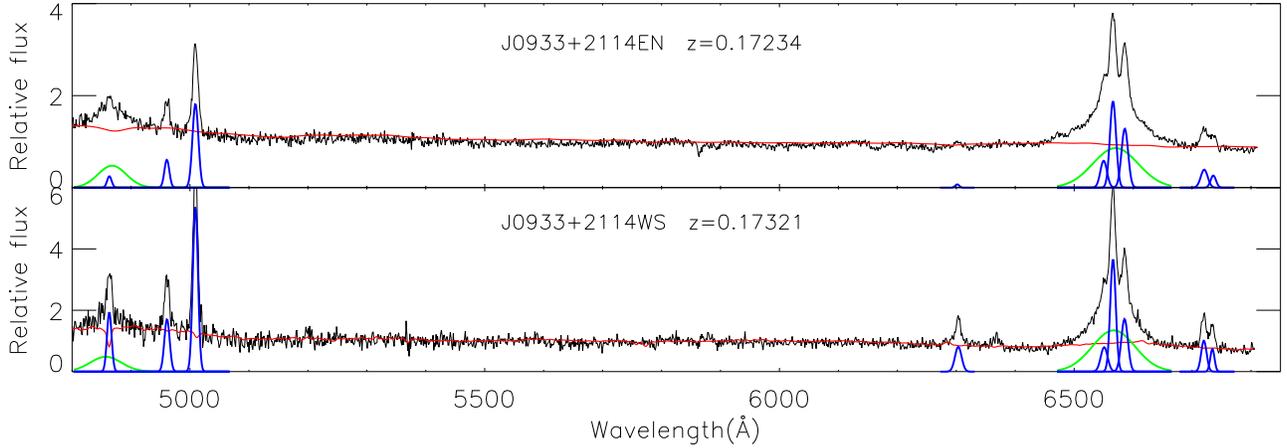}
\caption{The LJT spectra of dual AGN: J0933+2114 respectively for J0933+2114EN (top) and J0933+2114WS (bottom) showing emission lines H$\beta$, [O\,{\sc iii}]$\lambda\lambda$4959,5007, [O\,{\sc i}]$\lambda$6300, H$\alpha$, [N\,{\sc ii}] $\lambda\lambda$6549,6583 and [S\,{\sc ii}]$\lambda\lambda$6717,6731 at rest-frame wavelengths. We used pPXF to subtract continuous spectrum and got emission lines. The red lines represent the continuum component of host galaxy, the blue lines represent the narrow emission lines component and the green lines represent the broad emission lines component.} 
\label{spectra_fitting:J0933+2114}
\end{figure*}

To account for the effects of possible slit tilt, the whole long-slit spectrum is divided into six slices\footnote{The tilt angle of YFOSC slits are largely adjusted to 0.3-0.5 degrees and thus the tilt changes are within 1-3 pixels along the whole slices (each with a length about 300 pixels). In addition, six slices are enough to account for the point spread function (PSF) along the wavelength direction (mostly within 1-2 pixels).} along the wavelength direction and the 1D spectra in each slice are extracted with the above procedure.
Given the relatively large angular separations of our selected candidates, the spectra of the two cores in merging galaxies are well separated in general and their 1D spectra can be extracted with minimum flux loss and contamination.
To quantitatively show the robustness of our 1D spectral extraction procedure, we calculate the averaged flux loss and contamination ratios of the six slices for our observed targets.
The distributions of flux loss and contamination of the 1D spectra of our dual AGN candidates are shown in Fig.\,\ref{flux_loss_cont}. 
In general, the flux loss and contamination amount to only a few per cent.
This is corroborated by the high-quality example spectra presented in Fig.\,\ref{spectra_fitting:J0933+2114}.
Part of our targets have multiple observations and their 1D spectra are stacked via their signal-to-noise ratios (SNRs).

  
\subsubsection{Subtraction of contribution of host galaxy}
 
The continuum and absorption features of host galaxy should be subtracted properly from the extracted spectra before deriving the properties of emission lines.
To subtract the continuum and absorption features, the penalised PiXel Fitting software (pPXF, \citealt{Cappellari2004}, \citealt{Cappellari2017}) was applied to our long-slit spectra.
This code allows one to fully decompose gas emission from stellar absorption features, using a maximum penalized likelihood approach.
In pPXF, the MILES library\footnote{This library contains 985 flux well-calibrated stellar spectra with wavelength range from 3525 \AA\, to 7500 \AA, with a spectral resolution of FWHM $\sim$ 2.51 \AA, $\sigma$ $\sim$ 64 km $\mathrm{s}^{-1}$.} (\citealt{Sanchez2016}) with wide ranges of physical parameters was adopted as template to model the contribution from host galaxy.
Before fitting, all the template spectra were properly convolved to the resolution of our long-slit spectra and all the prominent emission lines were masked.
As an example, the performance of this subtraction was shown in Fig.\,3 for J0933+2114 system (as well as in the Appendix\,B for more examples).
The consistent results of flux ratio and FWHM measurements from LJT long-slit spectra and SDSS fiber spectra as checked in Section 3.3 show that the host subtraction by pPXF algorithm is generally reasonable for our low-resolution LJT spectra.

\subsubsection{Detections of emission lines}

\begin{figure*}[htp]
  \centering
  \includegraphics[width=8.8cm,height=4.1cm]{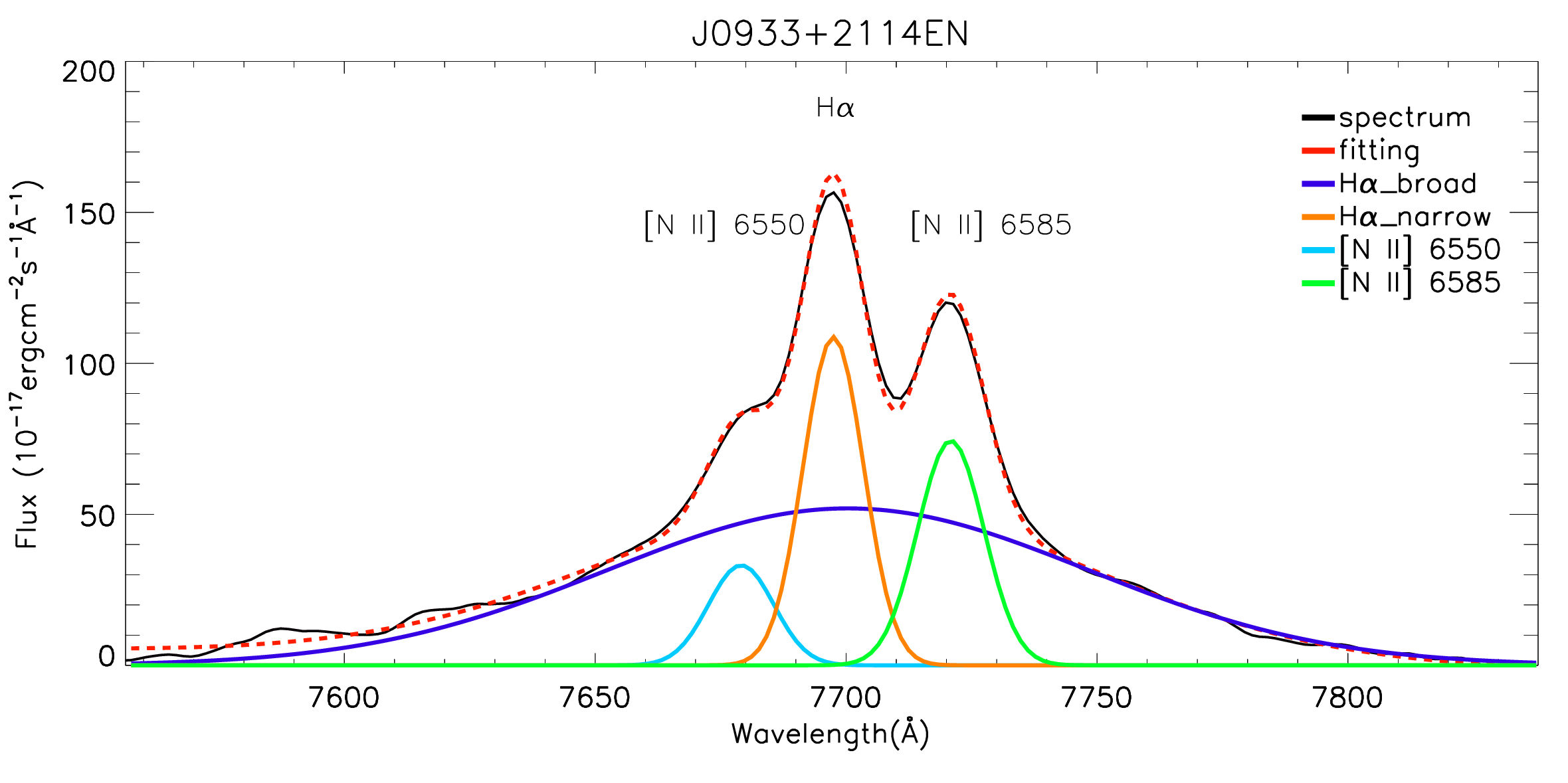}
  \includegraphics[width=8.8cm,height=4.1cm]{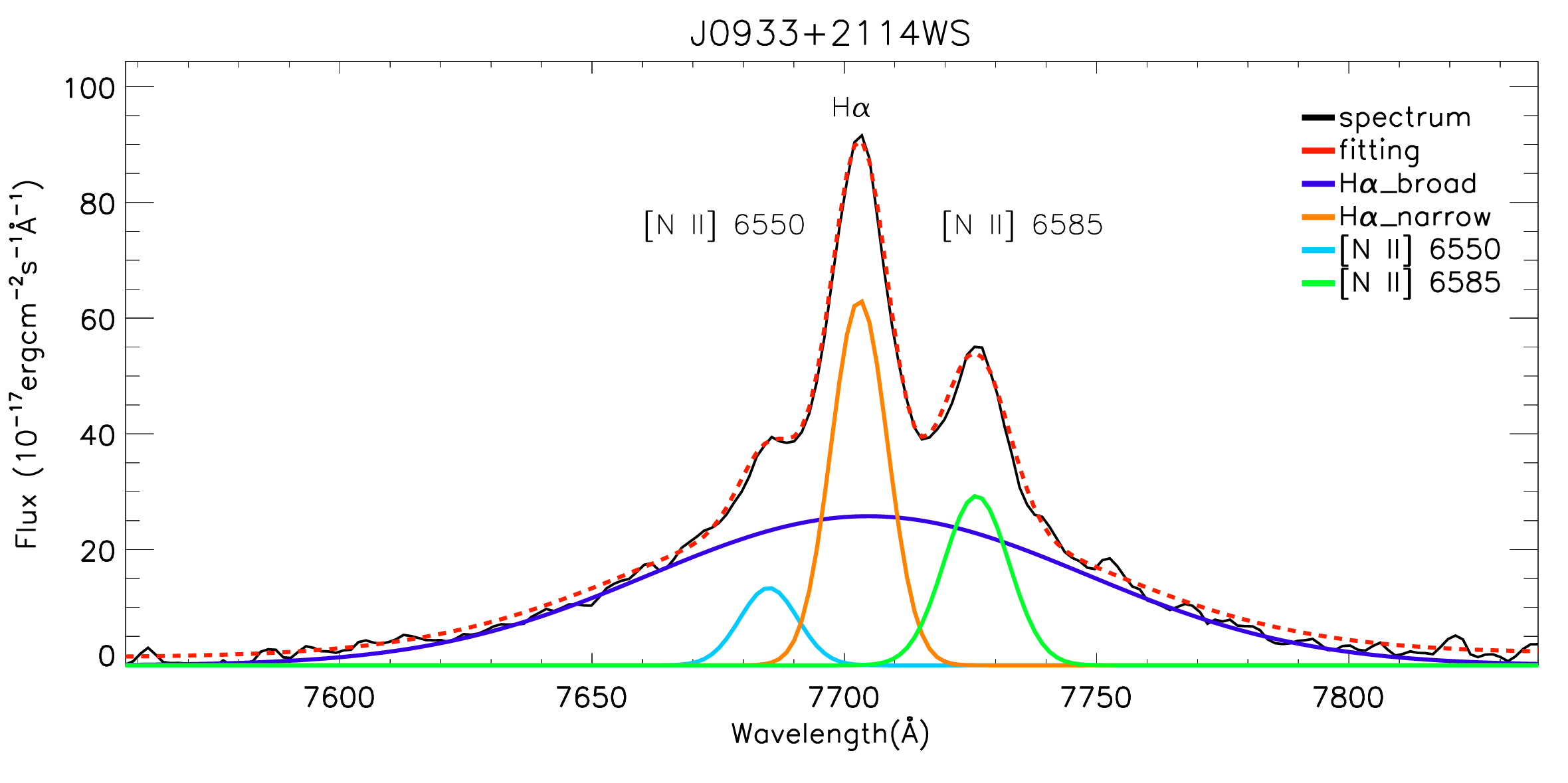}
\caption{Spectra fitting for J0933+2114EN (left) and J0933+2114WS (right) near the H$\alpha$ region. Both galaxies have a broad ${\mathrm{H}\alpha}$ component. We use one Gaussian for [N~{\sc ii}] $\lambda$6549, another Gaussian for [N~{\sc ii}] $\lambda$6583 and a pair of Gaussians for the H${\alpha}$ broad and narrow components that have the same central wavelength, to fit the whole spectral region.}
\label{spectrum fitting of J0933+2114}
\end{figure*}

Typically, 9 strong optical emission lines (i.e., H${\beta}$, [O\,{\sc iii}] $\lambda\lambda$4959, 5007, [O\,{\sc i}] $\lambda$6300, H${\alpha}$, [N\,{\sc ii}] $\lambda\lambda$6549, 6583, [S\,{\sc ii}] $\lambda\lambda$6717, 6731) can be detected from the spectra with continuum and absorption features properly subtracted by pPXF described in above section.
Line properties (e.g., flux, FWHM, central wavelength) can be derived by an IDL fitting procedure \texttt{MPFIT}\footnote{https://pages.physics.wisc.edu/~craigm/idl/fitting.html} (Craig Markwardt 2009).
We fit each component of detected emission lines by single or multiple Gaussian(s) to obtain their fluxes, central wavelengths and FWHMs\footnote{The FWHM values reported in Table\,\ref{finally DAGN} have been corrected for the instrumental broadening effect.}.   
Here we note that a emission line could be detected only with signal-to-noise ratio (SNR; defined as the ratio between the peak of emission line and the standard deviation of the continuum nearby the emission line) greater than 3.
For the case shown in Fig.\,\ref{spectrum fitting of J0933+2114}, we use one Gaussian for the [N~{\sc ii}] $\lambda$6550 line, one Gaussian for the [N~{\sc ii}] $\lambda$6585 line and a pair of Gaussians with the same central wavelength for the broad and narrow H${\alpha}$ components. 
If only a narrow component of the H${\alpha}$ emission detected, then only one Gaussian is used to fit it.
Similar fitting strategy is used for other detected emission lines. 
The error of each parameter is given by the covariance matrix provided by \texttt{MPFIT}.
The fitting results of all well detected emission lines are presented in Table\,\ref{emision flux} and Table\,\ref{emission_line:FWHM}.

\begin{figure*}[htp]
\centering
\includegraphics[width=18cm,height=5.6cm]{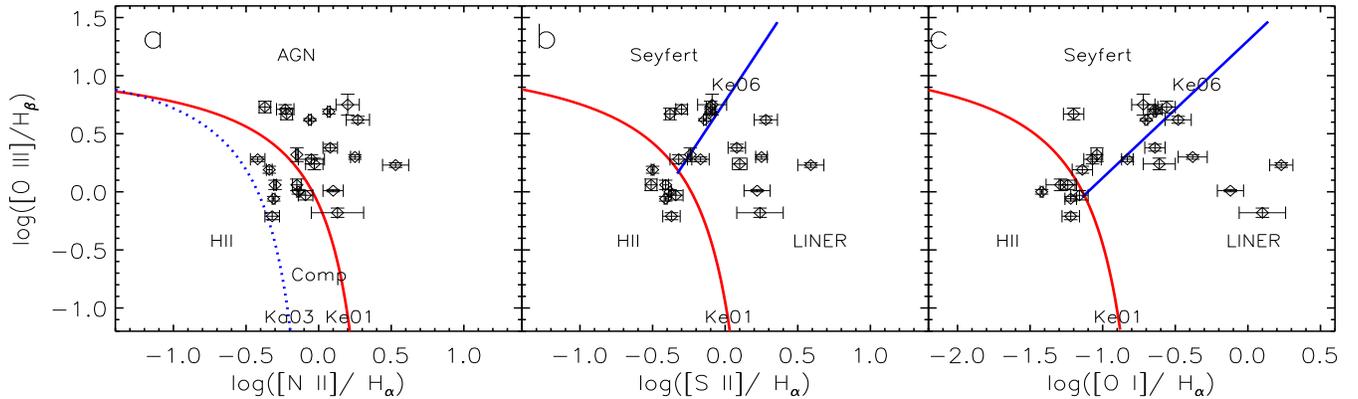}  
\caption{BPT diagrams of our dual AGNs. Detailed information about the classifications of the individual galaxies is shown in Table\,\ref{BPT classify}.}
\label{BPT_DAGN}
\end{figure*}

\begin{table*}[ht]
\scriptsize
\centering
\caption{Dual AGNs}
\begin{threeparttable}
\begin{tabular}{cccccccccccc}
\hline
\hline

Name &   AGN & $Z_{\rm m}$  & FWHM$_{\rm NLR}$ & FWHM$_{\rm BLR}$   & $V_{\rm offset}$   & Sep & $W_{1}$-$W_{2}$ & Radio  & Classification   \\
     &          &      &    (km $\mathrm{s}^{-1}$)  &    (km $\mathrm{s}^{-1}$)  &    (km $\mathrm{s}^{-1}$)  &    $\rm (''/kpc) $ & mag  &    \\

\hline  
\multicolumn{9}{c}{First phase (6 Dual AGNs)}\\
\hline 
\multirow{2}{*}{J015107.39-024526.87}   & J0151$-$0245WN   &$0.04774 \pm 0.00008 $  &$262 \pm 63 $  &$--$  & \multirow{2}{*}{$ 57 \pm 28 $}   & \multirow{2}{*}{ 5.1/4.8 }  &    \multirow{2}{*}{ 0.030 }    & $\rm N $ & LINER    \\
                            & J0151$-$0245ES   &$0.04793 \pm 0.00005 $  &$471 \pm 36 $  &$--$  &      &    &      & $\rm Y $  & LINER    \\
 \hline 

\multirow{2}{*}{J093347.76+211436.41} & J0933$+$2114EN    &$0.17234  \pm 0.00005 $  &$629 \pm 46 $  &$5091 \pm 153$  & \multirow{2}{*}{$ 261 \pm 21 $}   & \multirow{2}{*}{ 4.2/12.3 } &   0.714    & $\rm Y $ &  Type I AGN  \\
   & J0933$+$2114WS     &$0.17321 \pm 0.00005 $  &$548 \pm 53 $  &$4956 \pm 155$  &               &     &  0.873  & $\rm N $  & Type I AGN    \\

\hline  
 \multirow{2}{*}{J101757.07+344846.61}  &  J1017$+$3448WN    &$0.14321  \pm 0.00005 $  &$778 \pm 41 $  &$--$  & \multirow{2}{*}{$ 243 \pm 21 $}   & \multirow{2}{*}{ 5.6/14.2 } & 0.993   & $\rm Y $ &  Seyfert    \\
  & J1017$+$3448ES     &$0.14402 \pm 0.00005 $  &$614 \pm 48 $  &$3255 \pm 243 $  &               &     & 0.872 & $\rm N $  & Type I AGN     \\
  
\hline  
\multirow{2}{*}{J110544.48+195750.06}  &  J1105$+$1957EN   &$0.10421  \pm 0.00005 $  &$487 \pm 46 $  &$--$  & \multirow{2}{*}{$ 84 \pm 21 $}   & \multirow{2}{*}{ 3.9/7.4 } &  \multirow{2}{*}{ 0.329 }  & $\rm Y $ &  Comp  \\
  &  J1105$+$1957WS   &$0.10449 \pm 0.00005 $  &$399 \pm 50 $  &$--$  &               &     &      & $\rm N $  & Seyfert    \\
\hline  
\multirow{2}{*}{J163323.58+471858.95}   & J1633$+$4718N   &$ 0.11525 \pm 0.00005 $  &$413 \pm 71 $  &$--$  & \multirow{2}{*}{$ 153 \pm 21 $}   & \multirow{2}{*}{ 3.8/8.0 } &  \multirow{2}{*}{ 0.812 }  & $\rm N $ &  Ambiguous AGN    \\
   & J1633$+$4718S    &$0.11576 \pm 0.00005 $  &$517 \pm 57 $  &$ 1989 \pm 248 $  &          &    &       & $\rm Y $  & Type I AGN \\
\hline  
\multirow{2}{*}{J225810.01-011516.26}  & J2258$-$0115EN   &$0.11609  \pm 0.00005 $  &$738 \pm 43 $  &$--$  & \multirow{2}{*}{$ 300 \pm 21 $}   & \multirow{2}{*}{ 3.2/6.9 } & \multirow{2}{*}{ 0.652 }  & $\rm Y $ &  Comp    \\
   & J2258$-$0115WS   &$0.11709 \pm 0.00005 $  &$650 \pm 46 $  &$2929 \pm 347 $  &               &     &      & $\rm N $  & Type I AGN    \\
\hline
\multicolumn{9}{c}{Second phase (10 Dual AGNs)}\\ 
 
\hline  
\multirow{2}{*}{J021703.81-084515.97}  & J0217$-$0845EN   &$0.1078 \pm 0.0001 $  &$339 \pm 237 $  &$--$  & \multirow{2}{*}{$90 \pm 40 $}   & \multirow{2}{*}{ 6.0/12.0 } & 0.191  & $\rm N $ & LINER    \\
  & J0217$-$0845WS   &$0.1081 \pm 0.0001 $  &$670 \pm 164 $  &$--$  &               &     &  0.202   & $\rm Y $  & LINER    \\
  
\hline 	
\multirow{2}{*}{J075621.37+234043.97}  &   J0756$+$2340EN    &$0.0745 \pm 0.0001 $  &$806 \pm 151 $  &$--$  & \multirow{2}{*}{$120 \pm 40 $}   & \multirow{2}{*}{ 6.8/9.6 } & \multirow{2}{*}{ 0.255 }  & $\rm Y $ & Comp    \\
   &   J0756$+$2340WS    &$0.0741 \pm 0.0001 $  &$565 \pm 195 $  &$--$  &               &    &        & $\rm N $       & LINER    \\
\hline 
\multirow{2}{*}{J081347.49+494109.83}  &  J0813$+$4941WN    &$0.0942 \pm 0.0001 $  &$804 \pm 138 $  &$2640 \pm 123 $  & \multirow{2}{*}{$150 \pm 40 $}   & \multirow{2}{*}{ 3.3/5.8 } & \multirow{2}{*}{ 0.935 }  & $\rm Y $ & Type I AGN    \\
   &  J0813$+$4941ES    &$0.0947 \pm 0.0001 $  &$521 \pm 212 $  &$--$  &               &     &       & $\rm N $       & Comp \\
\hline
\multirow{2}{*}{J083355.49+153236.62}   & J0833$+$1532WN    &$0.1514 \pm 0.0001 $  &$427 \pm 285 $  &$--$  & \multirow{2}{*}{$60 \pm 40 $}   & \multirow{2}{*}{ 4.5/11.9 } &  \multirow{2}{*}{ 0.768 } & $\rm Y $ & Seyfert  \\
   & J0833$+$1532ES    &$0.1516 \pm 0.0001 $  &$486 \pm 227 $  &$ 3322 \pm 167 $  &               &      &      & $\rm N $       &  Type I AGN \\
\hline
\multirow{2}{*}{J084809.69+351532.12} & J0848$+$3515EN   &$0.0573 \pm 0.0001 $  &$496 \pm 272$  &$--$  & \multirow{2}{*}{$60 \pm 40 $}   & \multirow{2}{*}{ 5.6/6.2 } &  \multirow{2}{*}{ 0.632 }  & $\rm N $ & Seyfert  \\
   &   J0848$+$3515WS   &$0.0571 \pm 0.0001 $  &$498 \pm 221 $  &$2494 \pm 126 $  &               &     &       & $\rm Y $       &  Type I AGN\\
\hline
\multirow{2}{*}{J090714.61+520350.61}  &   J0907$+$5203EN    &$0.0601 \pm 0.0001 $  &$815 \pm 102$  &$--$  & \multirow{2}{*}{$150 \pm 40 $}   & \multirow{2}{*}{ 7.4/8.5 } & 0.429   & $\rm Y $ & Ambiguous AGN    \\
   &  J0907$+$5203WS   &$0.0596 \pm 0.0001 $  &$709 \pm 117 $  &$--$  &               &     & 0.368   & $\rm N $       &  Seyfert \\

\hline
\multirow{2}{*}{J121418.25+293146.70}  &  J1214$+$2931EN    &$0.0633  \pm 0.0001 $  &$576 \pm 191 $  &$--$  & \multirow{2}{*}{$ 60 \pm 40 $}   & \multirow{2}{*}{ 7.9/9.3 } &  0.059  & $\rm N $ &  Seyfert   \\
  &  J1214$+$2931WS    &$0.0635 \pm 0.0001 $  &$648 \pm 170 $  &$2417 \pm 117 $  &               &          &  1.274 & $\rm Y $  & Type I AGN    \\
  
\hline  
\multirow{2}{*}{J164507.91+205759.43}   &   J1645$+$2057WN    &$0.1301 \pm 0.0001 $  &$644 \pm 85$  &$--$  & \multirow{2}{*}{$270 \pm 40 $}   & \multirow{2}{*}{ 4.2/9.8 } &   \multirow{2}{*}{ 0.569 }  & $\rm N $ & Seyfert   \\
   &   J1645$+$2057ES    &$0.1310 \pm 0.0001 $  &$833 \pm 64 $  &$--$  &               &      &      & $\rm Y $   &  Comp \\
\hline
\multirow{2}{*}{J220634.97+000327.57}   &   J2206$+$0003WN     &$0.0466 \pm 0.0001 $  &$889 \pm 157 $  &$--$  & \multirow{2}{*}{$120 \pm 40 $}   & \multirow{2}{*}{ 4.7/4.3 } &  \multirow{2}{*}{ 0.107 }  & $\rm N $ & LINER    \\
   &   J2206$+$0003ES     &$0.0462 \pm 0.0001 $  &$650 \pm 185 $  &$--$  &               &      &      & $\rm Y $       &  LINER \\
\hline
\multirow{2}{*}{J231439.21+065312.97}   &   J2314$+$0653WN    &$0.0876 \pm 0.0001 $  &$393 \pm 141$  &$--$  & \multirow{2}{*}{$90 \pm 40 $}   & \multirow{2}{*}{ 4.1/6.7 } &  0.679  & $\rm Y $ & Seyfert   \\
   &   J2314$+$0653ES     &$0.0879 \pm 0.0001 $  &$419 \pm 133 $  &$--$  &               &      & 0.280  & $ \rm N $       &  Comp \\
\hline

\end{tabular}
\label{finally DAGN} 
\begin{flushleft}

   Notes: $Z_{\rm m}$: Median redshift of main emission lines detected with good SNRs.
          FWHM$_{\rm NLR}$: FWHM of narrow line region represented by the FWHM of [O\,{\sc iii}] $\lambda$5007 line.
          FWHM$_{\rm BLR}$: FWHM of broad line region represented by the FWHM of H${\alpha}$ broad component.
          $V_{\rm offset}$: Velocity offset of the two AGN cores.
          Sep: Separation of the two AGN optical cores in arcsec and in kpc.
          $W_{1}$-$W_{2}$: Magnitude difference in WISE $W_{1}$ and $W_{2}$ bands for the two resolved/unresolved cores.
          Radio: N\,(NO), Y\,(YES), whether detected as the radio-excess AGN result from radio power versus H$\alpha$ luminosity.
          SyII: Seyfert 2.
          Classification: Classifications of the two AGN optical cores.

\end{flushleft}
\end{threeparttable}
\end{table*}

  
\begin{figure*}[htp]
\centering

\subfigure{\includegraphics[scale=0.56]{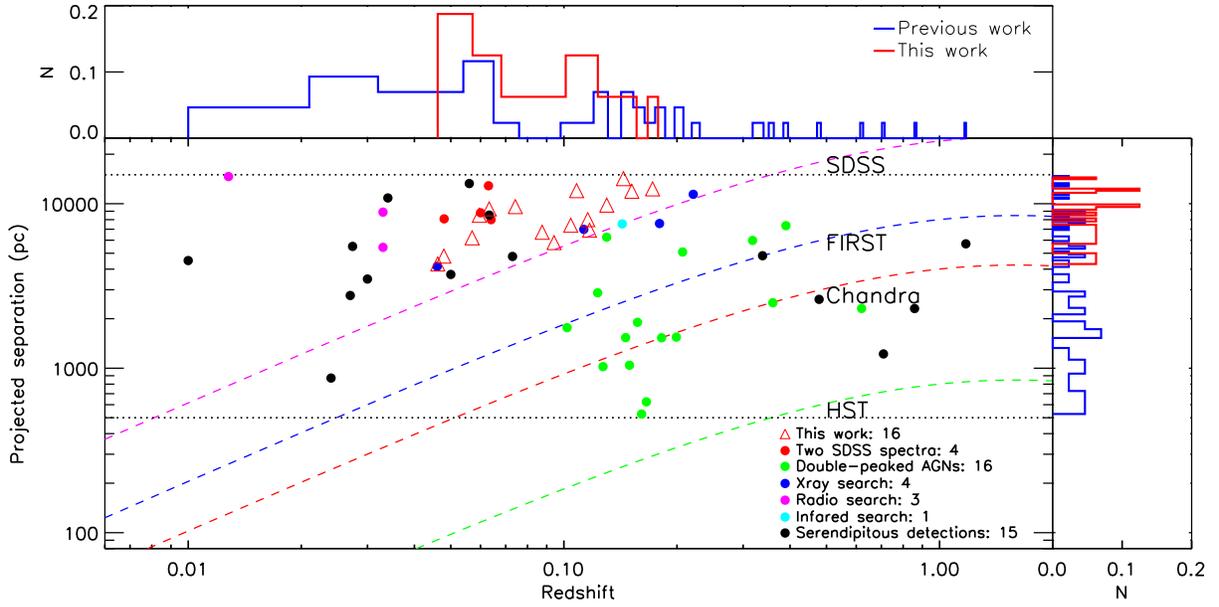}}

\caption{Dual AGNs from the current and previous work. The magenta, blue, red, green dashed lines, respectively, represent the critical resolutions of SDSS spectroscopy, FIRST, Chandra, HST catalogs, i.e., 3.0\arcsec, 1.0\arcsec, 0.5\arcsec, 0.1\arcsec, respectively. The two horizontal dashed lines are the typical separations, i.e., 0.5 and 15\,kpc, of dual AGNs, respectively. The top sub-panel shows the number distribution along the redshift for the dual AGNs found previously (blue line) and in this work (red line), respectively. 
A gap of the distribution of the dual AGNs from previous work now is filled by our sample significantly.
The right sub-panel shows the number distribution along the projected separation for the dual AGNs found previously (blue line) and in this work (red line), respectively.} 

\label{The_known_DAGNs}
\end{figure*}

  
\begin{figure}[htp]
\centering

\subfigure{\includegraphics[width=8.5cm,height=7.5cm]{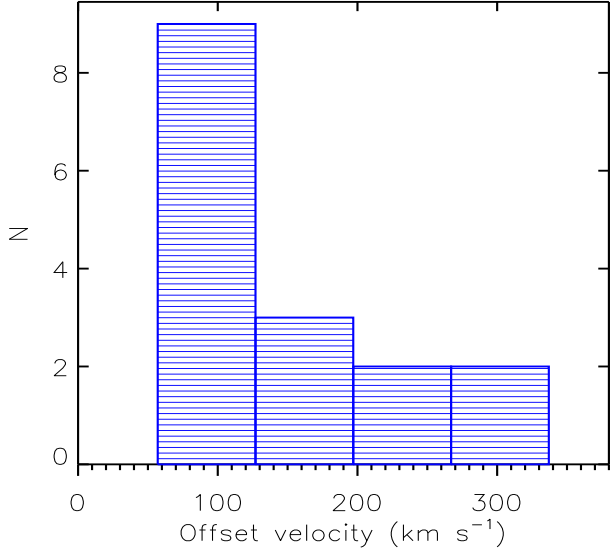}}

\caption{Velocity offsets of the two cores for our dual AGN sample.} 

\label{offset_velocity}
\end{figure}


\begin{figure*}[htp]
  \centering
  \includegraphics[scale=0.58]{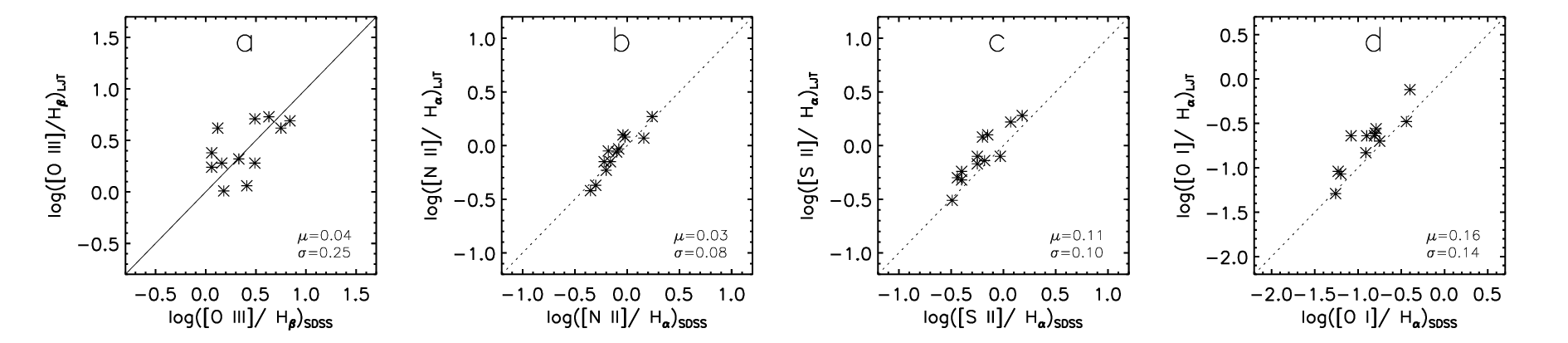}

\caption{Comparisons of flux ratios as measured from our long-slit and SDSS fiber spectra, for, from left to right, a: log([O\,{\sc iii}]/${\mathrm{H}\beta})_{\rm LJT}$ v.s. log([O\,{\sc iii}]/${\mathrm{H}\beta})_{\rm SDSS}$. b: log([N\,{\sc ii}]/${\mathrm{H}\alpha})_{\rm LJT}$ v.s. log([N\,{\sc ii}]/${\mathrm{H}\alpha})_{\rm SDSS}$. c: log([S\,{\sc ii}]/${\mathrm{H}\alpha})_{\rm LJT}$ v.s. log([S\,{\sc ii}]/${\mathrm{H}\alpha})_{\rm SDSS}$. d: log([O\,{\sc i}]/${\mathrm{H}\alpha})_{\rm LJT}$ v.s. log([O\,{\sc i}]/${\mathrm{H}\alpha})_{\rm SDSS}$.} 
 
\label{flux_ratio_compare}
\end{figure*}

  
\begin{figure}[htp]
\centering

\subfigure{\includegraphics[width=8.5cm,height=7.5cm]{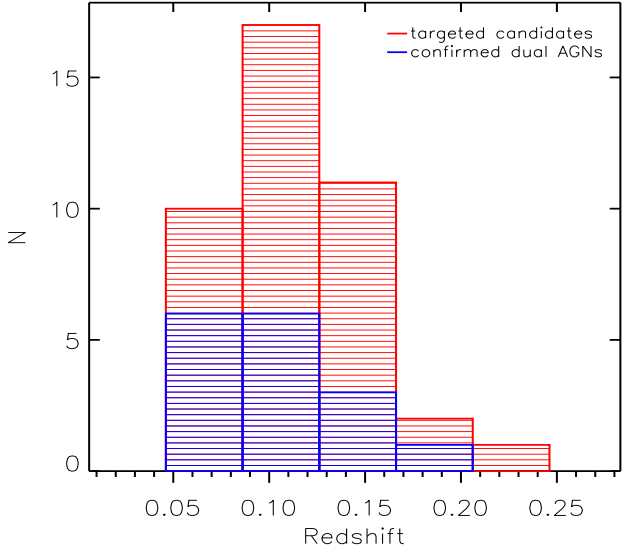}}

\caption{Redshift distribution for our targeted (red) and dual AGNs (blue).} 

\label{The_redshift_distribution}
\end{figure}

\begin{figure}[htp]
\centering

\subfigure{\includegraphics[width=8.5cm,height=7.5cm]{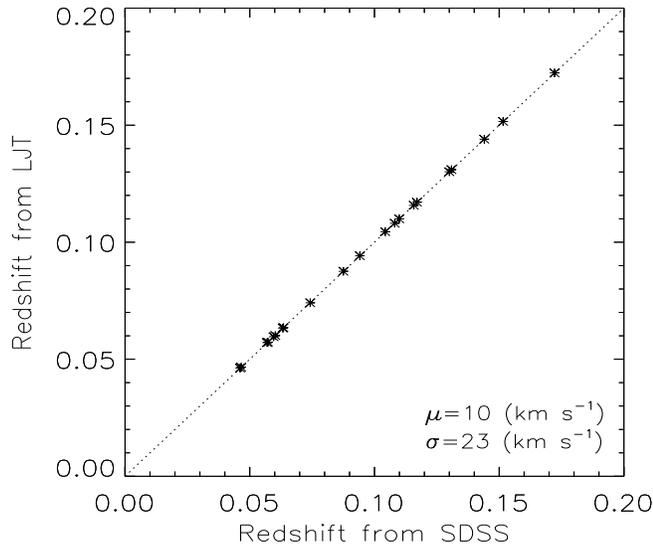}}

\caption{Comparison of redshifts yielded by our long-slit and SDSS spectra for part of cores in our dual AGN sample.}

\label{redshift_compare}
\end{figure}

\begin{figure}[htp]
\centering

\subfigure{\includegraphics[width=8.5cm,height=7.5cm]{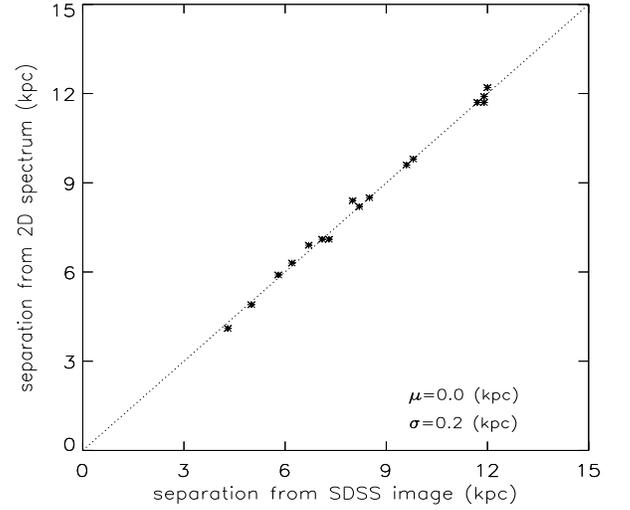}}

\caption{Comparison of separations measured from 2D spectra and from SDSS images for the 16 identified dual AGNs.}

\label{separation_compare}
\end{figure}


\begin{figure}[htp]
\centering

\subfigure{\includegraphics[width=8.5cm,height=7.5cm]{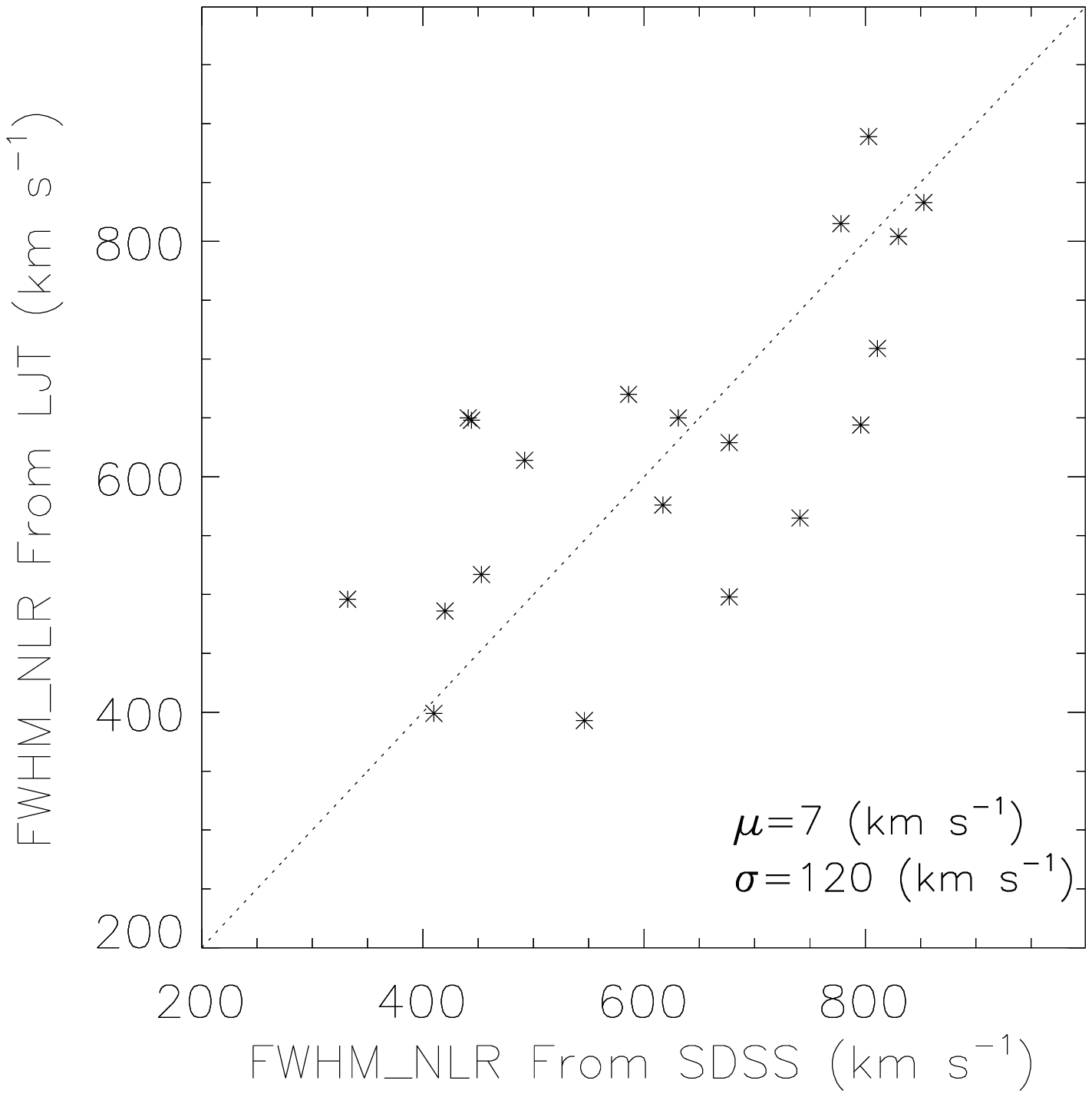}}

\caption{Comparison of FWHMs of NLRs measured from our long-slit and SDSS spectra for part of cores in our dual AGN sample.} 

\label{FWHM_compare}
\end{figure}


\section{Results}
\label{Result}

\subsection{Identification and classification of AGNs}
\label{criteria}


In this Section, we attempt to identify AGNs and classify them from the observed targets based on the analysis of the measured emission lines in the optical range. The details are given in the following. We also indicate the caveats of the current AGN identifications purely based on optical data at the end of this Section.

Here, we adopt the criteria developed by \citealt{hao2005} to identify Type I AGN:
1) FWHM of H${\alpha}$ $>$ 1200 km $\mathrm{s}^{-1}$ and h(H${\alpha}$ broad)/h(H${\alpha}$ narrow) $>$ 0.1; or 2) FWHM of H${\alpha}$ $>$ 2200 km $\mathrm{s}^{-1}$, where h(H${\alpha}$ broad) and h(H${\alpha}$ narrow) are the heights of H${\alpha}$ broad and narrow line components, respectively. The height is given by the peak of the Gaussian fits.

For Type II AGN (i.e., those with narrow emission lines), we adopt the so-called BPT diagrams (see \citealt{Baldwin1981}, \citealt{Veilleux1987}, \citealt{Kewley2001}, hereafter Ke01, \citealt{Kauffmann2003}, hereafter Ka03, \citealt{Kewley2006}, hereafter Ke06) to identify them, based on the four optical emission line flux ratios, i.e., [O\,{\sc iii}] $\lambda$5007/H$\beta$, [N\,{\sc ii}] $\lambda$6583/H$\alpha$, [S\,{\sc ii}] $\lambda\lambda$6717,6731/H$\alpha$, and [O\,{\sc i}] $\lambda$6300/H$\alpha$.
The fluxes used have not been corrected for any reddening and extinction effects.
As the lines involved in those ratios have close central wavelengths, the non-consideration of reddening corrections has minor impact on the results (e.g., \citealt{Kauffmann2003}).
The so-called ``maximum starburst line" on the the BPT diagram is derived from the upper limit of the theoretical pure stellar photoionization models (Ke01, Ka03, Ke06).
As Fig.\,\ref{BPT_DAGN} shows, sources above the red solid line (Ke01) are likely to be dominated by AGNs, below the blue dashed line (Ka03) are purely star-forming galaxies, between the red solid line (Ke01) and the blue dashed line (Ka03) are AGNs/star-forming composite galaxies (Comp). In the AGN region, Seyferts and LINERs populate dominantly above and below the blue solid line (Ke06), respectively. 
More details of the BPT diagrams are described in \cite{Kewley2006}.
Based on those BPT diagrams, the galaxies are classified into star-forming galaxies (H\,{\sc ii}), AGN/star-forming composite galaxies (Comp), Seyfert galaxies (Seyfert), low ionization nuclear emission line region (LINER) and ambiguous galaxies (ambiguous AGN).
Ambiguous galaxies are those that classified as one subtype of AGN in one or two of the diagram(s) but classified as another subtype of AGN in the remaining diagram(s). 
The detail classifications of our dual AGNs is presented in Table\,\ref{BPT classify}.

The BPT diagrams have been widely used to identify optical AGNs. 
However, some caveats remain when using the BPT diagrams.
We note that the LINERs, Comps and ambiguous galaxies identified by the BPT technique may suffer contamination from the H\,{\sc ii} or star-forming galaxies, given the current line ratio measurements errors and the potential systematics of the theoretical criteria (e.g., \citealt{Kauffmann2009, Stern2013, Azadi2017}).
Some studies also indicate that the distinct regions of AGN and H\,{\sc ii} on the BPT diagrams have some overlaps (e.g., \citealt{Kauffmann2009, Heckman2014}).
This empirical identification will miss the lineless AGNs, due to their extremely weak, sometimes completely undetected emission lines (\citealt{Cid Fernandes2010, Netzer2015}).
Not all of the AGN populations can be recovered by only using a single waveband observation (e.g., \citealt{Azadi2017}).
The further follow-up observations by either radio or X-ray could provide more vital constraints on the nature of the identified dual AGNs.
It is worth mentioning that identifying outflows in these optical emission line profiles could further provide indirect evidence on the nature of our dual AGN candidates, since the outflows around NLR can be driven by AGN feedback (e.g., \citealt{Di_Matteo2005, Hopkins2005, Crenshaw2015, Muller-Sanchez2016, Nevin2018}).
However, as shown in \cite{Nevin2018}, the velocity offset of the outflows are typically within 1000 km\,s$^{-1}$, and the current spectroscopy observations ($R \sim 300$) can not be used to identify such outflow features. Our ongoing long-slit spectroscopy observations by the Double Spectrograph (DBSP) with resolving power around 2000 mounted on Hale 5.1m telescope will allow us to do such analysis and provide more constraints on the nature of the dual AGN candidates.


\subsection{Dual AGN sample}
\label{Dual AGN samples}

Based on long-slit observations of 41 candidates with the YFOSC and the identification criteria presented in Section\,\ref{criteria}, we have found 16 likely dual AGNs. 
For the remaining 25 targets, 12 of them are AGN and normal galaxy pairs, 5 of them are AGN and star-forming galaxy pairs, 1 target is an AGN and star nonphysical pair and other 7 targets are hard to identify either due to the unresolved 2D long-slit spectra or low quality spectra.
In our dual AGN sample, 15 are newly found and only one (J1214+2931) has been revealed with X-ray image by \cite{Secrest2017}.
The detailed properties of the 16 likely dual AGNs are presented in Table\,\ref{finally DAGN}.
The high efficiency of our systematic searching method is about 39 per cent (16/41).
Amongst our identified dual AGNs, 28.1 per cent (9/32) are Type I AGNs, 25.0 per cent (8/32) Seyfert 2, 21.9 per cent (7/32) LINERs, 18.8 per cent (6/32) Comp and 6.3 per cent (2/32) ambiguous classification.

As an example, we briefly discuss the properties of dual AGN J0933+2114. 
As shown in Fig.\,\ref{slitimages of J0933+2114}, the two spectra of the dual cores, i.e. J0933+2114EN and J0933+2114WS, are spatially well resolved, allowing robust identification and classification of the system.

J0933+2114EN has a the redshift of $0.17234  \pm 0.00005$ and the FWHM of the NLR is $629 \pm 46 $ km $\mathrm{s}^{-1}$ as measured from the [O\,{\sc iii}] $\lambda$5007 line. 
The spectrum of J0933+2114EN has a broad line component as shown in Fig.\,\ref{spectrum fitting of J0933+2114}. 
The FWHM of the BLR is $5091 \pm 153$ km $\mathrm{s}^{-1}$ as given by the ${\mathrm{H}\alpha}$ broad line component. 
This value of BLR is much greater than 2200 km $\mathrm{s}^{-1}$, thus J0933+2114EN is clearly a Type I AGN.

J0933+2114WS has a redshift of $0.17321 \pm 0.00005 $ and the FWHM of the NLR is $548 \pm 53 $ km $\mathrm{s}^{-1}$, again measured from the [O\,{\sc iii}] $\lambda$5007 line. 
J0933+2114WS also has a broad line component as shown in Fig.\,\ref{spectrum fitting of J0933+2114}. 
The FWHM of the the BLR is $4956 \pm 155$ km $\mathrm{s}^{-1}$ as measured from the ${\mathrm{H}\alpha}$ broad line component. 
The velocity of the BLR is again greater than 2200 km $\mathrm{s}^{-1}$, confirming it a Type I AGN.

J0933+2114 is thus identified as a dual AGN composed of two Type I AGNs. The two optical cores have a separation of 12.3 kpc and a velocity offset of $ 261 \pm 21 $ km $\mathrm{s}^{-1}$.

The distribution of dual AGNs identified in the current work in the plane of redshift versus projection distance is presented in Fig.\,\ref{The_known_DAGNs}, together with those previously known in the literature (Paper I).
Dual AGNs detected with hard X-ray or radio observations have redshifts systematically smaller than those found by the double peak technique.
Dual AGNs identified in the current work have redshifts ranging from 0.0461 to 0.1722 and separations ranging from 4.3 kpc to 14.2 kpc.
As Fig.\,\ref{The_known_DAGNs} shows, our dual AGN found in this work make the dual AGN sample more complete in redshift space.
For our sample, the velocity offsets of the two cores range from 57 km $\mathrm{s}^{-1}$ to 300 km $\mathrm{s}^{-1}$ (Fig.\,\ref{offset_velocity}), in good agreement with previous results (50-300 km $\mathrm{s}^{-1}$; \citealt{Comerford2013, Koss2016}).

Detailed analyses of the remaining 15 likely dual AGNs in the current work are presented in Appendix\,A. 
In total, we have found 16 likely dual AGNs and significantly increased the number of known dual AGNs.

\subsection{Validation of measurements}
\label{Validations of the measurements}

Given the relatively low resolution of our spectra (especially those of G3), the accuracies of flux ratio, redshift, separation and FWHM measurements presented here should be examined.
Doing so, we compare our results with those derived from the SDSS fiber spectra that have a higher spectral resolution ($R \sim 2200$).

\subsubsection{Flux ratios}
\label{Compare_SDSS_LJT}

To check the reliability of our BPT diagram results, the flux ratios of our likely dual AGNs are compared to those derived from the SDSS spectra for part of optical cores.
First, we compare the flux ratios measured from our long-slit spectra with those from the SDSS fiber spectra.
For most of the two cores in dual AGNs, only one core has been identified as AGN in SDSS and another core has no SDSS spectrum.  
There are 32 optical cores in 16 dual AGNs, but only 20 cores have SDSS spectra (Table\,\ref{BPT classify}).
As Fig.\,\ref{flux_ratio_compare} shows, flux ratio measurements from our spectra are generally consistent with those from the SDSS.
Here we note that the flux measurements of emission lines from SDSS spectra are similar to those of our long-slit spectra as described in Sections 2.3.3 and 2.3.4.
As shown in Table\,\ref{BPT classify}, BPT classifications based on the flux ratio measurements from our long-slit spectra are totally in agreement with those based on the SDSS spectra, implying the reliability of the former measurements.

\subsubsection{Redshifts}
\label{Compare redshift}
To check the wavelength calibration accuracy of our long-slit spectra, the derived redshifts of those likely dual AGNs are compared to those deduced from the SDSS spectra. 
The distribution of redshifts of the targeted candidates and our dual AGNs is displayed in Fig.\,\ref{The_redshift_distribution}.
As shown in Fig.\,\ref{redshift_compare}, the results agree with each other very well.
The median value and standard deviation of the redshift differences are only 10 and 23 km s$^{-1}$, respectively, confirming the good wavelength calibration of our long-slit spectra. 

\subsubsection{Separations}
\label{Compare separation}
To check whether the 2D spectra are actually from the two optical cores of our selected merging galaxies, the separations between the two cores from SDSS images are compared to those from the 2D spectra for the 16 identified dual AGNs.
Fig.\,\ref{separation_compare} clearly shows the separations fitted from the 2D spectra are in excellent agreement with those from SDSS images.
The median value and standard deviation of the separation differences are just 0 and 0.2 kpc, respectively, confirming that our observed 2D spectra are actually from the two cores of merging galaxies.

\subsubsection{FWHMs}
\label{Compare FWHM of NLR}

FWHM is an important parameter that characterizes the NLR and BLR of an AGN.
The velocity width of BLR is large enough to be resolved by our spectra and is thus well measured.
For the NLRs, the instrumental broadening of our spectra is comparable or slightly higher than their velocity widths.
By comparing with the FWHMs deduced from the SDSS spectra, we examine the accuracy of the FWHMs of NLRs measured from our spectra.
The results plotted in Fig.\,\ref{FWHM_compare} show reasonable agreement.
The median value and standard deviation of the differences are only 7 and 120 km\,s$^{-1}$, respectively. 

\begin{figure}[thp]
\centering
\subfigure{\includegraphics[width=8.2cm,height=7.0cm]{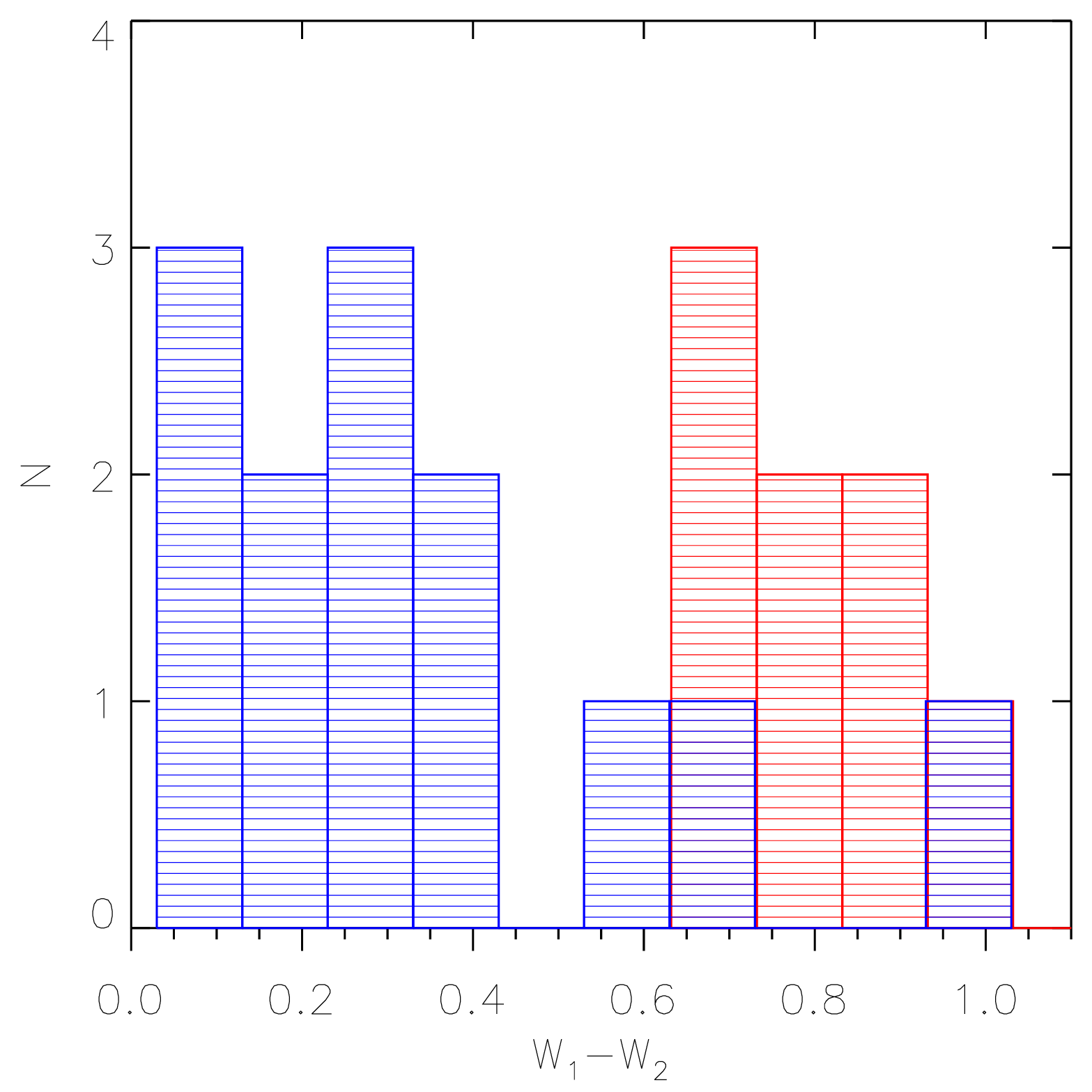}}
\caption{Distribution of $ W_{1}$-$W_{2}$ colors of dual AGNs. The red shaded columns represent Type I AGNs or dual AGNs that contain at least one Type I AGN, while the blue columns represent Type II AGNs or dual AGNs that contain only Type II AGN.}
\label{w1_w2_compare}
\end{figure} 
   
           
\begin{figure}[thp]
\centering

\subfigure{\includegraphics[width=8.0cm,height=7.0cm]{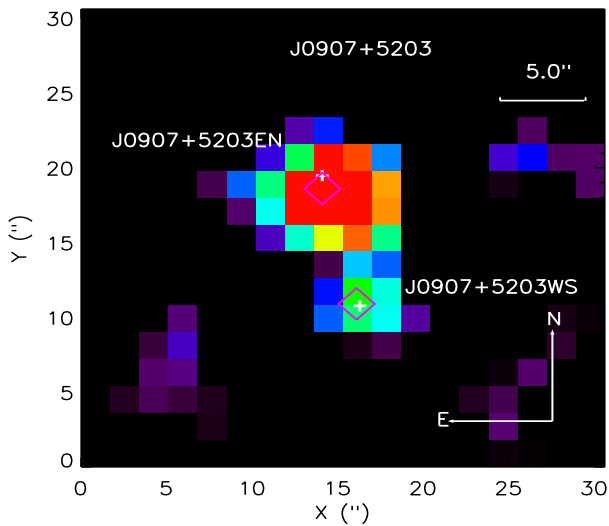}}

\caption{Two radio detections of dual AGN: J0907+5203 from the FIRST catalogue. 
The positions of the two radio detections are marked by magenta diamonds.
The white pluses indicate the positions of the two optical cores.
} 

\label{Radio image of J0907+5203}
\end{figure}

  
\begin{figure}[thp]
\centering
\subfigure{\includegraphics[width=8.0cm,height=7.0cm]{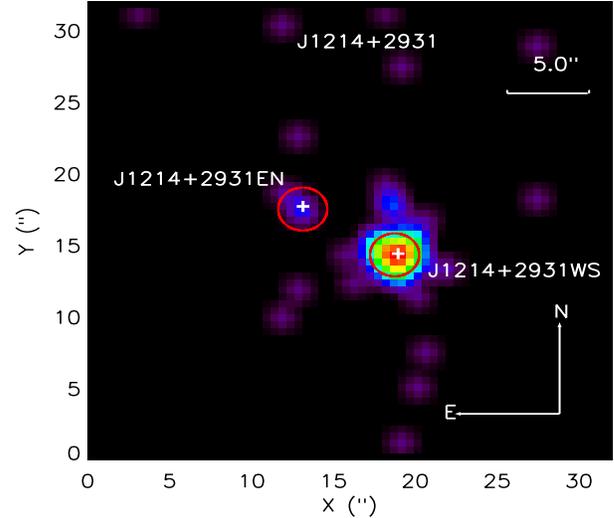}}
\caption{Double X-ray cores of dual AGN: J1214+2931 from the Chandra catalog. 
In this image, the positions of the two X-ray cores from the Chandra observation are marked by the red circles.
The white pluses indicate the positions of the two optical cores.} 
\label{X-ray image of J1214+2931}
\end{figure}

\section{Discussion}
\label{Discussions}

\subsection{Implications for dual AGN systematic searching}
\label{Implications}

A systematic search for dual AGNs generally includes three steps.
To start with, a large sample of potential candidates of dual AGNs is required.  
Then, follow-up spectroscopy is carried out to exclude most contamination.
As a final step, radio or hard X-ray observations provide vital evidence of activity.
In this process, the first step is the key to obtaining a large sample of candidate dual AGNs.
As mentioned in Section 1, several previous attempts proposed to systematically search for dual AGNs, including that based on DPAGNs (e.g., \citealt{Liu2010a,Liu2011, Shen2011,Comerford2012}) and the infrared colors of AGNs (e.g., \citealt{Satyapal2014,Satyapal2017}).

Hundreds of DPAGNs have been selected from the DEEP2 Galaxy Redshift Survey (e.g., \citealt{Gerke2007}) and from the SDSS survey (e.g., \citealt{wang2009, Liu2010a, Smith2010}).
However, follow-up spectroscopic observations reveal that only 2 to 5 per cent of the selected DPAGNs are bona fide dual AGNs (e.g., \citealt{Liu2011, Shen2011,Comerford2012}).
The main reason underlying this low efficiency is the large contamination of rotating disks or bi-conical outflows of the NLR gas surrounding single AGN that also produce double peaked emission line profiles (e.g., \citealt{Smith2012, Comerford2014}).

\cite{Satyapal2017} propose an alternative new method to search for dual AGN candidates based on the mid-infrared color $W1-W2$ from the WISE survey (e.g., \citealt{Wright2010}).
They argue that this method could discover the buried population of dual AGNs missed by optical observations. 
This method is highly powerful for luminous and highly obscured AGN systems, especially those systems including Type I AGNs (e.g., \citealt{Hickox2017, Satyapal2017}).
However, this method may not be efficient at identifying dual AGNs composed of only Type II AGNs.
In Fig.\,\ref{w1_w2_compare}, we show the $W_1 - W_2$ color distribution of our dual AGN sample.
The Type I AGN or dual AGN systems that contain at least one Type I AGN show redder $W_1 - W_2$ colors ($>$ 0.6 mag) compared to typical values of stars and galaxies.
In comparison, the dual AGNs that contain only Type II AGNs have a wide distribution of $W_1 - W_2$ colors, ranging from 0 to 1.0 mag that overlap significantly with those of stars or galaxies. 
In this sense, selection based on the mid-infrared color alone may exclude the dual AGN that only contain Type II AGNs and can not be used to construct a dual AGN sample without selection biases.
But we note that those dual AGN candidates containing only Type II AGNs (especially those classified as Comp and Ambiguous AGN) in the current work require further solid confirmation (e.g., X-ray or radio follow-up observations).
The method also suffers from the relatively low angular resolution of the WISE survey (\citealt{Wright2010}).
As such, the yield of dual AGN sample from this method are all quite limited.

\begin{figure}[htp]
\centering
\subfigure{\includegraphics[scale=0.35,angle=0]{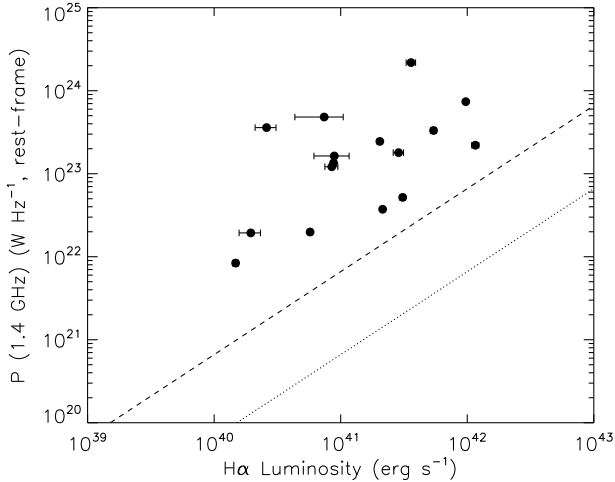}}
\caption{Rest-fram 1.4\,GHz radio power versus H$\alpha$ luminosity for the radio detected cores.
Dotted line indicate radio-derived star formation rates (SFRs) equal to the H$\alpha$-derived SFRs (using the relations from \citealt{Hopkins2001}), while the dashed line mark the former 10 times larger than the latter.}
\label{haradio}
\end{figure}

In addition, the systematic search for dual AGNs based on hard X-ray observations (e.g., \citealt{Koss2011, liu2013}) and VLA Stripe 82 radio imaging (\citealt{Fu2015}) has also been carried out.
Restricted by the limited angular resolution and survey area, the number of dual AGNs thus identified is relatively less than theoretically predicted (e.g., \citealt{Komossa_Zensus_review2016}).
Higher angular resolution all-sky radio and X-ray surveys are required in the near future.

Here, we newly note that one system (J0907+5203; Fig.\,\ref{Radio image of J0907+5203}) in our dual AGN sample has two radio detections from the FIRST survey and their positions are in excellent agreement with the optical cores from the SDSS imaging.
The radio detection of J0907+5203EN has a integrated flux density of 2.33 mJy with a typical RMS of 0.15 mJy\footnote{http://sundog.stsci.edu/index.html} given by the FIRST survey catalog (e.g., \citealt{White1997}).
By assuming a general power law index\footnote{The assumed spectral index doesn't significantly affect the value of the radio luminosity given the small redshift (e.g., \citealt{Pushkarev2012}, \citealt{Hovatta2014}).} of $-0.7$, the luminosity of J0907$+$5203EN is $P_{1.4 {\rm GHz}} = 1.9 \times 10^{22}$\,W\,Hz$^{-1}$.
As the 1 mJy source detection threshold, the radio detection of J0907+5203WS doesn't have a integrated flux density in the FIRST survey catalog (e.g., \citealt{White1997, Helfand2015}). 
Another one (J1214+2931; Fig.\,\ref{X-ray image of J1214+2931}) shows two cores in the X-ray image obtained with the Chandra (also reported in \citealt{Secrest2017}) and their positions are again in line with the optical positions from the SDSS imaging. 
The dual AGN J1214+2931 is composed with a strong X-ray core (J1214+2931WS) and a weak X-ray core (J1214+2931EN) in 0.5-7 keV. 
The strong X-ray core (J1214+2931WS) is extremely luminous, with an intrinsic X-ray luminosity of\, L$_{2-10 \rm keV}$ = 4.0 $\times$ $10^{43} \rm erg\, s^{-1}$.
The weak X-ray core (J1214+2931EN) is only 2.6$\sigma$ detection with only 7 counts in 0.5-2 keV. 
More detail information is described in \cite{Secrest2017}.

In addition to the J0907+5203 system, the remaining 15 dual AGN candidates also have one radio detection from the FIRST catalog. 
The radio powers are then calculated for all the dual AGN candidates again assuming a power law index of $-0.7$.
To explore those radio radiation natures, the diagram between $P_{\rm 1.4 GHz}$ and $L_{H_{\alpha}}$ is shown in Fig.\,\ref{haradio}. 
The plot clearly show that all the systems have radio powers over 10 times larger than those expected from the $H_{\alpha}$ traced star formation rates, implying significant radio excesses from AGN activities. 
The results partly show that the candidates found in the current work are promising dual AGN systems.

To conclude, compared to the previous methods, the systematic method of searching for dual AGNs employed in the current work is relatively efficient, cost-effective, and capable to build a large sample of dual AGNs.

\subsection{Potential applications of the sample}
\label{Applications}

As mentioned earlier, this newly identified dual AGN sample, together with those known dual AGN in the literature, could provide vital information for understanding the co-evolution of SMBHs and their host galaxies. 
For example, by measuring the black hole masses and the velocity dispersions of the individual AGNs in those dual AGN systems, one can derive the $M_{\rm BH}$-$\sigma$ relation for the dual AGN systems and test whether the relation is consistent with that found for normal galaxies and single AGNs (e.g., \citealt{Komossa2007,Fu2011a}). 
In addition, one can explore the kinematics of gas and stellar components in those systems and investigate whether their behaviors are the same or not (e.g., \citealt{Villforth2015, Zhang2016}). 
In the forthcoming papers of this series, we will present results demonstrating the power of our dual AGN sample to understand the above issues.

\section{Conclusion}
\label{Conclusions}

Using an innovative method of systematically searching for and identifying dual-AGN systems amongst kpc scale merging galaxies, a total of 222 candidates are selected. 
As the first results of our ASTRO-DARING project, we have obtained long-slit spectroscopic observations for 41 targets between November 2014 and February 2017 using the YFOSC mounted on the LJT of Yunnan Observatories.

By careful data reduction, 1D spectra extraction, emission line profile fitting and AGN classification (based on the Balmer emission line widths and emission line ratios), 16 likely dual AGNs are finally identified (15 of them are found for the first time).

Our new searching approach is efficient (about 40 per cent, 16/41) and cost effective.
With this new method, we plan to construct the current largest sample of dual AGNs ($>$ 50) and better understand the co-evolution of host galaxies and their SMBHs.

\section*{Acknowledgements}

We acknowledge the support of staff of the Lijiang 2.4-m telescope. 
Fund for the telescope has been provided by the Chinese Academy of Sciences and the People's Government of Yunnan Province. 
The work of J. M. Bai is supported by the NSFC (grants 11133006, 11361140347) and the Strategic Priority Research Program ``The Emergence of Cosmological Structures'' of the Chinese Academy of Sciences (grant No. XDB09000000).
Y. Huang and X.-W. Liu acknowledge the supports by National Natural Science Foundation of China grants 11833006, 11811530289 and 11903027.
It is a pleasure to thank Dr. Sarah Bird for a thorough read of the manuscript and improving the language significantly.
We thank Hai-Cheng Feng, Kai-Xing Lu, Ding-Rong Xiong for assistance. 

This work has made use of data products from the LJT (Lijiang 2.4-m telescope), SDSS, FIRST and WISE.


\bibliographystyle{apjs}

\newpage

\begin{appendices} 

\appendix

\section{FWHM of well detected emission lines}

\begin{table*}[!thp]
\tiny
\centering
\caption{}
\begin{threeparttable}
\begin{tabular}{ccccccccccc}
\hline
\hline

Name  & $\rm FWHM$  & $\rm FWHM$ & $\rm FWHM$ & $\rm FWHM$  & $\rm FWHM$ & $\rm FWHM$  & $\rm FWHM$ & $\rm FWHM$ & $\rm FWHM$ \\
\hline  
Emission line  & $\rm \mathrm{H}\beta$  & $\rm \OIIIL$ & $\rm \OIIIB$ & $\rm \OI$  & $\rm \NIIL$ & $\rm \mathrm{H}\alpha$  & $\rm \NIIB$ & $\rm \SIIL$ & $\rm \SIIB$ \\
\hline  
\multicolumn{10}{c}{First phase (6 Dual AGNs)}\\

\hline  

J0151$-$0245WN & $ 204 \pm 88 $ & $ 221 \pm 77 $ & $ 262 \pm 63 $ & $ 495 \pm 21 $ & $ 177 \pm 63 $ & $ 257 \pm 39 $ & $ 344 \pm 32 $ & $ 456 \pm 27 $ & $ 396 \pm 31 $ \\ 
J0151$-$0245ES & $ 259 \pm 69 $ & $ 211 \pm 102 $ & $ 471 \pm 36 $ & $ 557 \pm 55 $ & $ 562 \pm 25 $ & $ 213 \pm 51 $ & $ 416 \pm 23 $ & $ 704 \pm 35 $ & $ 621 \pm 38 $ \\
\hline

J0933$+$2114EN & $ 192 \pm 99 $ & $ 525 \pm 58 $ & $ 629 \pm 46 $ & $ 555 \pm 47 $ & $ 566 \pm 43 $ & $ 519 \pm 34 $ & $ 588 \pm 31 $ & $ 660 \pm 47 $ & $ 382 \pm 68 $ \\ 
J0933$+$2114WS & $ 522 \pm 59 $ & $ 545 \pm 55 $ & $ 548 \pm 53 $ & $ 404 \pm 54 $ & $ 541 \pm 36 $ & $ 449 \pm 38 $ & $ 758 \pm 23 $ & $ 365 \pm 50 $ & $ 240 \pm 78 $ \\ 
\hline

J1017$+$3448NW & $ 899 \pm 47 $ & $ 736 \pm 63 $ & $ 778 \pm 41 $ & $ 779 \pm 178 $ & $ 509 \pm 43 $ & $ 368 \pm 47 $ & $ 392 \pm 44 $ & $ 590 \pm 27 $ & $ 395 \pm 41 $ \\ 
J1017$+$3448SE & $ 514 \pm 77 $ & $ 406 \pm 75 $ & $ 614 \pm 48 $ & $ 179 \pm 122 $ & $ 468 \pm 66 $ & $ 486 \pm 39 $ & $ 425 \pm 43 $ & $ 110 \pm 165 $ & $ 91 \pm 57 $ \\
\hline

J1105$+$1957EN & $ 131 \pm 55 $ & $ 261 \pm 57 $ & $ 487 \pm 46 $ & $ 413 \pm 22 $ & $ 516 \pm 37 $ & $ 236 \pm 35 $ & $ 292 \pm 28 $ & $ 400 \pm 20 $ & $ 246 \pm 32 $ \\ 
J1105$+$1957WS & $ 450 \pm 47 $ & $ 120 \pm 78 $ & $ 399 \pm 50 $ & $ 428 \pm 27 $ & $ 480 \pm 27 $ & $ 378 \pm 22 $ & $ 332 \pm 25 $ & $ 486 \pm 36 $ & $ 321 \pm 24 $ \\ 
\hline

J1633$+$4718N & $ 680 \pm 48 $ & $ 333 \pm 94 $ & $ 413 \pm 71 $ & $ 466 \pm 54 $ & $ 121 \pm 83 $ & $ 284 \pm 60 $ & $ 303 \pm 56 $ & $ 316 \pm 51 $ & $ 294 \pm 55 $ \\ 
J1633$+$4718S & $ 772 \pm 41 $ & $ 317 \pm 95 $ & $ 517 \pm 57 $ & $ 591 \pm 40 $ & $ 132 \pm 68 $ & $ 496 \pm 36 $ & $ 355 \pm 49 $ & $ 339 \pm 50 $ & $ 369 \pm 47 $ \\ 
\hline

J2258$-$0115EN & $ 711 \pm 48 $ & $ 405 \pm 96 $ & $ 738 \pm 43 $ & $ 406 \pm 54 $ & $ 326 \pm 57 $ & $ 539 \pm 31 $ & $ 481 \pm 35 $ & $ 615 \pm 37 $ & $ 473 \pm 37 $ \\ 
J2258$-$0115WS & $ 624 \pm 55 $ & $ 790 \pm 38 $ & $ 650 \pm 46 $ & $ 634 \pm 33 $ & $ 519 \pm 33 $ & $ 369 \pm 47 $ & $ 397 \pm 43 $ & $ 340 \pm 51 $ & $ 164 \pm 61 $ \\ 
\hline  

\multicolumn{10}{c}{Second phase (10 Dual AGNs)}\\
\hline  
  
J0217$-$0845EN & $ 548 \pm 213 $ & $ 512 \pm 224 $ & $ 339 \pm 237 $ & $ 573 \pm 153 $ & $ 244 \pm 173 $ & $ 483 \pm 164 $ & $ 344 \pm 206 $ & $ 308 \pm 238 $ & $ 605 \pm 119 $ \\ 
J0217$-$0845WS & $ 657 \pm 178 $ & $ 641 \pm 175 $ & $ 670 \pm 164 $ & $ 651 \pm 149 $ & $ 193 \pm 166 $ & $ 418 \pm 162 $ & $ 402 \pm 164 $ & $ 479 \pm 139 $ & $ 147 \pm 132 $ \\ 
\hline

J0756$+$2340EN & $ 780 \pm 150 $ & $ 718 \pm 160 $ & $ 806 \pm 151 $ & $ 622 \pm 115 $ & $ 617 \pm 105 $ & $ 604 \pm 106 $ & $ 554 \pm 115 $ & $ 536 \pm 115 $ & $ 546 \pm 111 $ \\ 
J0756$+$2340WS & $ 728 \pm 161 $ & $ 731 \pm 159 $ & $ 565 \pm 195 $ & $ 563 \pm 129 $ & $ 490 \pm 136 $ & $ 454 \pm 142 $ & $ 419 \pm 152 $ & $ 568 \pm 123 $ & $ 531 \pm 116 $ \\ 
\hline

J0813$+$4941WN & $ 525 \pm 245 $ & $ 709 \pm 162 $ & $ 804 \pm 138 $ & $ 575 \pm 278 $ & $ 434 \pm 149 $ & $ 383 \pm 173 $ & $ 381 \pm 168 $ & $ 617 \pm 100 $ & $ 589 \pm 104 $ \\ 
J0813$+$4941ES & $ 458 \pm 255 $ & $ 795 \pm 153 $ & $ 521 \pm 212 $ & $ 648 \pm 114 $ & $ 334 \pm 276 $ & $ 616 \pm 104 $ & $ 691 \pm 192 $ & $ 615 \pm 199 $ & $ 563 \pm 115 $ \\ 
\hline

J0833$+$1532WN & $ 379 \pm 249 $ & $ 391 \pm 297 $ & $ 427 \pm 285 $ & $ 508 \pm 138 $ & $ 380 \pm 138 $ & $ 466 \pm 159 $ & $ 733 \pm 165 $ & $ 450 \pm 139 $ & $ 296 \pm 151 $ \\ 
J0833$+$1532ES & $ 561 \pm 210 $ & $ 535 \pm 210 $ & $ 486 \pm 227 $ & $ 771 \pm 190 $ & $ 702 \pm 194 $ & $ 611 \pm 107 $ & $ 477 \pm 144 $ & $ 494 \pm 124 $ & $ 418 \pm 146 $ \\ 
\hline

J0848$+$3515EN & $ 352 \pm 238 $ & $ 631 \pm 179 $ & $ 496 \pm 272 $ & $ 367 \pm 172 $ & $ 475 \pm 136 $ & $ 398 \pm 224 $ & $ 384 \pm 250 $ & $ 609 \pm 100 $ & $ 299 \pm 136 $ \\ 
J0848$+$3515WS & $ 545 \pm 215 $ & $ 370 \pm 203 $ & $ 498 \pm 221 $ & $ 599 \pm 117 $ & $ 437 \pm 149 $ & $ 544 \pm 117 $ & $ 619 \pm 103 $ & $ 515 \pm 118 $ & $ 515 \pm 118 $ \\ 
\hline

J0907$+$5203EN & $ 761 \pm 118 $ & $ 821 \pm 104 $ & $ 815 \pm 102 $ & $ 420 \pm 127 $ & $ 696 \pm 171 $ & $ 623 \pm 178 $ & $ 369 \pm 178 $ & $ 538 \pm 114 $ & $ 511 \pm 190 $ \\ 
J0907$+$5203WS & $ 581 \pm 162 $ & $ 702 \pm 121 $ & $ 709 \pm 117 $ & $ 485 \pm 119 $ & $ 277 \pm 208 $ & $ 455 \pm 108 $ & $ 543 \pm 162 $ & $ 494 \pm 194 $ & $ 463 \pm 199 $ \\
\hline

J1214$+$2931EN & $ 371 \pm 242 $ & $ 490 \pm 230 $ & $ 576 \pm 191 $ & $ 588 \pm 119 $ & $ 398 \pm 216 $ & $ 586 \pm 144 $ & $ 332 \pm 191 $ & $ 444 \pm 139 $ & $ 611 \pm 100 $ \\ 
J1214$+$2931WS & $ 556 \pm 156 $ & $ 683 \pm 164 $ & $ 648 \pm 170 $ & $ 430 \pm 164 $ & $ 550 \pm 199 $ & $ 562 \pm 114 $ & $ 398 \pm 161 $ & $ 445 \pm 137 $ & $ 535 \pm 117 $ \\ 
\hline

J1645$+$2057WN & $ 511 \pm 111 $ & $ 524 \pm 103 $ & $ 644 \pm 85 $ & $ 443 \pm 74 $ & $ 210 \pm 159 $ & $ 280 \pm 126 $ & $ 856 \pm 80 $ & $ 212 \pm 150 $ & $ 175 \pm 106 $ \\ 
J1645$+$2057ES & $ 668 \pm 199 $ & $ 526 \pm 124 $ & $ 833 \pm 64 $ & $ 547 \pm 63 $ & $ 359 \pm 151 $ & $ 386 \pm 101 $ & $ 521 \pm 58 $ & $ 280 \pm 123 $ & $ 378 \pm 83 $ \\ 
\hline

J2206$+$0003WN & $ 777 \pm 175 $ & $ 376 \pm 257 $ & $ 889 \pm 157 $ & $ 407 \pm 144 $ & $ 581 \pm 96 $ & $ 342 \pm 102 $ & $ 165 \pm 118 $ & $ 243 \pm 127 $ & $ 384 \pm 118 $ \\ 
J2206$+$0003ES & $ 255 \pm 237 $ & $ 757 \pm 178 $ & $ 650 \pm 185 $ & $ 293 \pm 190 $ & $ 342 \pm 98 $ & $ 353 \pm 123 $ & $ 592 \pm 55 $ & $ 268 \pm 118 $ & $ 110 \pm 79 $ \\
\hline

J2314$+$0653WN & $ 680 \pm 187 $ & $ 427 \pm 132 $ & $ 393 \pm 141 $ & $ 449 \pm 178 $ & $ 613 \pm 56 $ & $ 270 \pm 119 $ & $ 189 \pm 107 $ & $ 207 \pm 149 $ & $ 251 \pm 122 $ \\ 
J2314$+$0653ES & $ 232 \pm 162 $ & $ 880 \pm 179 $ & $ 419 \pm 133 $ & $ 313 \pm 134 $ & $ 374 \pm 87 $ & $ 295 \pm 109 $ & $ 370 \pm 186 $ & $ 461 \pm 167 $ & $ 265 \pm 197 $ \\ 
\hline
   
\end{tabular}
\label{emission_line:FWHM} 
\begin{flushleft}
  Notes: FWHM in units of km $\mathrm{s}^{-1}$. The FWHMs of $\rm \mathrm{H}\beta$ and $\rm \mathrm{H}\alpha$ emission lines are both for the narrow line component. The FWHM of $\rm \mathrm{H}\alpha$ broad line component shown in Table\,5.
\end{flushleft}
\end{threeparttable}
\end{table*}


\newpage

\section{Notes For Individual Dual AGNs}
 

\begin{figure*}[ht]
  \centering
  \includegraphics[width=5.20cm,height=5cm]{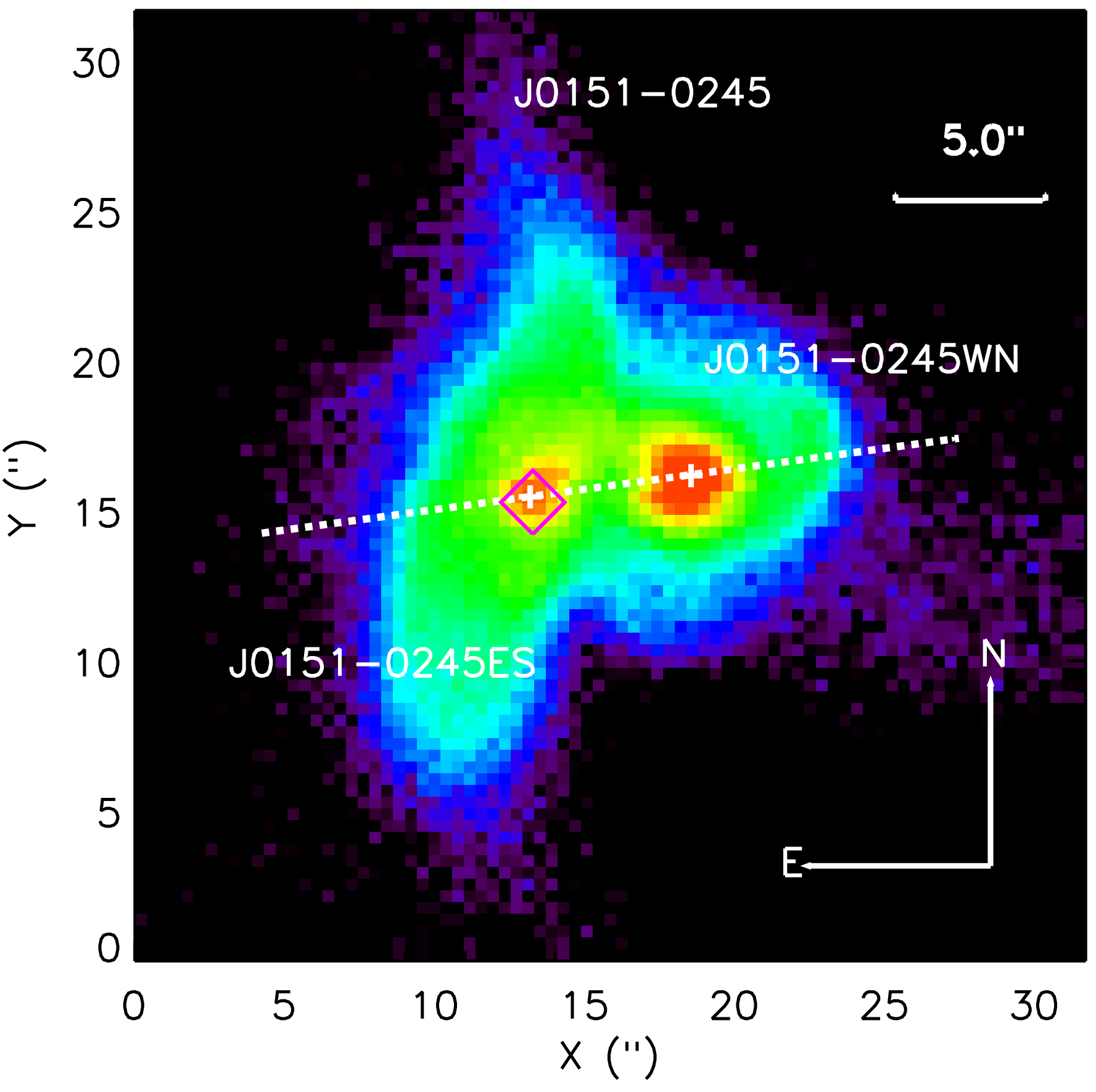}
  \includegraphics[width=12.2cm,height=5cm]{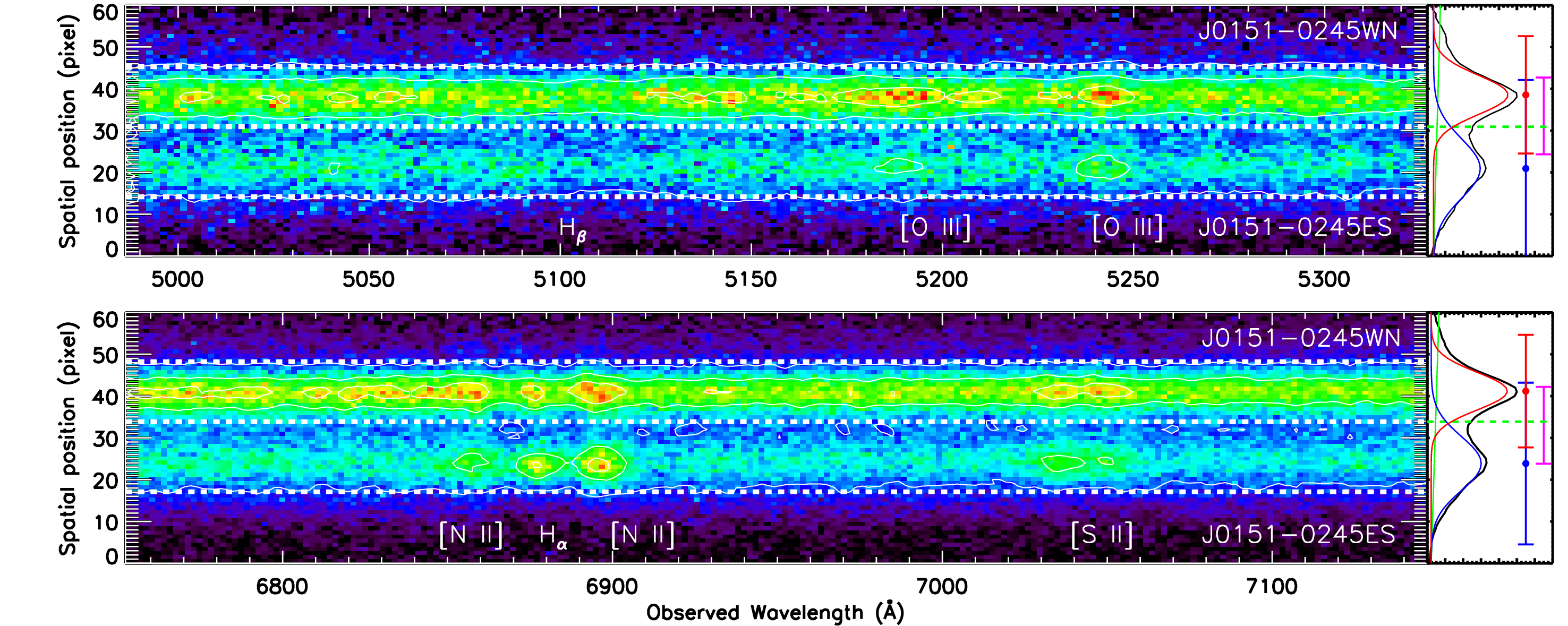}
\caption{Same as Fig.\,\ref{slitimages of J0933+2114} but for J0151-0245. The spectra from LJT. }
\label{slitimages of J0151-0245}
\end{figure*}


\begin{figure*}[ht]
\centering
  \includegraphics[scale=0.51]{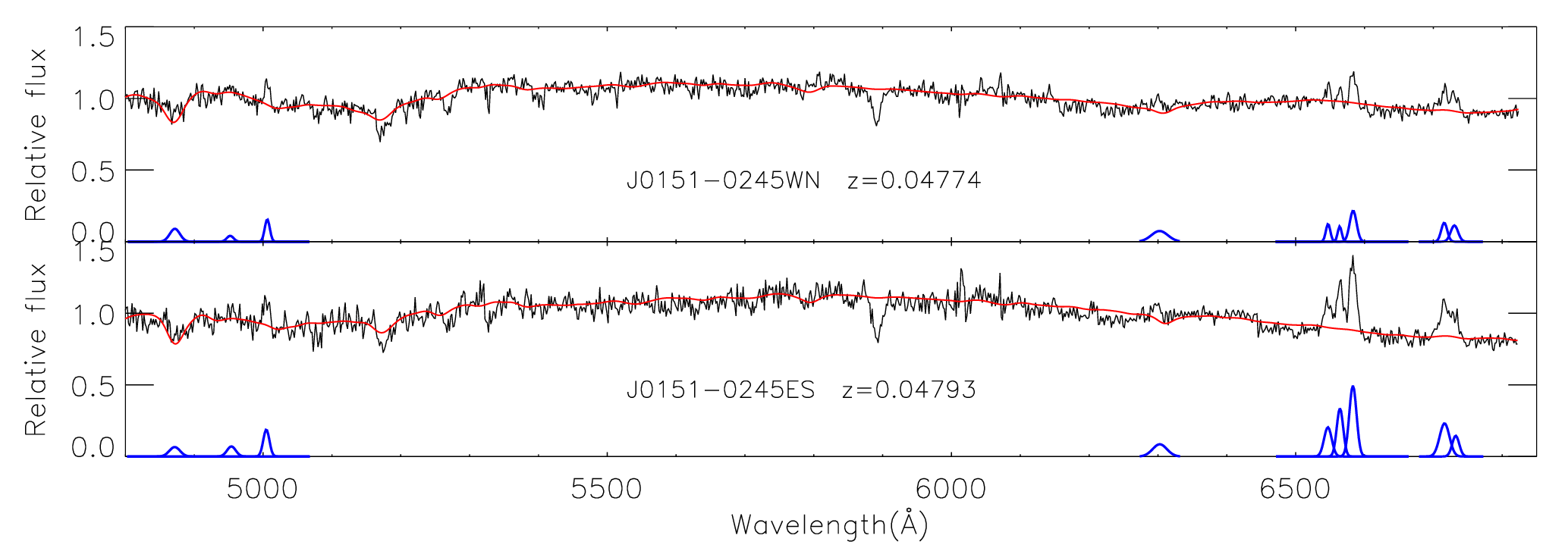}
\caption{Same as Fig.\,\ref{spectra_fitting:J0933+2114} but for J0151-0245.}
\label{spectra_fitting:J0151-0245}
\end{figure*}

\begin{figure*}[ht]
  \centering
  \includegraphics[scale=0.58]{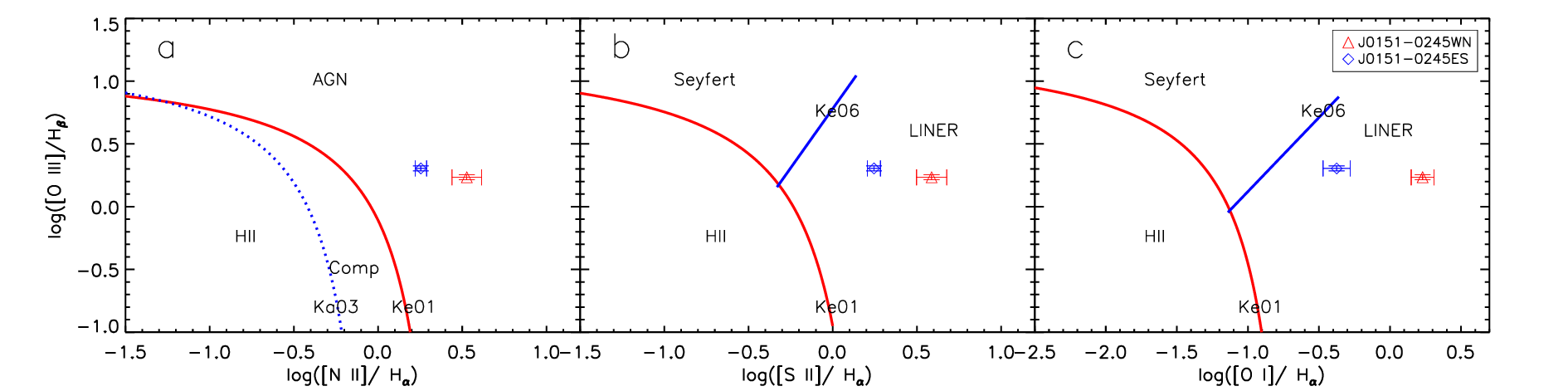} 

\caption{Same as Fig.\,\ref{BPT_DAGN} but for J0151-0245.}

\label{BPT diagrams of J0151-0245}
\end{figure*}

\subsection*{\rm Dual AGN: J015107.39-024526.87}
\label{DAGN of J0151-0245}

%

Two sets of AGN spectra are spatially resolved as shown in Fig.\,\ref{slitimages of J0151-0245}, so the two cores, i.e. J0151$-$0245WN and J0151$-$0245ES can be identified separately.

The fitting of extracted 1D spectra of the two cores are shown in Fig.\,\ref{spectra_fitting:J0151-0245}. The redshifts, FWHMs of emission lines and emission line flux ratios of the two cores, measured from the 1D spectra, are presented in Tables\,\ref{BPT classify} and \ref{finally DAGN}.
For the two cores, no broad line components are detected, we therefore use BPT diagram to classsify their types (Fig.\,\ref{BPT diagrams of J0151-0245}). According to the diagnosis, both cores are classified as LINER. 

The object J015107.39-024526.87 has been revealed as a dual AGN composed of two LINERs. This dual AGN has a separation of 4.8 kpc and a velocity offset of $ 57 \pm 28 $ km $\mathrm{s}^{-1}$.

\clearpage

\begin{figure*}[ht]
  \centering
  \includegraphics[width=5.20cm,height=5.0cm]{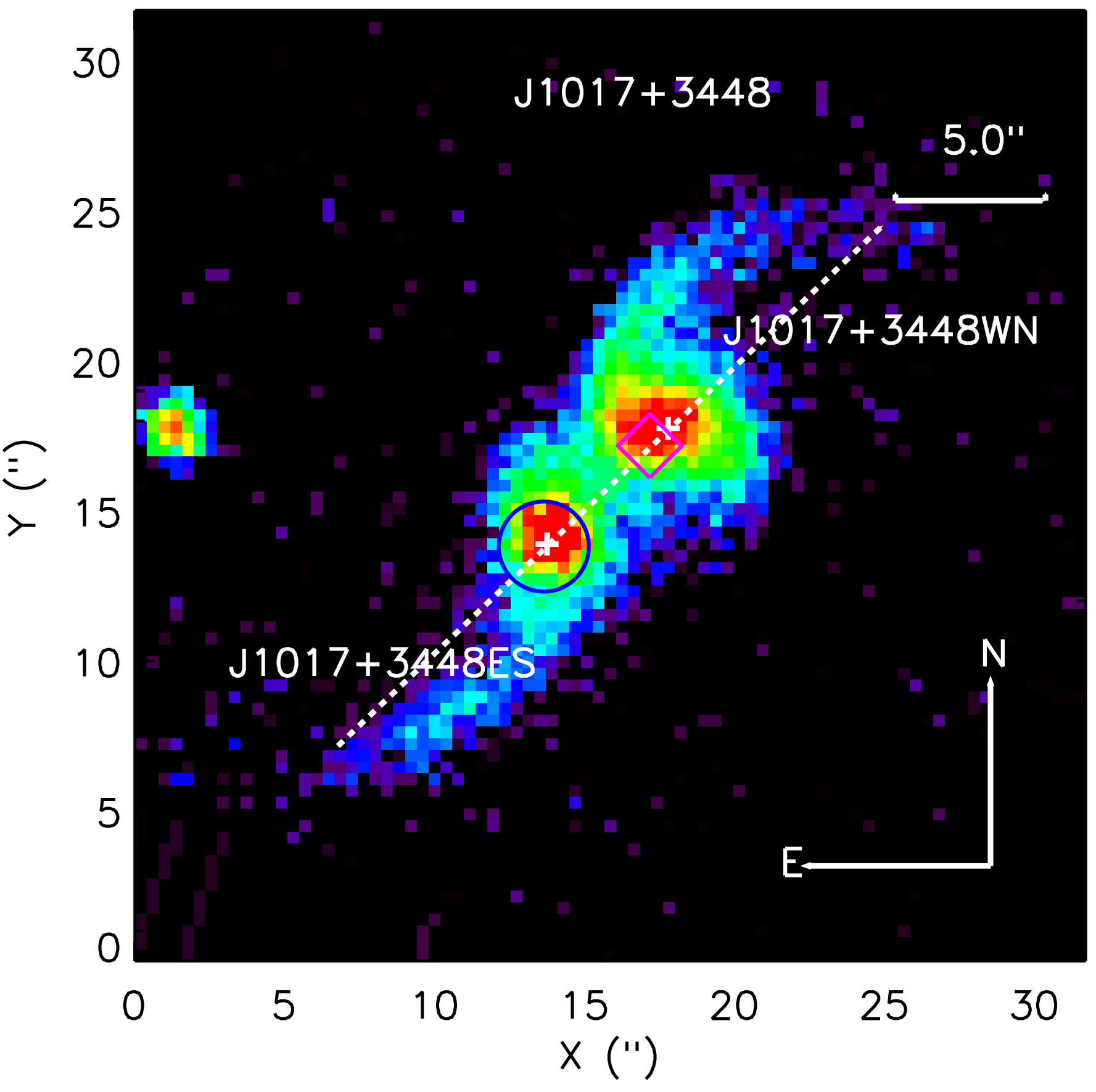}
  \includegraphics[width=12.2cm,height=5.0cm]{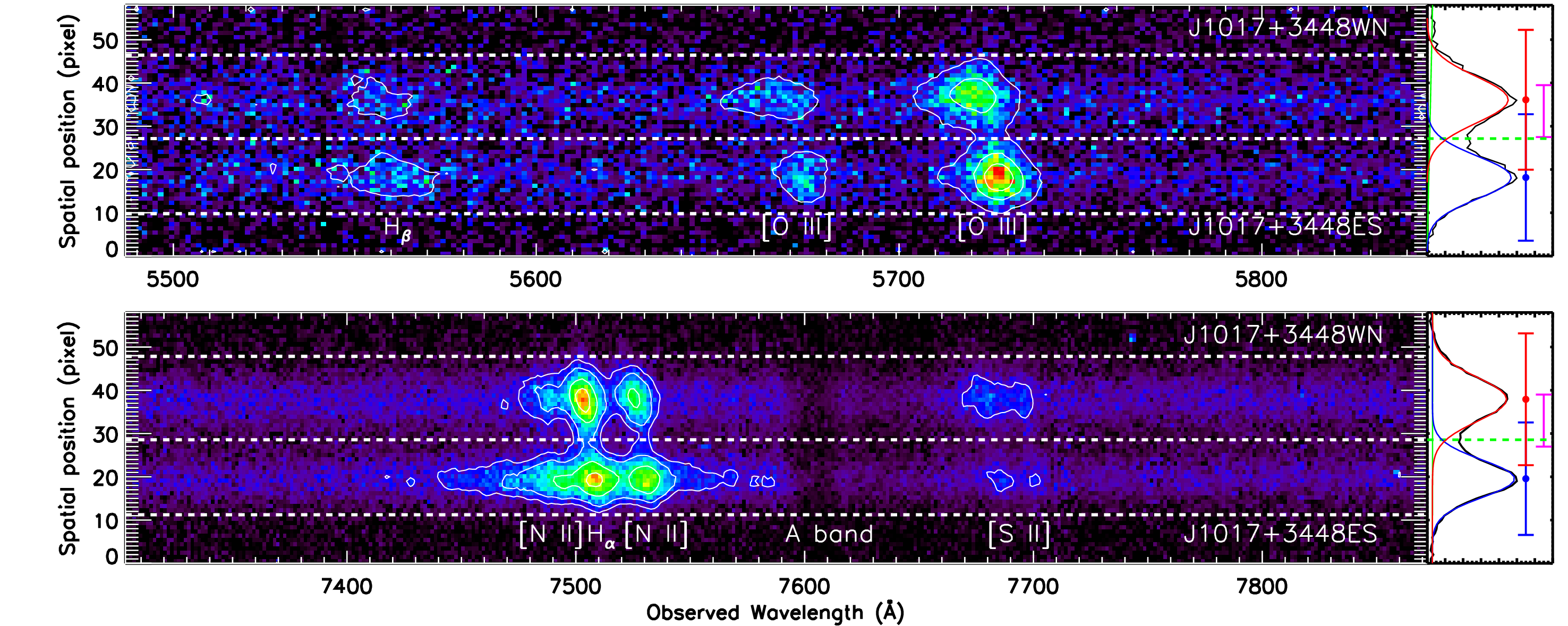}
\caption{ Same as Fig.\,\ref{slitimages of J0933+2114} but for J1017+3448.}
\label{slitimages of J1017+3448}
\end{figure*}

\begin{figure*}[ht]
\centering
\includegraphics[scale=0.51]{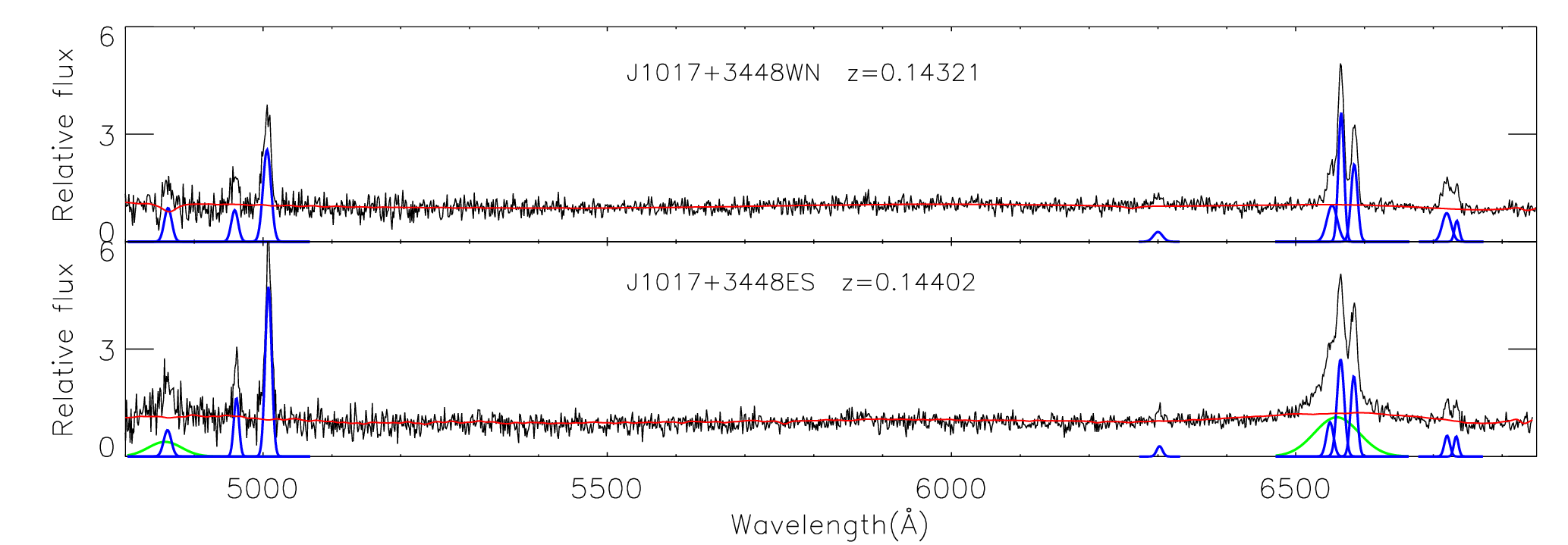}
\caption{Same as Fig.\,\ref{spectra_fitting:J0933+2114} but for J1017+3448. The spectra from LJT.}
\label{spectra_fitting:J1017+3448}
\end{figure*}

\begin{figure*}[ht]
  \centering
  \includegraphics[scale=0.58]{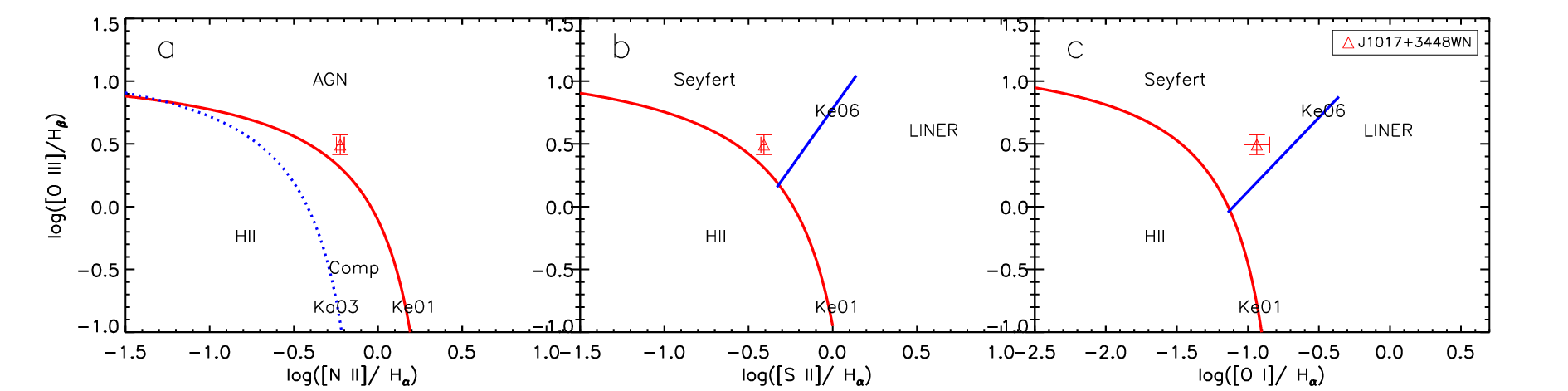}

\caption{Same as Fig.\,\ref{BPT_DAGN} but for J1017+3448.}

\label{BPT diagrams of J1017+3448}
\end{figure*}

\subsection*{\rm Dual AGN: J101757.07+344846.61} 
\label{DAGN of J1017+3448}


%
%

Two sets of AGN spectra are spatially resolved as shown in Fig.\,\ref{slitimages of J1017+3448}, so the two cores, i.e. J1017+3448WN and J1017+3448ES can be identified separately.

The fitting of extracted 1D spectra of the two cores are shown in Fig.\,\ref{spectra_fitting:J1017+3448}. The redshifts, FWHMs of emission lines and emission line flux ratios of the two cores, measured from the 1D spectra, are presented in Tables\,\ref{BPT classify} and \ref{finally DAGN}.
The spectrum of J1017+3448WN does not show a broad line (Fig.\,\ref{spectra_fitting:J1017+3448}). 
We use the BPT diagram to distinguish this AGN shown in Fig.\,\ref{BPT diagrams of  J1017+3448}. 
It is classified as a Seyfert (AGN).
The spectrum of J1017+3448ES has broad line (Fig.\,\ref{spectra_fitting:J1017+3448}, FWHM\,$> 2000$\,km\,s$^{-1}$ as measured from H$\alpha$ broad line component) and thus it is a Type I AGN.

The object J101757.07+344846.61 has been revealed as a dual AGN composed of Seyfert (J1017+3448WN) and Type I AGN (J1017+3448ES). This dual AGN has a separation of 14.2 kpc and a velocity offset of $ 243 \pm 21 $ km $\mathrm{s}^{-1}$.

\clearpage


\begin{figure*}[ht]
  \centering
  \includegraphics[width=5.20cm,height=5.0cm]{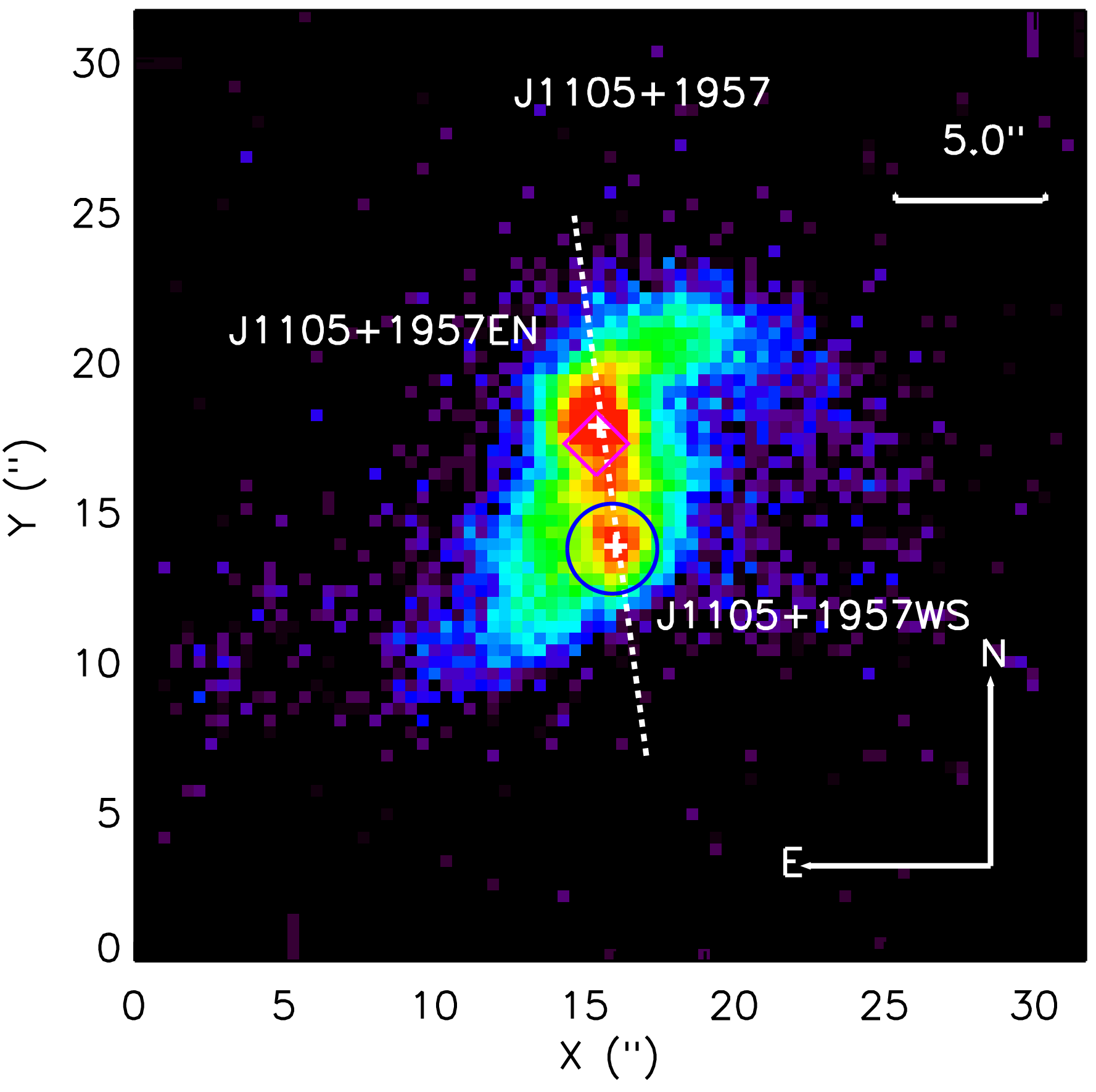}
  \includegraphics[width=12.2cm,height=5.0cm]{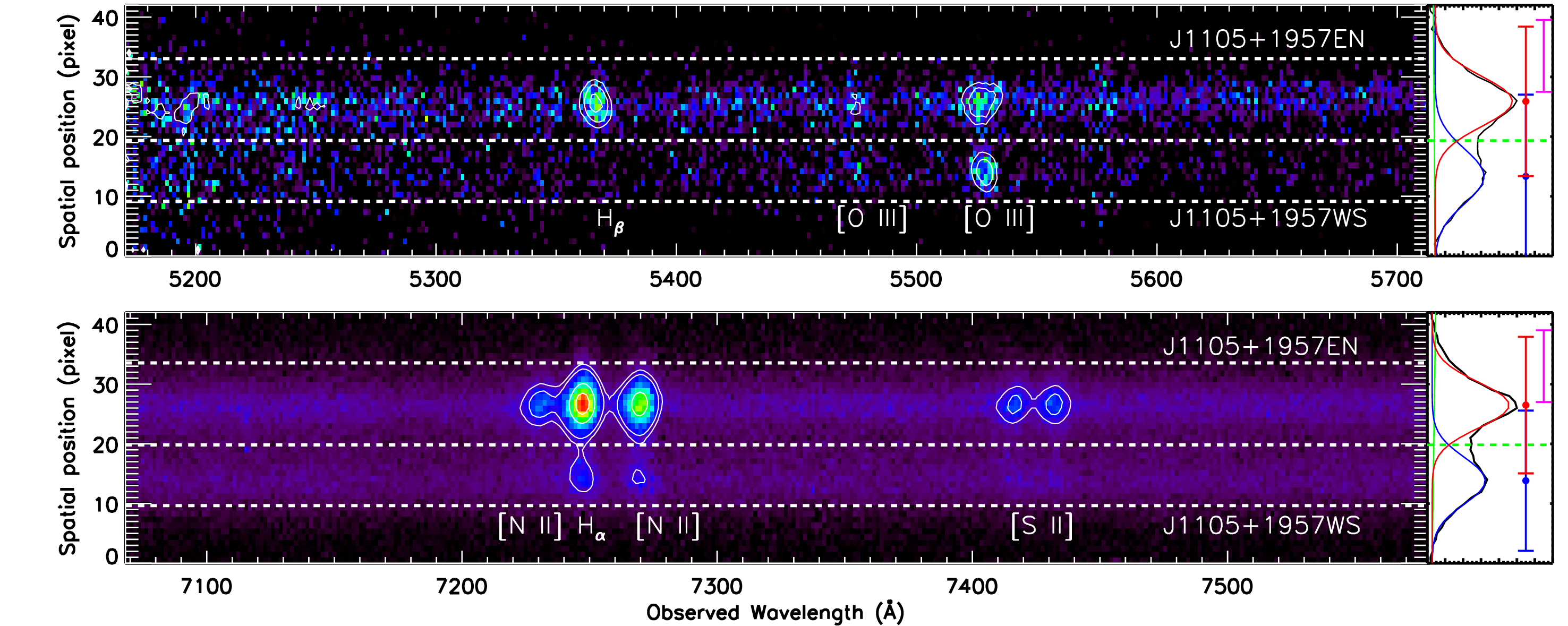}
\caption{Same as Fig.\,\ref{slitimages of J0933+2114} but for J1105+1957. }
\label{slitimages of J1105+1957}
\end{figure*}

\begin{figure*}[ht]
\centering
\includegraphics[scale=0.51]{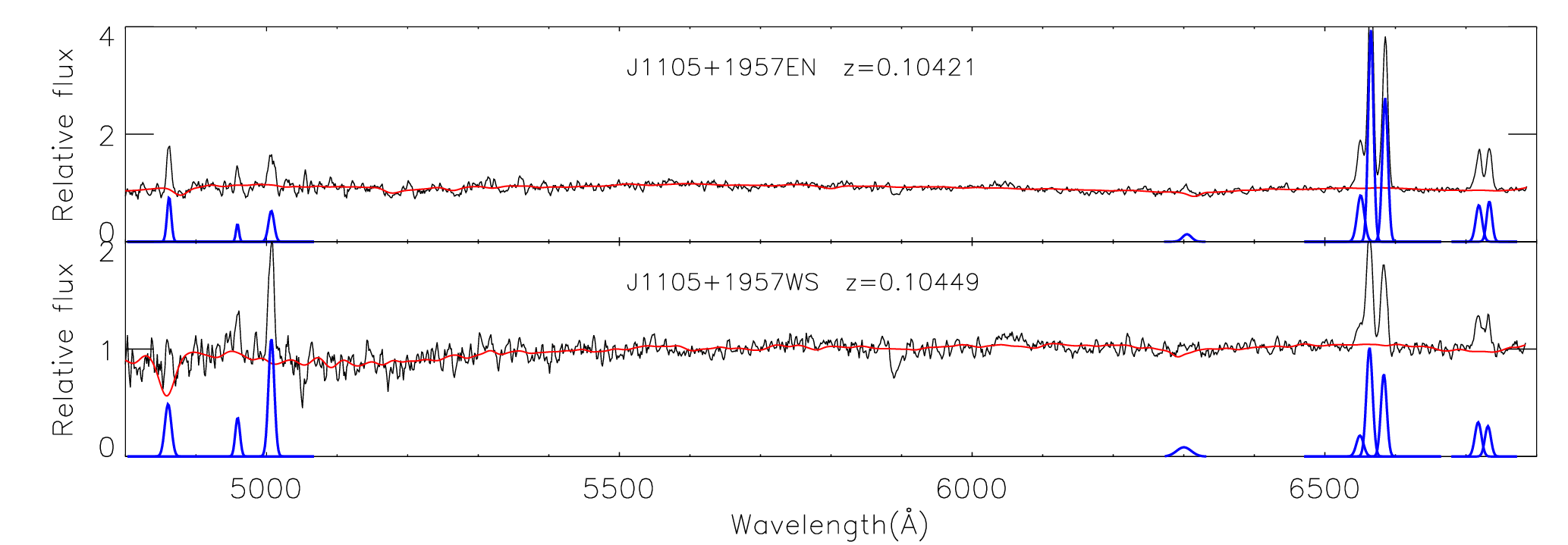}
\caption{Same as Fig.\,\ref{spectra_fitting:J0933+2114} but for J1105+1957. The spectra from LJT.}
\label{spectra_fitting:J1105+1957}
\end{figure*}

\begin{figure*}[ht]
  \centering
  \includegraphics[scale=0.58]{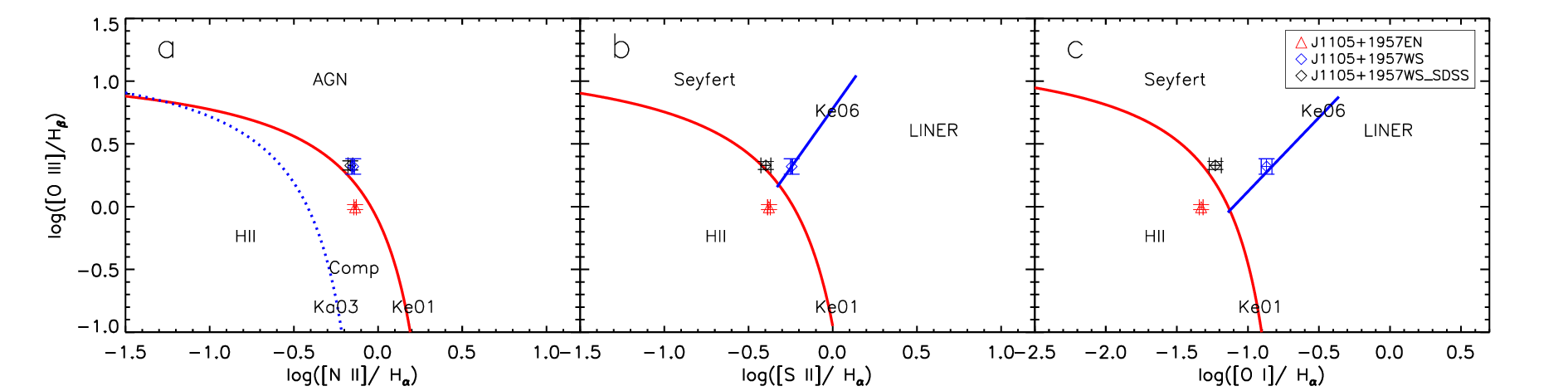}

\caption{Same as Fig.\,\ref{BPT_DAGN} but for J1105+1957.} 
 
\label{BPT diagrams of J1105+1957}
\end{figure*}

\subsection*{\rm Dual AGN: J110544.48+195750.06} 
\label{DAGN of J1105+1957}

%
%

Two sets of AGN spectra are spatially resolved as shown in Fig.\,\ref{slitimages of J1105+1957}, so the two cores, i.e. J1105+1957EN and J1105+1957WS can be identified separately.

The fitting of extracted 1D spectra of the two cores are shown in Fig.\,\ref{spectra_fitting:J1105+1957}. 
The redshifts, FWHMs of emission lines and emission line flux ratios of the two cores, measured from the 1D spectra, are presented in Tables\,\ref{BPT classify} and \ref{finally DAGN}. 
For the two cores, no broad line components are detected, we therefore use BPT diagram to classsify their types (Fig.\,\ref{BPT diagrams of J1105+1957}). 
According to the diagnosis, J1105+1957EN is classified as Comp (AGN) and J1105+1957WS is classified as Seyfert (AGN). 

The object J110544.48+195750.06 has been revealed as a dual AGN composed of ambiguous galaxy (AGN) (J1105+1957EN) and Seyfert (J1105+1957WS). This dual AGN has a separation of 7.4 kpc and a velocity offset of $ 84 \pm 21 $ km $\mathrm{s}^{-1}$.

\clearpage


\begin{figure*}[ht]
  \centering
  \includegraphics[width=5.20cm,height=5.0cm]{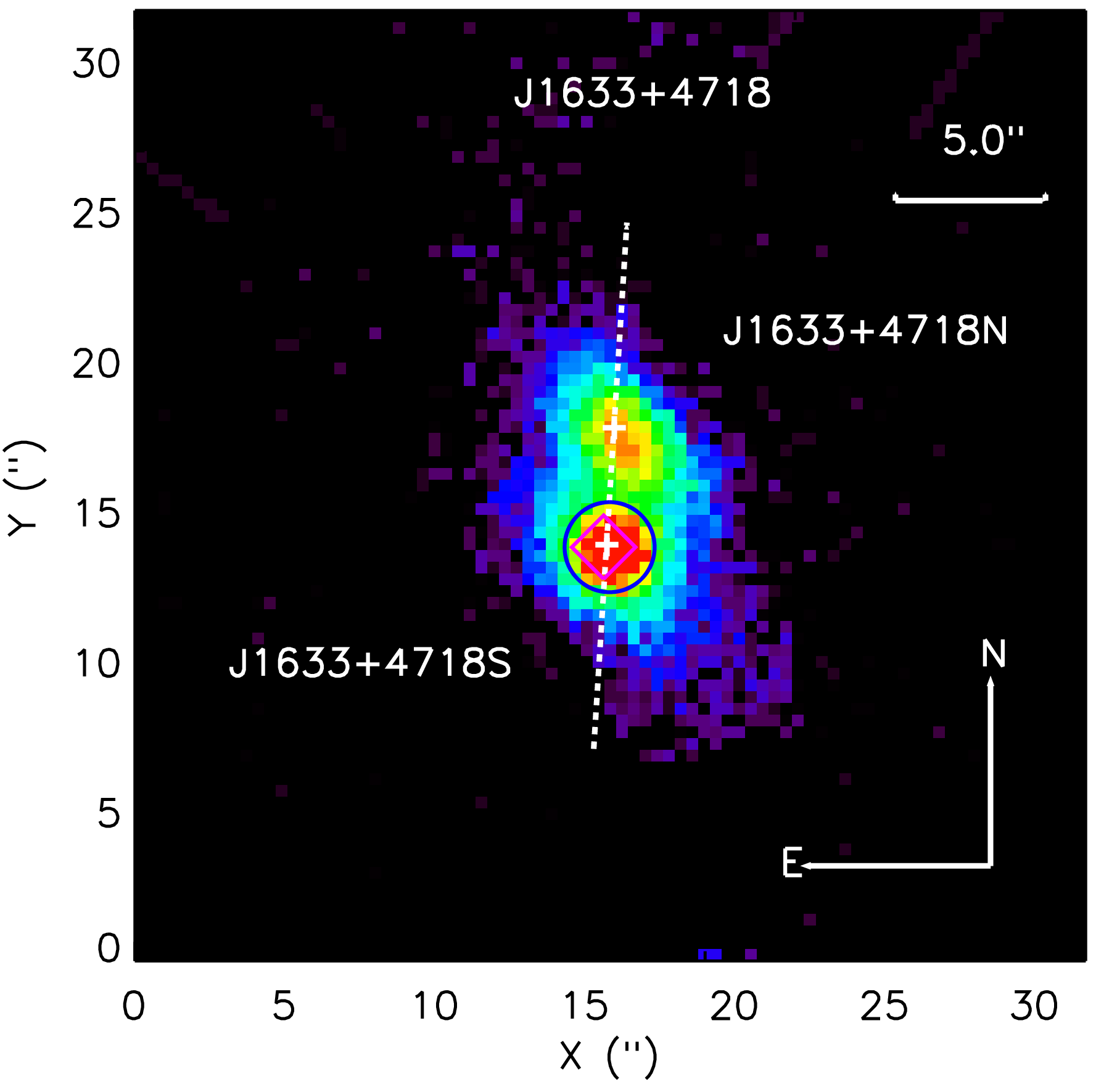}
  \includegraphics[width=12.2cm,height=5.0cm]{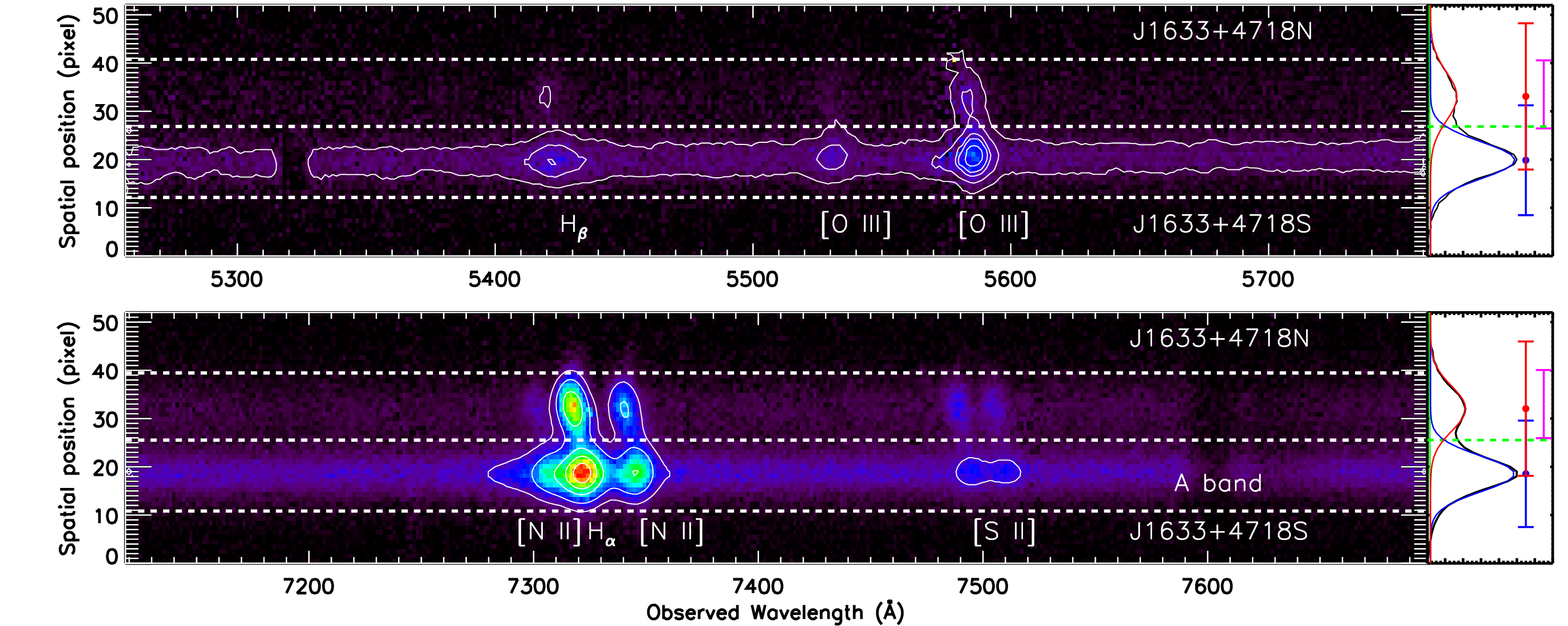}
\caption{Same as Fig.\,\ref{slitimages of J0933+2114} but for J1633+4718. }
\label{slitimages of J1633+4718}
\end{figure*}


\begin{figure*}[ht]
\centering
\includegraphics[scale=0.51]{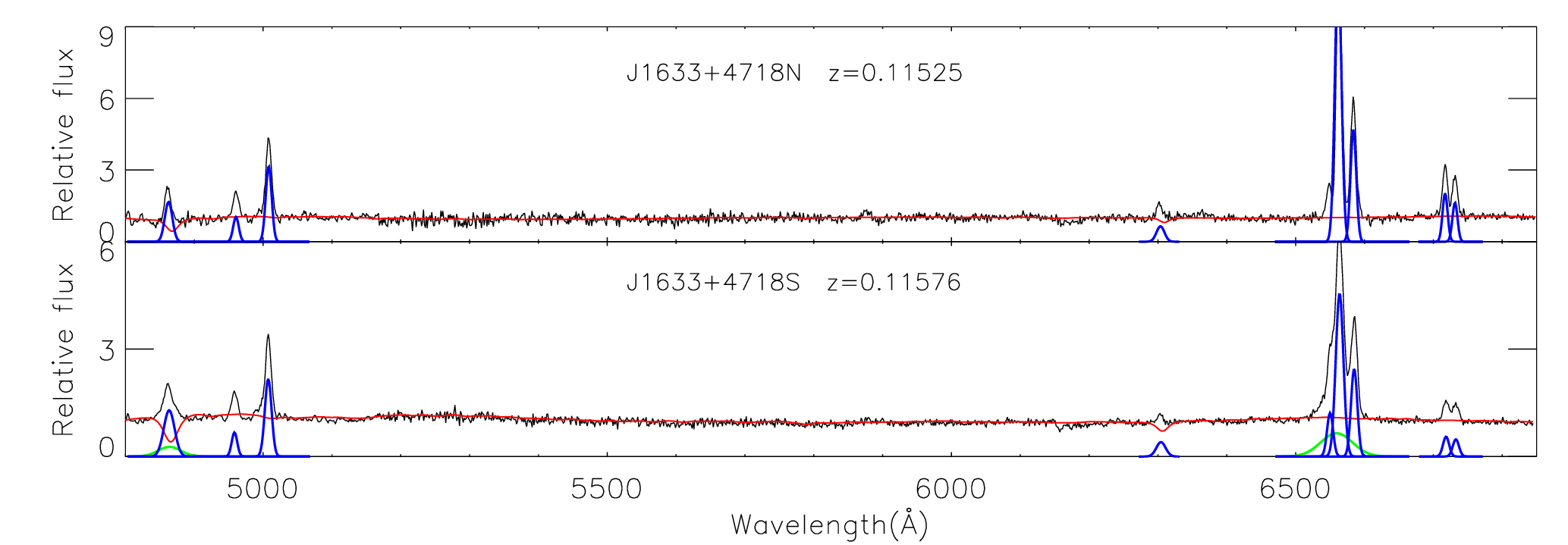}
\caption{Same as Fig.\,\ref{spectra_fitting:J0933+2114} but for J1633+4718. The spectra from LJT.}
\label{spectra_fitting:J1633+4718}
\end{figure*}

\begin{figure*}[ht]
  \centering
  \includegraphics[scale=0.58]{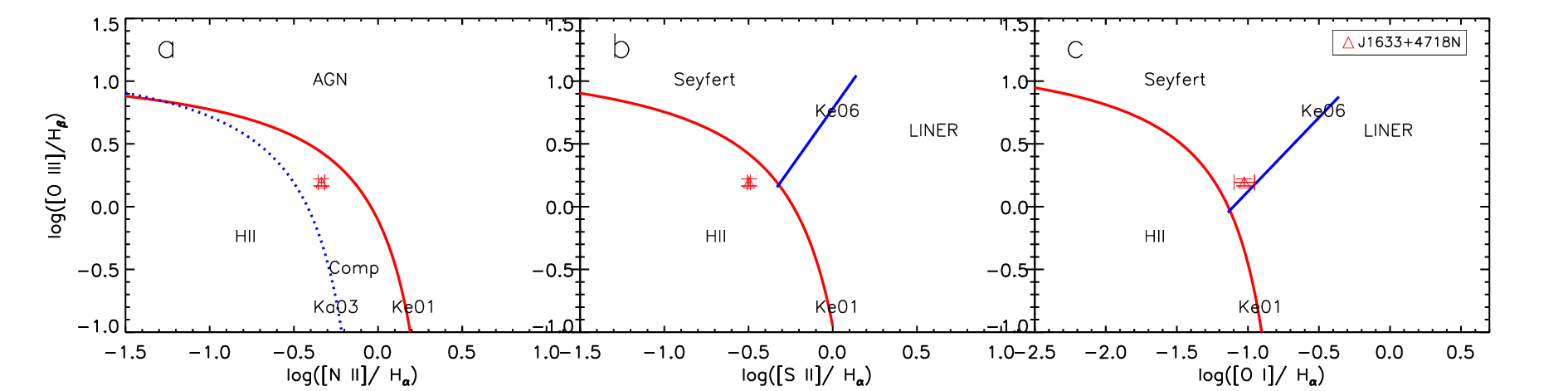}

\caption{Same as Fig.\,\ref{BPT_DAGN} but for J1633+4718.}

\label{BPT diagrams of J1633+4718}
\end{figure*}

\subsection*{\rm Dual AGN: J163323.58+471858.95} 
\label{DAGN of J1633+4718}

%
%

Two sets of AGN spectra are spatially resolved as shown in Fig.\,\ref{slitimages of J1633+4718}, so the two cores, i.e. J1633+4718N and J1633+4718S can be identified separately. 

The fitting of extracted 1D spectra of the two cores are shown in Fig.\,\ref{spectra_fitting:J1633+4718}. The redshifts, FWHMs of emission lines and emission line flux ratios of the two cores, measured from the 1D spectra, are presented in Tables\,\ref{BPT classify} and \ref{finally DAGN}.
The spectrum of J1633+4718N doesn't show a broad line (Fig.\,\ref{spectra_fitting:J1633+4718}). 
We use the BPT diagram to distinguish this AGN shown in Fig.\,\ref{BPT diagrams of  J1633+4718}.
The J1633+4718N is revealed as ambiguous galaxy (AGN).
The spectrum of J1633+4718S has broad line (Fig.\,\ref{spectra_fitting:J1633+4718}). 
The FWHM of the BLR is $ 1965 \pm 66 $ km $\mathrm{s}^{-1}$ that measured from the ${\mathrm{H}\alpha}$ broad line component and h(Hbroad)/h(Hnarrow) is 0.29 ($>$ 0.1 ). 
According to Hao et al. (2005), J1633+4718S is a Type I AGN.

The object J163323.58+471858.95 has been revealed as a dual AGN composed of ambiguous galaxy (AGN) (J1633+4718N) and Type I AGN (J1633+4718S). This dual AGN has a separation of 8.0 kpc and a velocity offset of $ 153 \pm 21 $ km $\mathrm{s}^{-1}$.

\clearpage

\begin{figure*}[ht]
  \centering
  \includegraphics[width=5.20cm,height=5.0cm]{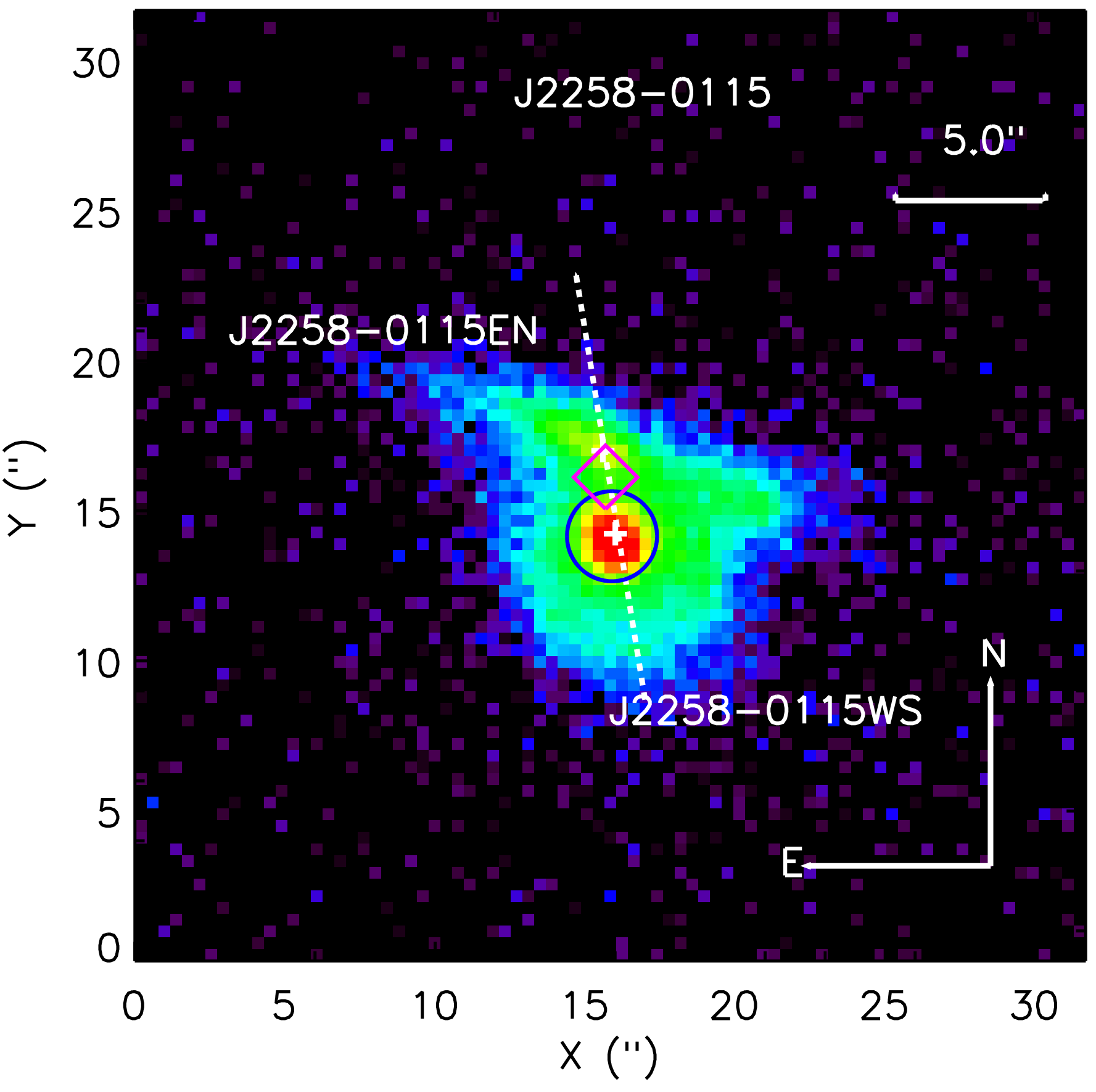}
  \includegraphics[width=12.2cm,height=5.0cm]{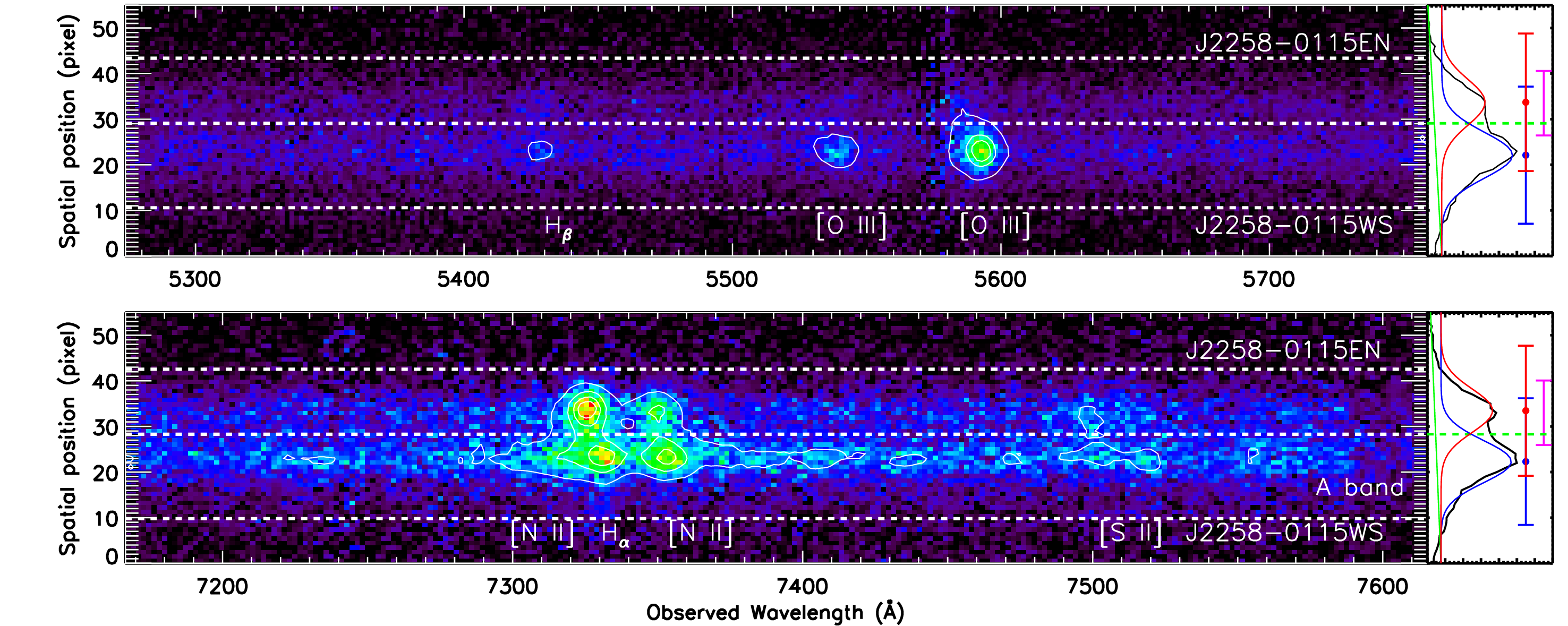}
\caption{Same as Fig.\,\ref{slitimages of J0933+2114} but for J2258-0115. }
\label{slitimages of J2258-0115}
\end{figure*}


\begin{figure*}[ht]
\centering
\includegraphics[scale=0.51]{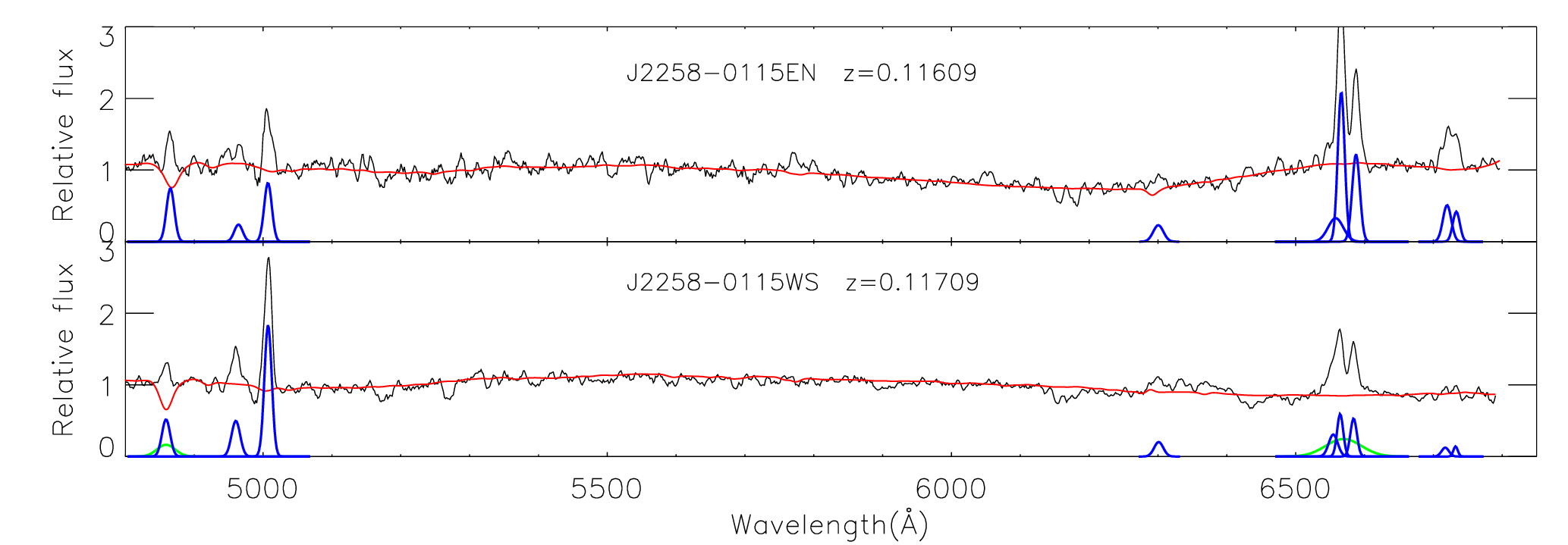}
\caption{Same as Fig.\,\ref{spectra_fitting:J0933+2114} but for J2258-0115. The spectra from LJT.}
\label{spectra_fitting:J2258-0115}
\end{figure*}

\begin{figure*}[ht]
  \centering
  \includegraphics[scale=0.58]{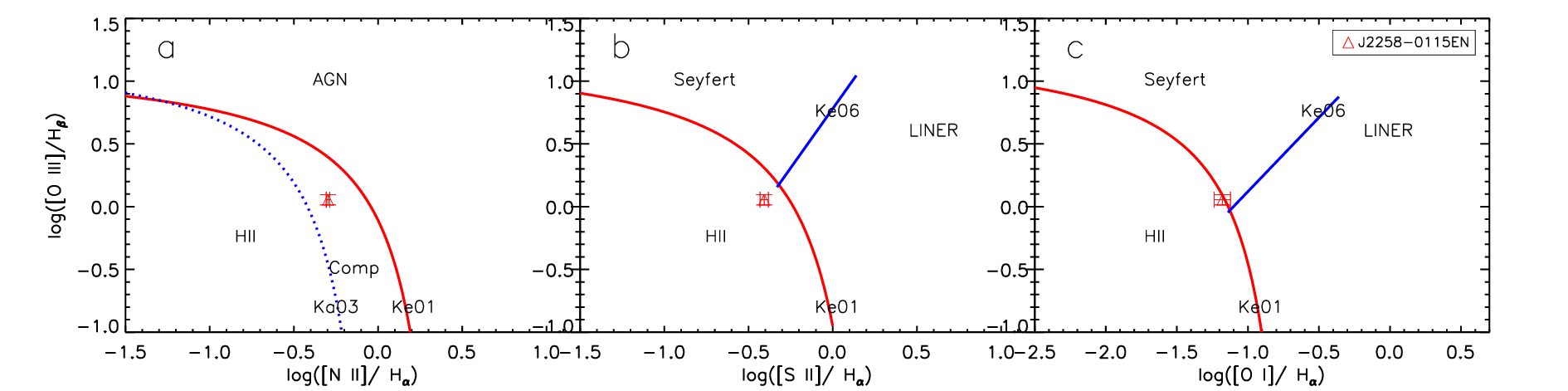}

\caption{Same as Fig.\,\ref{BPT_DAGN} but for J2258-0115.} 
 
\label{BPT diagrams of J2258-0115}
\end{figure*}

\subsection*{\rm Dual AGN: J225810.01-011516.26} 
\label{DAGN of J2258-0115}

%
%
%

Two sets of AGN spectra are spatially resolved as shown in Fig.\,\ref{slitimages of J2258-0115}, so the two cores, i.e. J2258$-$0115EN and J2258$-$0115WS can be identified separately.

The fitting of extracted 1D spectra of the two cores are shown in Fig.\,\ref{spectra_fitting:J2258-0115}. The redshifts, FWHMs of emission lines and emission line flux ratios of the two cores, measured from the 1D spectra, are presented in Tables\,\ref{BPT classify} and \ref{finally DAGN}.
The spectrum of J2258$-$0115EN doesn't show a broad line (Fig.\,\ref{spectra_fitting:J2258-0115}). 
We use the BPT diagram to distinguish this AGN shown in Fig.\,\ref{BPT diagrams of  J2258-0115} and it is classified as Comp (AGN).
The spectrum of J2258$-$0115WS has broad line (Fig.\,\ref{spectra_fitting:J2258-0115}, FWHM\,$> 2000$\,km\,s$^{-1}$ as measured from H$\alpha$ broad line component) and thus it is a Type I AGN.

The object J225810.01-011516.26 has been revealed as a dual AGN composed of ambiguous galaxy (AGN) (J2258$-$0115EN) and Type I AGN (J2258$-$0115WS). This dual AGN has a separation of 6.9 kpc and a velocity offset of $ 300 \pm 21 $ km $\mathrm{s}^{-1}$.

\clearpage



\begin{figure*}[ht]
  \centering
  \includegraphics[width=5.20cm,height=5.0cm]{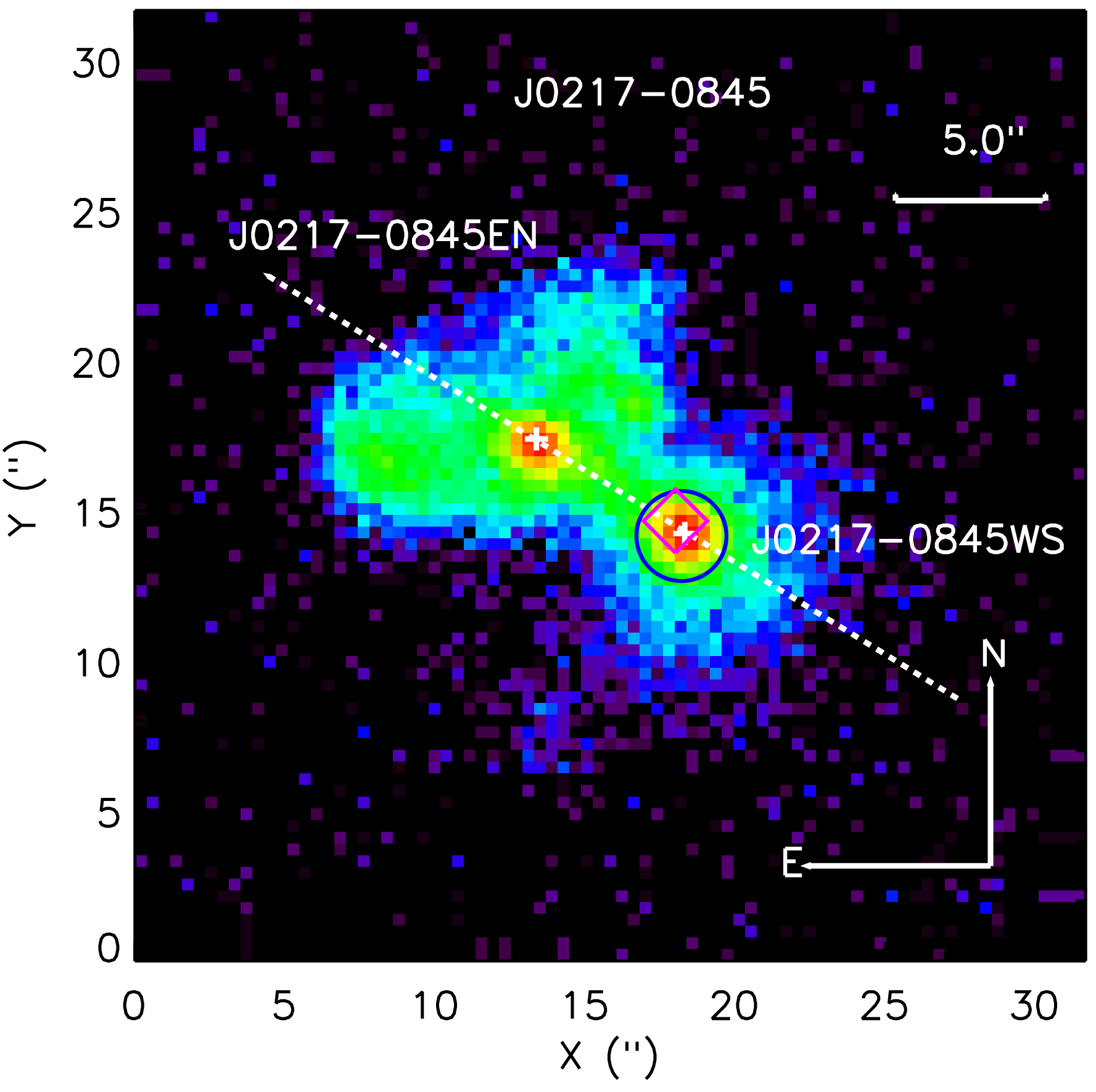}
  \includegraphics[width=12.2cm,height=5.0cm]{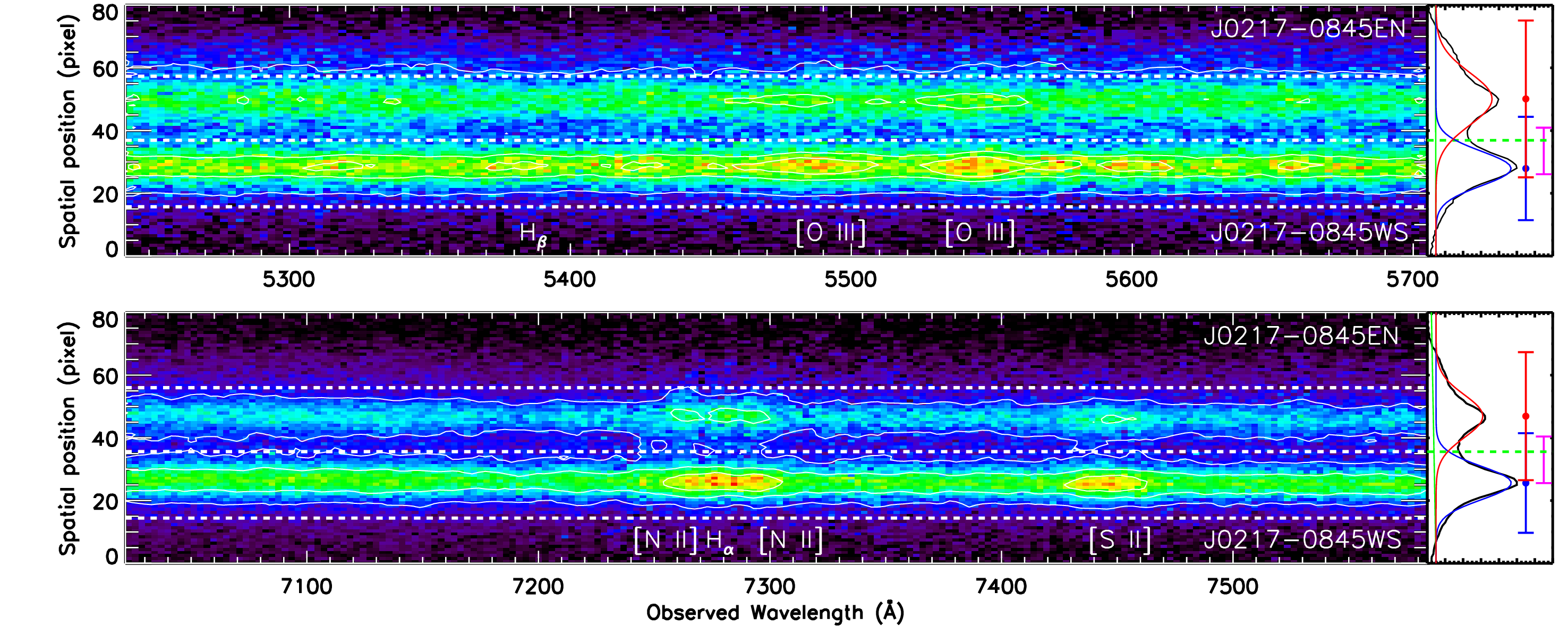}
\caption{Same as Fig.\,\ref{slitimages of J0933+2114} but for J0217-0845. }
\label{slitimages of J0217-0845}
\end{figure*}


\begin{figure*}[ht]
\centering
\includegraphics[scale=0.51]{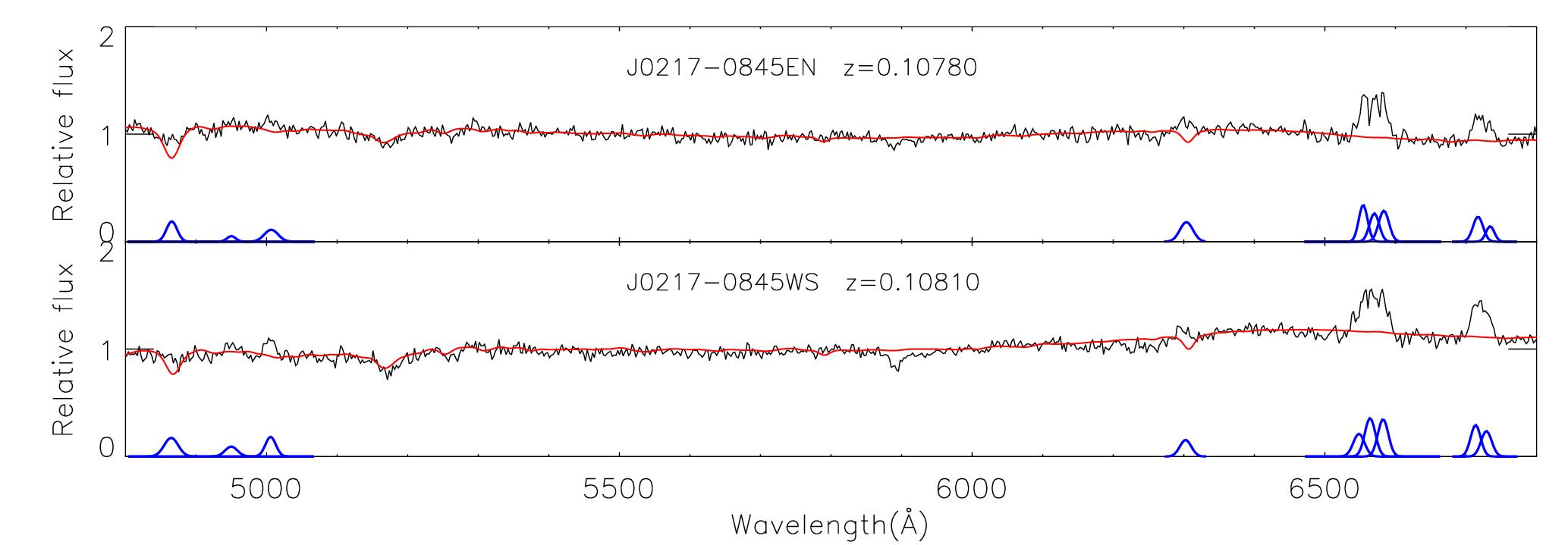}
\caption{Same as Fig.\,\ref{spectra_fitting:J0933+2114} but for J0217-0845. The spectra from LJT.}
\label{spectra_fitting:J0217-0845}
\end{figure*}

\begin{figure*}[ht]
  \centering
  \includegraphics[scale=0.58]{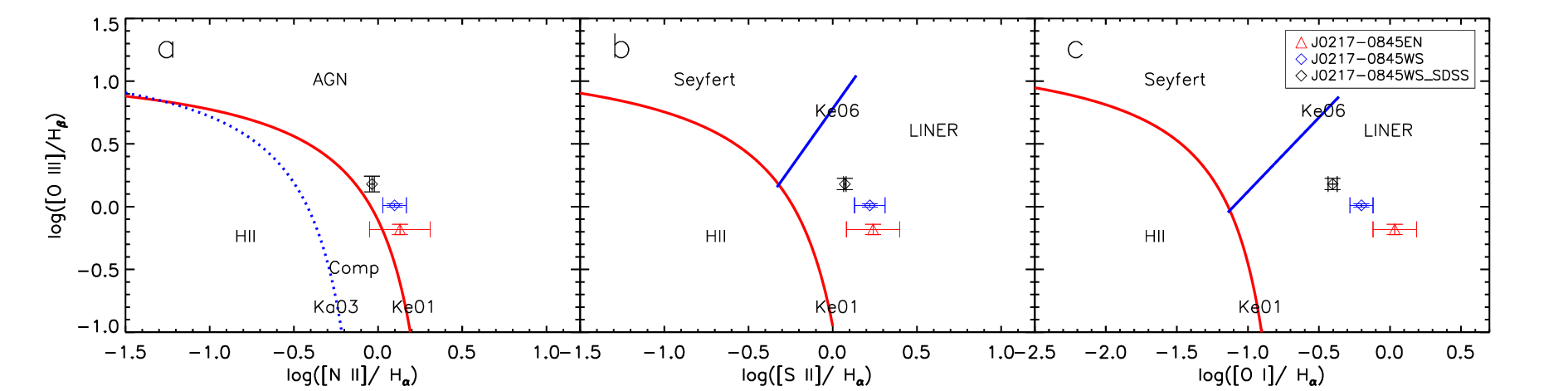}

\caption{Same as Fig.\,\ref{BPT_DAGN} but for J0217-0845.}

\label{BPT diagrams of J0217-0845}
\end{figure*}

\subsection*{\rm Dual AGN: J021703.81-084515.97}
\label{DAGN of J0217-0845}

%
%
%

Two sets of AGN spectra are spatially resolved as shown in Fig.\,\ref{slitimages of J0217-0845}, so the two cores, i.e. J0217-0845EN and J0217-0845WS can be identified separately. 

The fitting of extracted 1D spectra of the two cores are shown in Fig.\,\ref{spectra_fitting:J0217-0845}. The redshifts, FWHMs of emission lines and emission line flux ratios of the two cores, measured from the 1D spectra, are presented in Tables\,\ref{BPT classify} and \ref{finally DAGN}.
For the two cores, no broad line components are detected, we therefore use BPT diagram to classsify their types (Fig.\,\ref{BPT diagrams of J0217-0845}). According to the diagnosis, both cores are classified as LINER. 

The object J021703.81-084515.97 has been revealed as a dual AGN composed of LINER (J0217-0845EN) and LINER (J0217-0845WS). This dual AGN has a separation of 12.0 kpc and a velocity offset of $90 \pm 40 $ km $\mathrm{s}^{-1}$.

\clearpage


\begin{figure*}[ht]
  \centering
  \includegraphics[width=5.20cm,height=5.0cm]{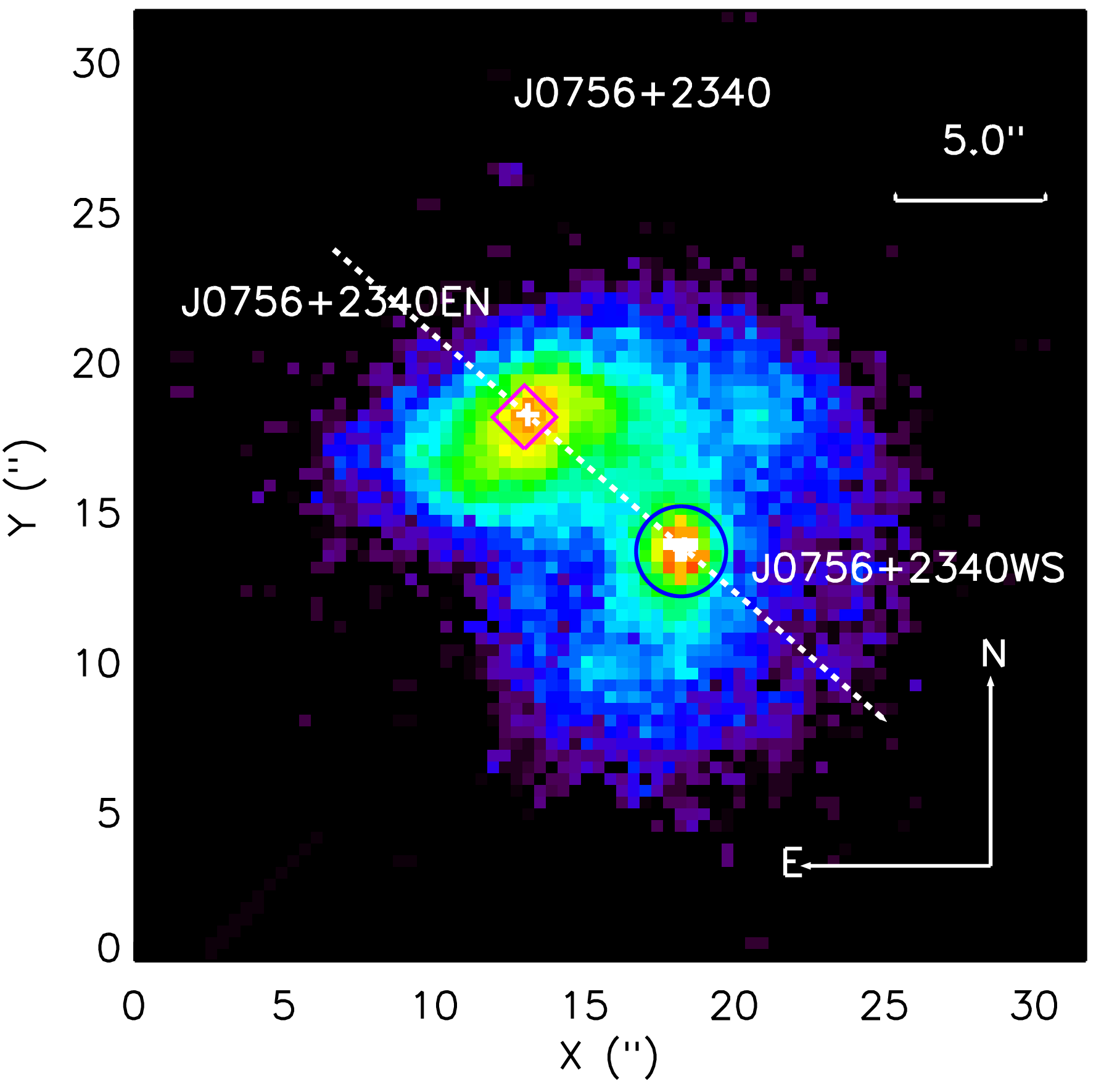}
  \includegraphics[width=12.2cm,height=5.0cm]{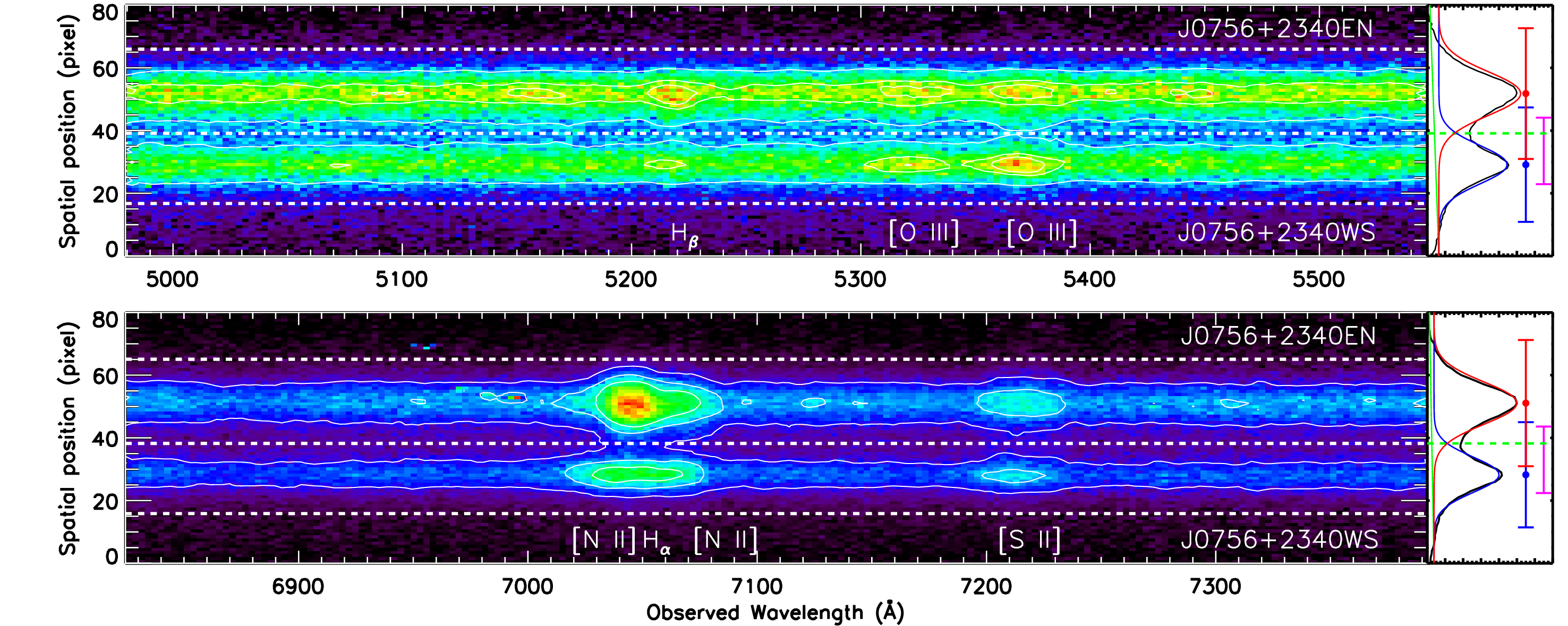}
\caption{Same as Fig.\,\ref{slitimages of J0933+2114} but for J0756+2340. }
\label{slitimages of J0756+2340}
\end{figure*}


\begin{figure*}[ht]
\centering
\includegraphics[scale=0.51]{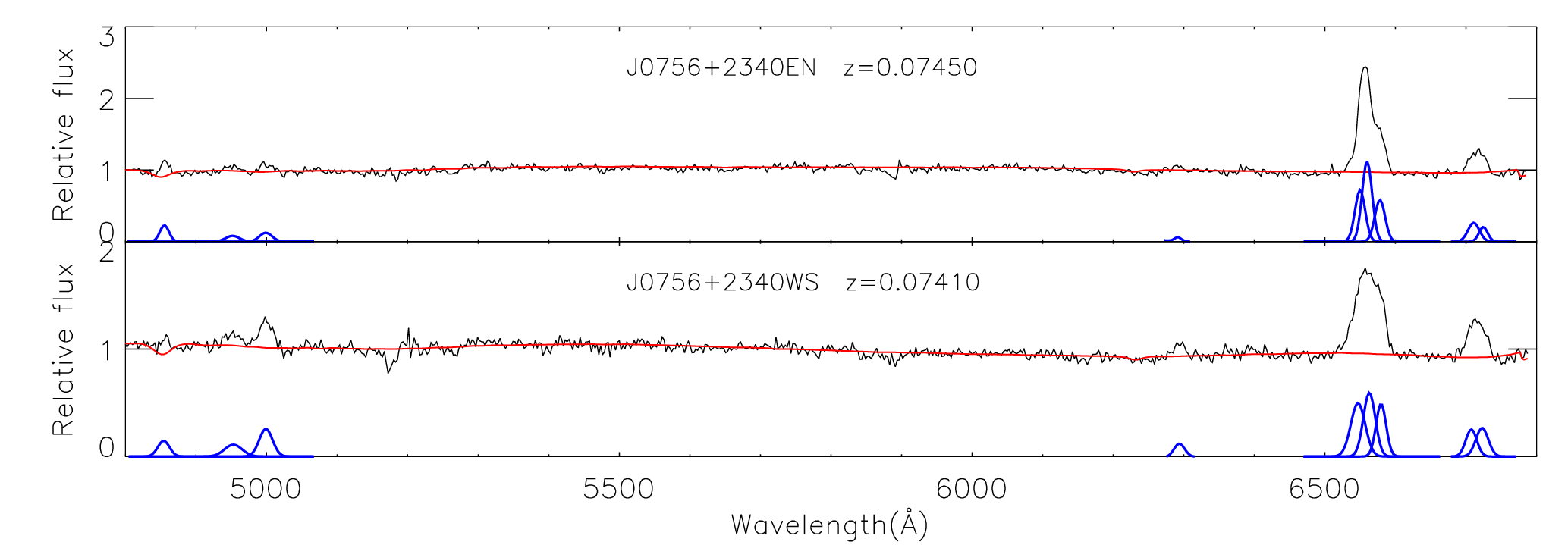}
\caption{Same as Fig.\,\ref{spectra_fitting:J0933+2114} but for J0756+2340. The spectra from LJT.}
\label{spectra_fitting:J0756+2340}
\end{figure*}

\begin{figure*}[ht]
  \centering
  \includegraphics[scale=0.58]{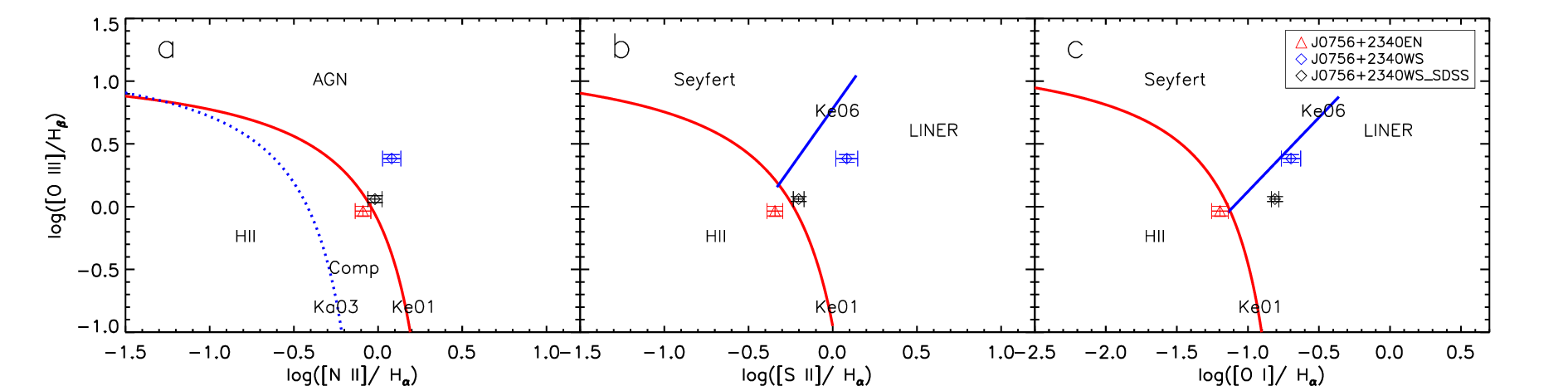}

\caption{Same as Fig.\,\ref{BPT_DAGN} but for J0756+2340.} 
 
\label{BPT diagrams of J0756+2340}
\end{figure*}

\subsection*{\rm Dual AGN: J075621.37+234043.97}
\label{DAGN of J0756+2340}


%
%

Two sets of AGN spectra are spatially resolved as shown in Fig.\,\ref{slitimages of J0756+2340}, so the two cores, i.e. J0756+2340EN and J0756+2340WS can be identified separately.

The fitting of extracted 1D spectra of the two cores are shown in Fig.\,\ref{spectra_fitting:J0756+2340}. 
The redshifts, FWHMs of emission lines and emission line flux ratios of the two cores, measured from the 1D spectra, are presented in Tables\,\ref{BPT classify} and \ref{finally DAGN}. 
For the two cores, no broad line components are detected, we therefore use BPT diagram to classsify their types (Fig.\,\ref{BPT diagrams of J0756+2340}). 
According to the diagnosis, J0756+2340EN is classified as Comp (AGN) and J0756+2340WS is classified as LINER (AGN).

The object J075621.37+234043.97 has been revealed as a dual AGN composed of ambiguous AGN (J0756+2340EN) and LINER (J0756+2340WS). This dual AGN has a separation of 9.6 kpc and a velocity offset of $120 \pm 40 $ km $\mathrm{s}^{-1}$.

\clearpage


\begin{figure*}[ht]
  \centering
  \includegraphics[width=5.20cm,height=5.0cm]{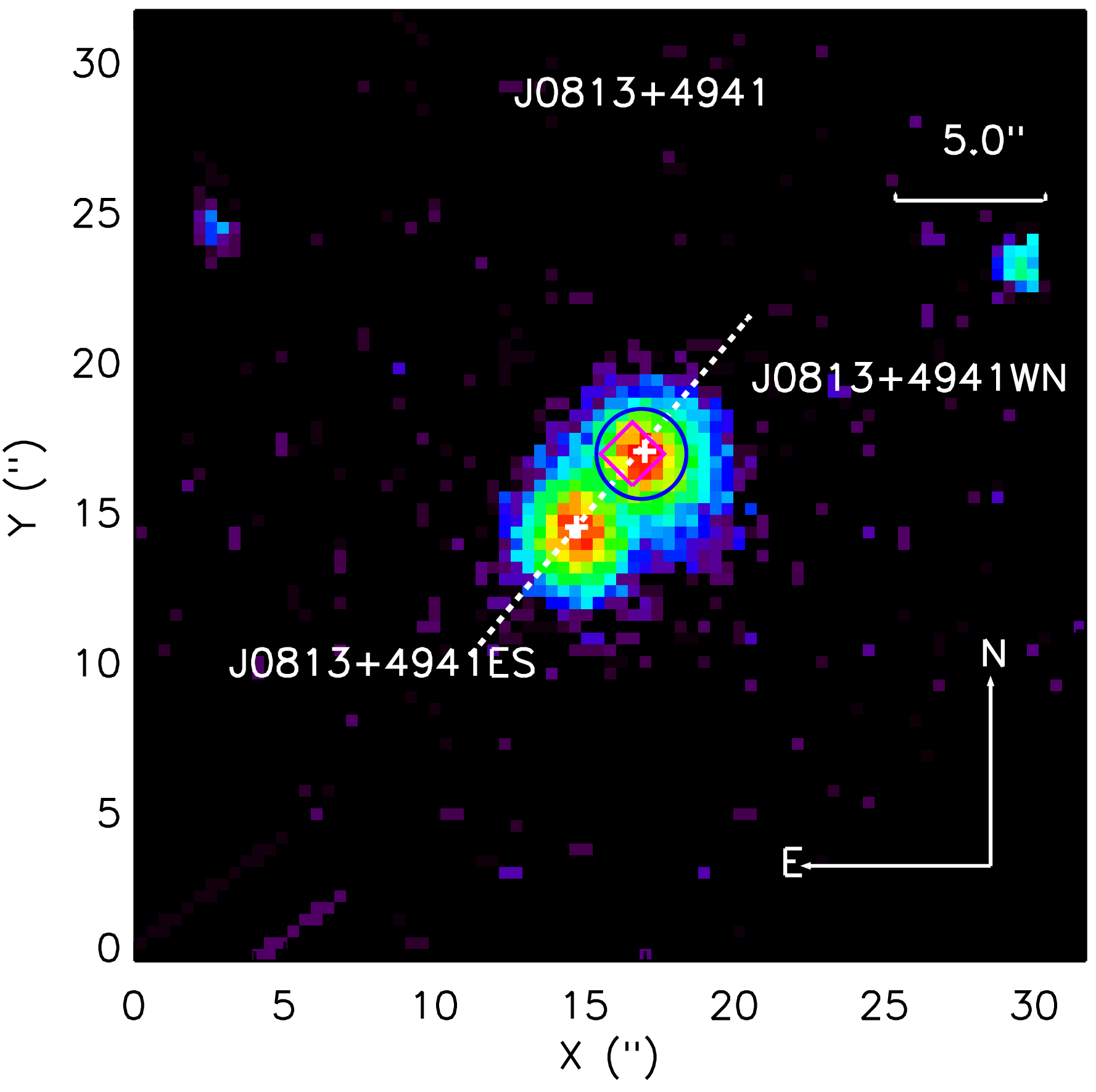}
  \includegraphics[width=12.2cm,height=5.0cm]{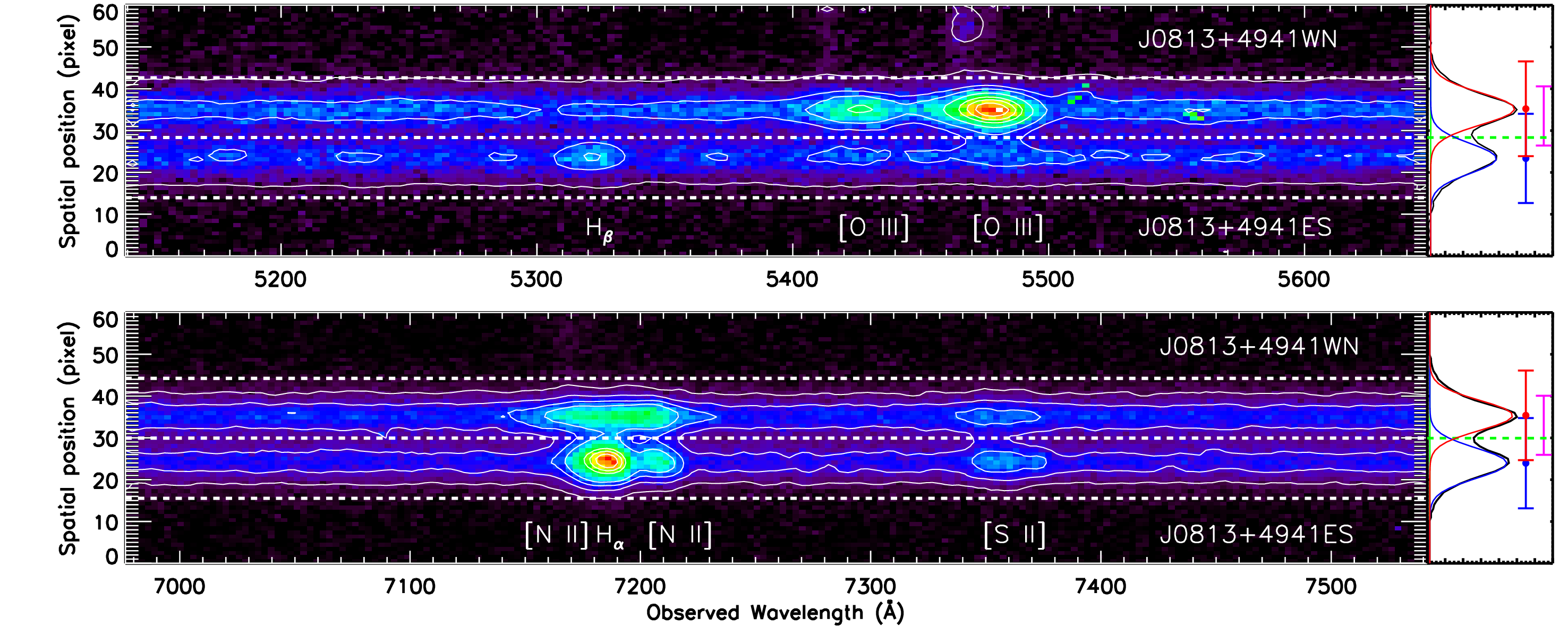}
\caption{Same as Fig.\,\ref{slitimages of J0933+2114} but for J0813+4941. }
\label{slitimages of J0813+4941}
\end{figure*}


\begin{figure*}[ht]
\centering
\includegraphics[scale=0.51]{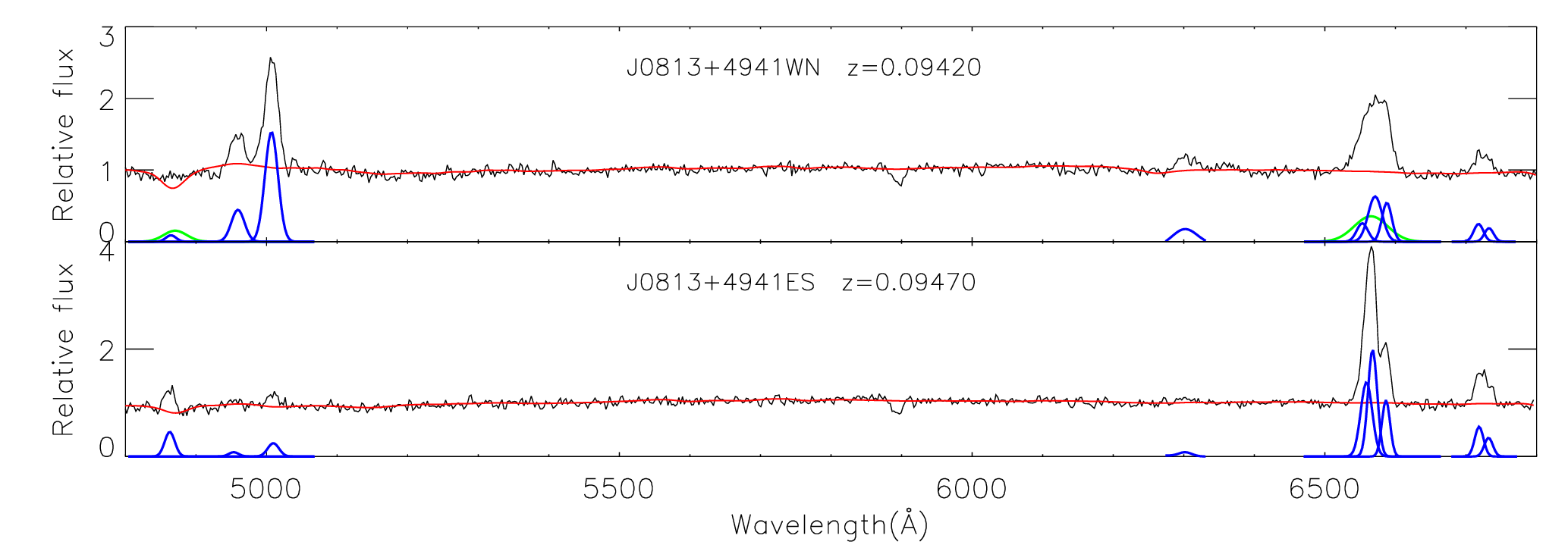}
\caption{Same as Fig.\,\ref{spectra_fitting:J0933+2114} but for J0813+4941. The spectra from LJT.}
\label{spectra_fitting:J0813+4941}
\end{figure*}

\begin{figure*}[ht]
  \centering
  \includegraphics[scale=0.58]{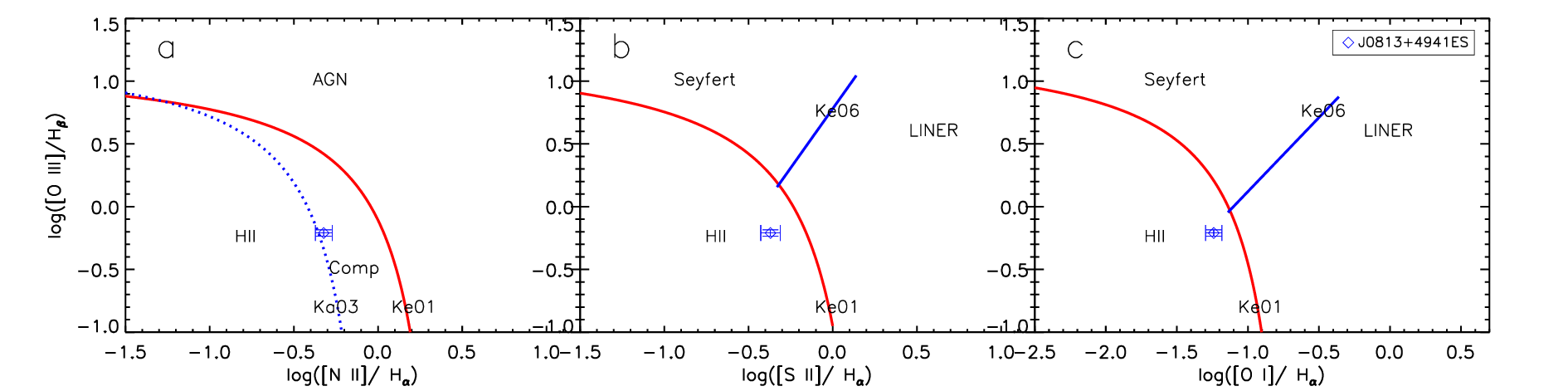}

\caption{Same as Fig.\,\ref{BPT_DAGN} but for J0813+4941.} 
 
\label{BPT diagrams of J0813+4941}
\end{figure*}

\subsection*{\rm Dual AGN: J081347.49+494109.83}
\label{DAGN of J0813+4941}

%
%
%

Two sets of AGN spectra are spatially resolved as shown in Fig.\,\ref{slitimages of J0813+4941}, so the two cores, i.e. J0813+4941WN and J0813+4941ES can be identified separately.

The fitting of extracted 1D spectra of the two cores are shown in Fig.\,\ref{spectra_fitting:J0813+4941}. The redshifts, FWHMs of emission lines and emission line flux ratios of the two cores, measured from the 1D spectra, are presented in Tables\,\ref{BPT classify} and \ref{finally DAGN}.
The spectrum of J0813+4941WN has broad line (Fig.\,\ref{spectra_fitting:J0813+4941}; FWHM\,$> 2000$\,km\,s$^{-1}$ as measured from H$\alpha$ broad line component) and thus it is a Type I AGN.
The spectrum of J0813+4941ES only has narrow line component.
We use the BPT diagram to distinguish this AGN shown in Fig.\,\ref{BPT diagrams of J0813+4941} and it is classified as Comp (AGN).

The object J081347.49+494109.83 has been revealed as a dual AGN composed of Type I AGN (J0813+4941WN) and Comp (J0813+4941ES). 
This dual AGN has a separation of 5.8 kpc and a velocity offset of $150 \pm 40 $ km $\mathrm{s}^{-1}$.

\clearpage

\begin{figure*}[ht]
  \centering
  \includegraphics[width=5.20cm,height=5.0cm]{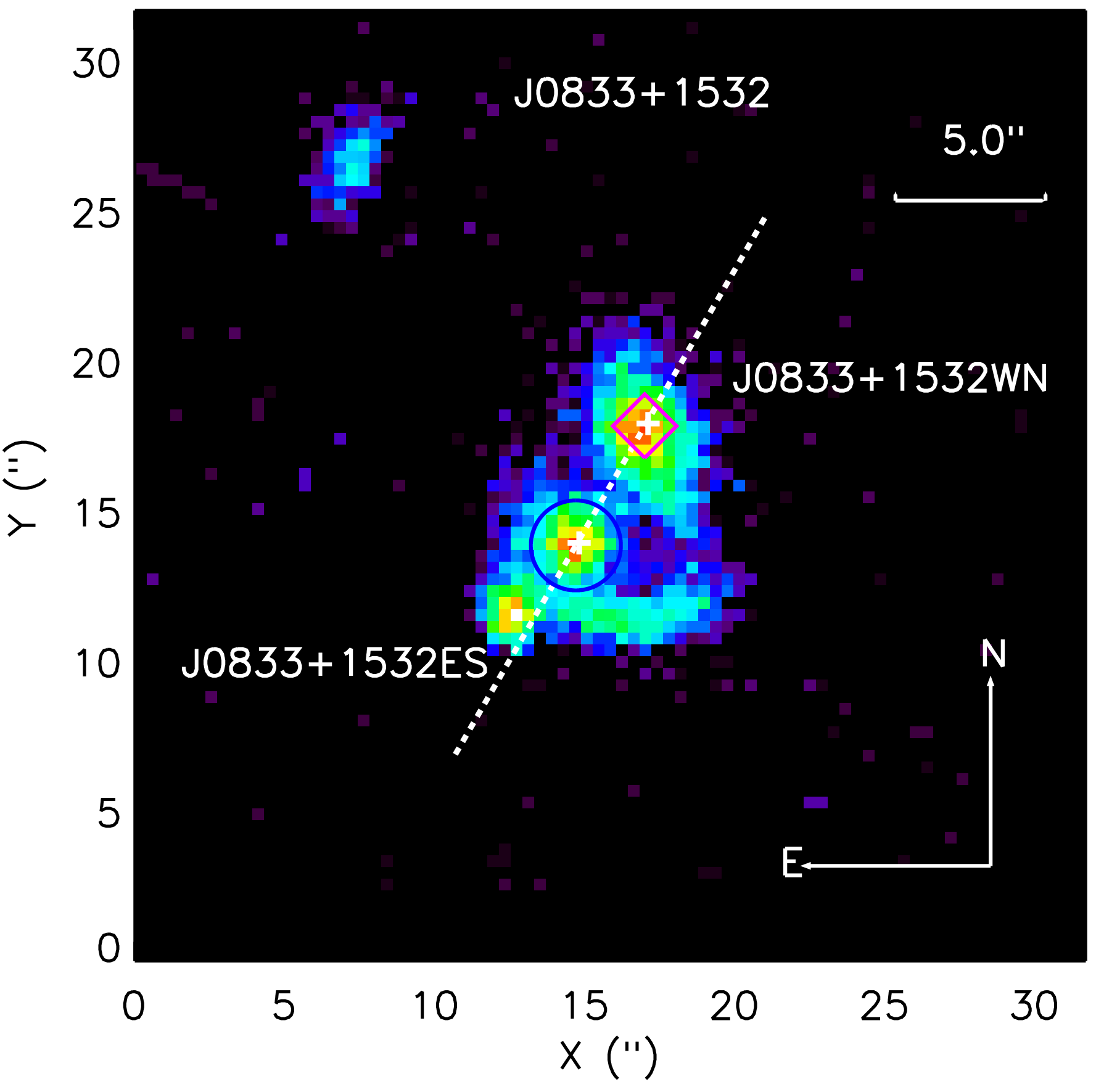}
  \includegraphics[width=12.2cm,height=5.0cm]{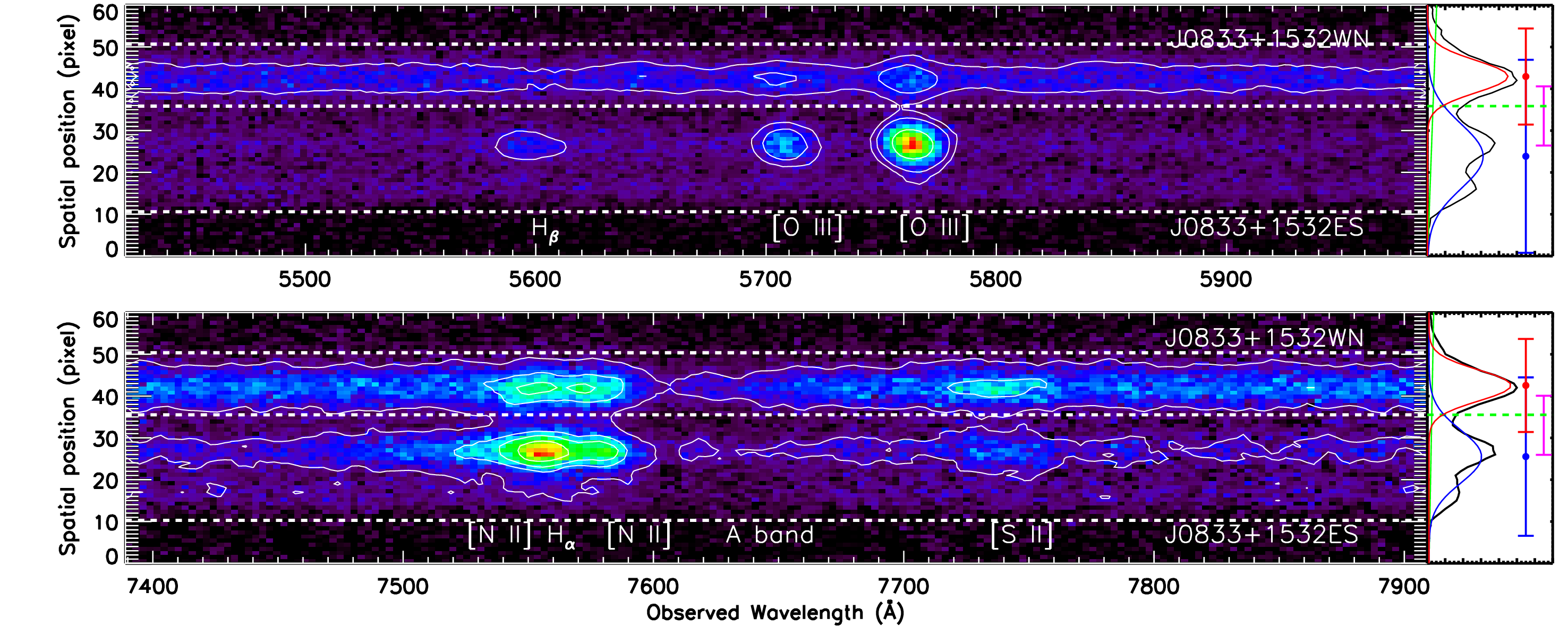}
\caption{Same as Fig.\,\ref{slitimages of J0933+2114} but for J0833+1532. }
\label{slitimages of J0833+1532}
\end{figure*}

\begin{figure*}[ht]
\centering
\includegraphics[scale=0.51]{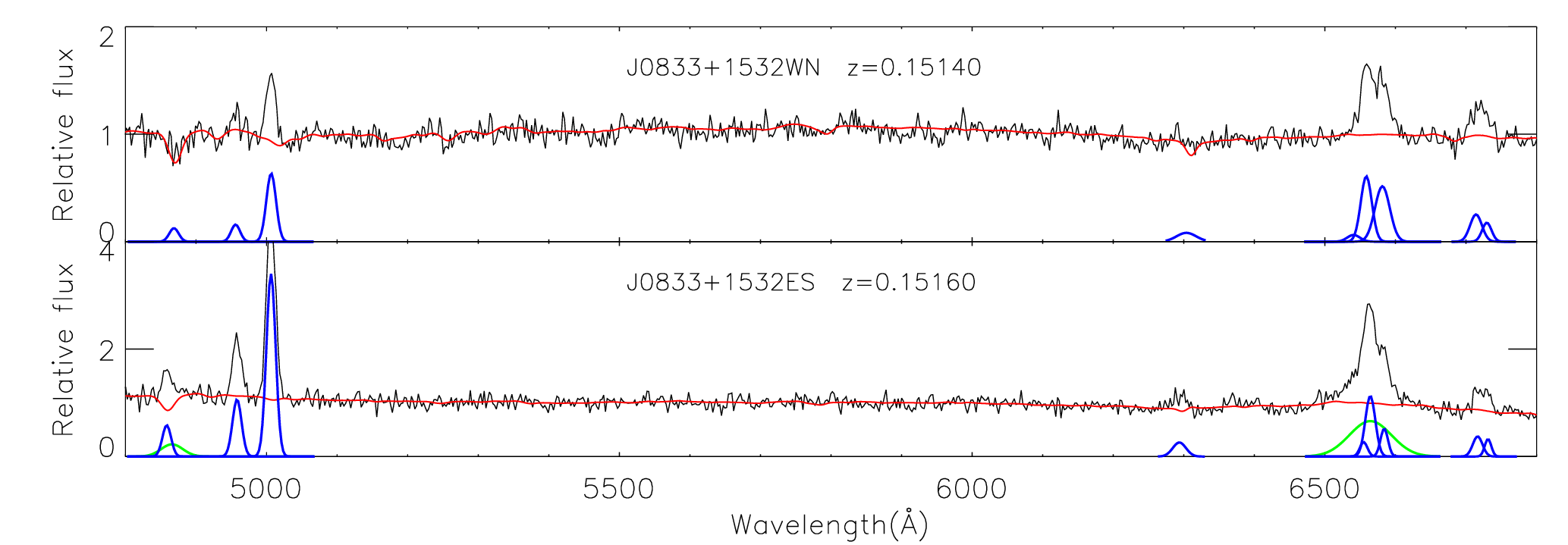}
\caption{Same as Fig.\,\ref{spectra_fitting:J0933+2114} but for J0833+1532. The spectra from LJT.}
\label{spectra_fitting:J0833+1532}
\end{figure*}

\begin{figure*}[ht]
  \centering
  \includegraphics[scale=0.58]{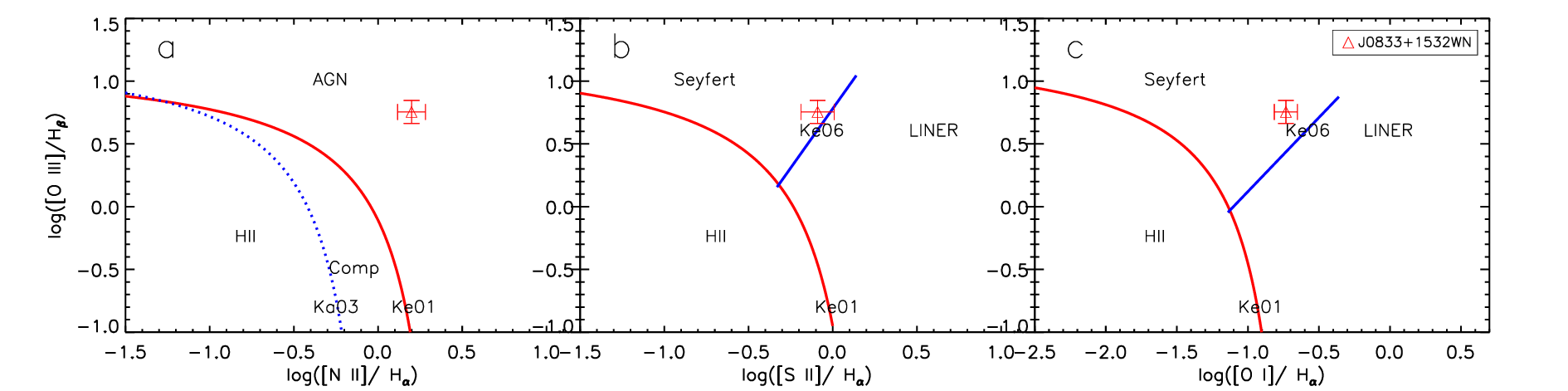}
\caption{Same as Fig.\,\ref{BPT_DAGN} but for J0833+1532.}

\label{BPT diagrams of J0833+1532}
\end{figure*}

\subsection*{\rm Dual AGN: J083355.49+153236.62}
\label{DAGN of J0833+1532}

%
%

Two sets of AGN spectra are spatially resolved, as shown in Fig.\,\ref{slitimages of J0833+1532}, so the two cores, i.e. J0833+1532WN and J0833+1532ES can be identified separately.

The fitting of extracted 1D spectra of the two cores are shown in Fig.\,\ref{spectra_fitting:J0833+1532}. The redshifts, FWHMs of emission lines and emission line flux ratios of the two cores, measured from the 1D spectra, are presented in Tables\,\ref{BPT classify} and \ref{finally DAGN}.
The spectrum of J0833+1532WN only has narrow line component.
We use the BPT diagram to distinguish this AGN shown in Fig.\,\ref{BPT diagrams of J0833+1532} and it is classified as Seyfert (AGN).
The spectrum of J0833+1532ES has broad line (Fig.\,\ref{spectra_fitting:J0833+1532}; FWHM\,$> 2000$\,km\,s$^{-1}$ as measured from H$\alpha$ broad line component) and thus it is a Type I AGN.

The object J083355.49+153236.62 has been revealed as a dual AGN composed of Seyfert (J0833+1532WN) and Type I AGN (J0833+1532ES). This dual AGN has a separation of 11.9 kpc and a velocity offset of $60 \pm 40 $ km $\mathrm{s}^{-1}$.

\clearpage


\begin{figure*}[ht]
  \centering
  \includegraphics[width=5.20cm,height=5.0cm]{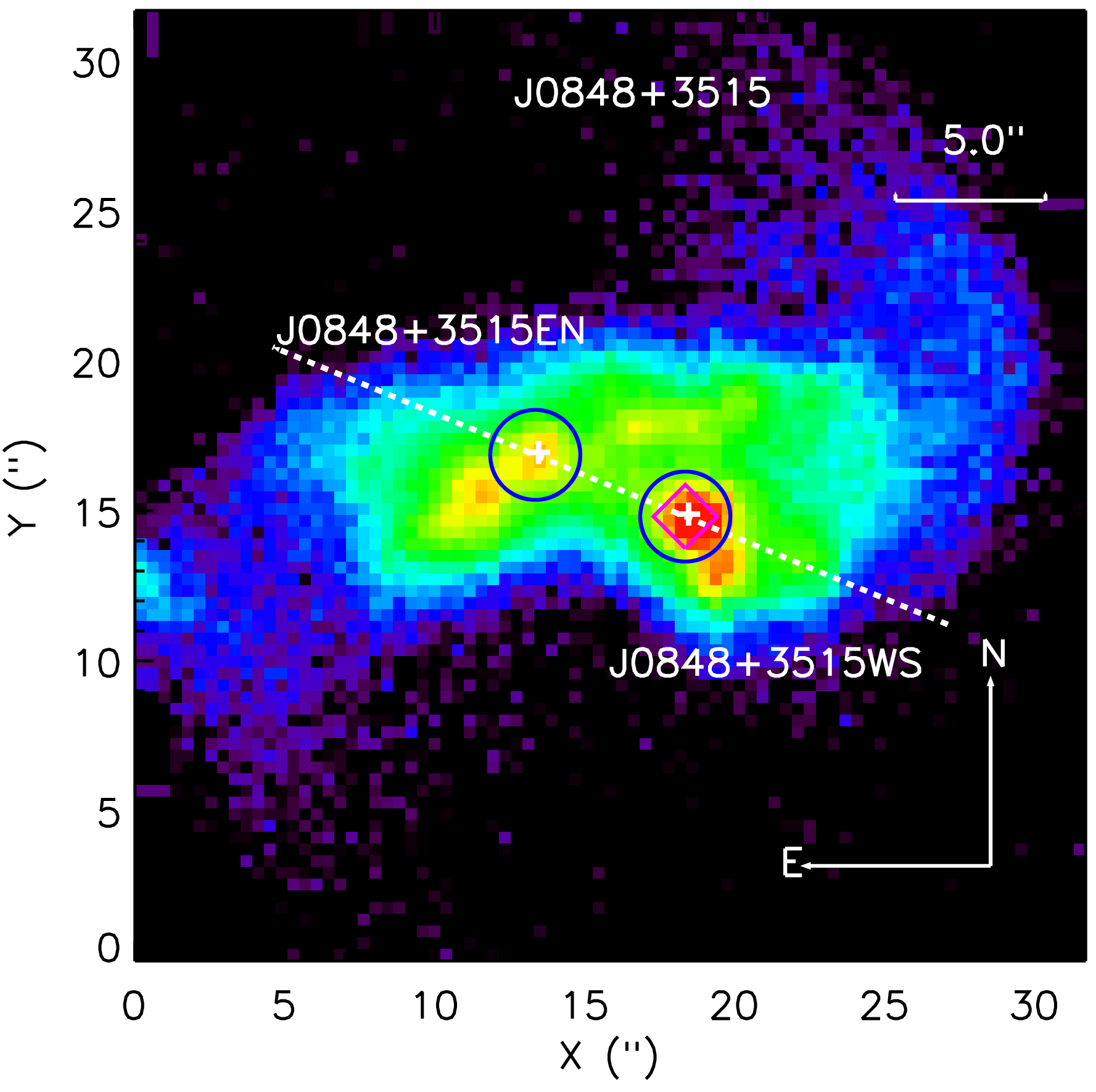}
  \includegraphics[width=12.2cm,height=5.0cm]{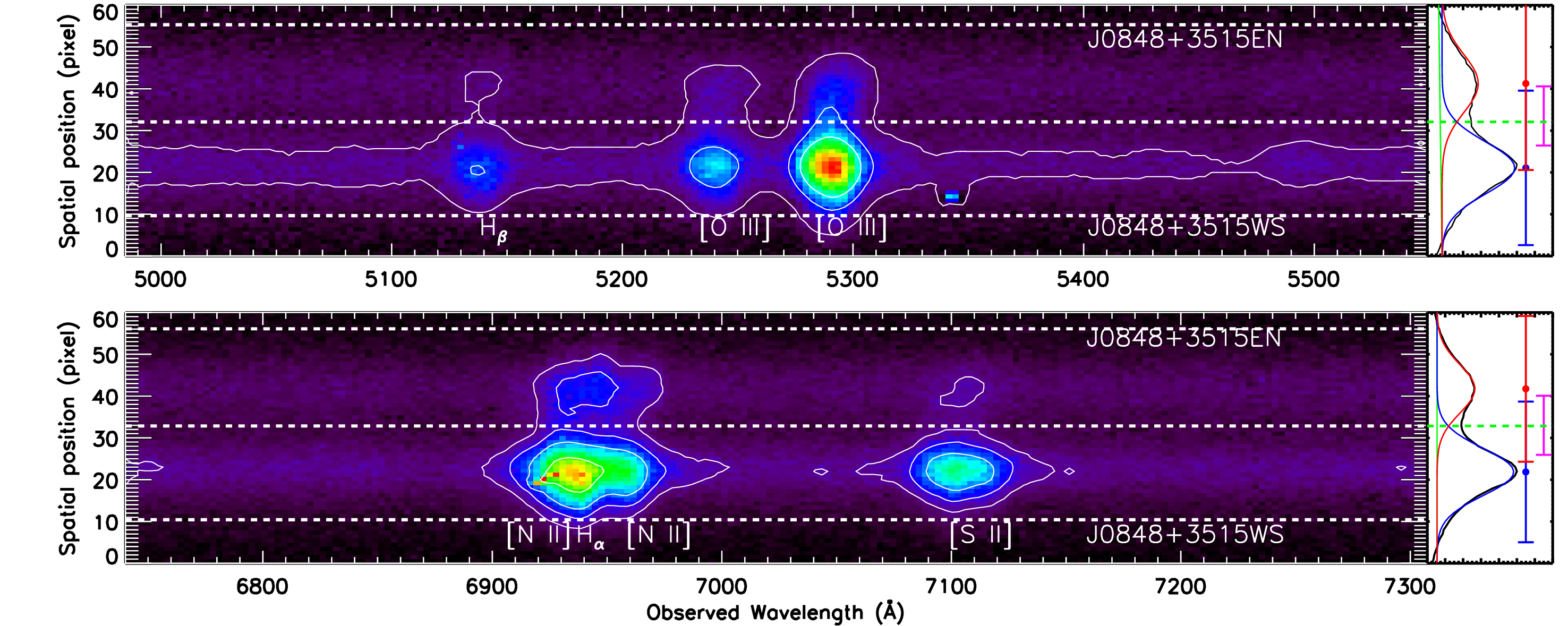}
\caption{Same as Fig.\,\ref{slitimages of J0933+2114} but for J0848+3515. }
\label{slitimages of J0848+3515}
\end{figure*}


\begin{figure*}[ht]
\centering
\includegraphics[scale=0.51]{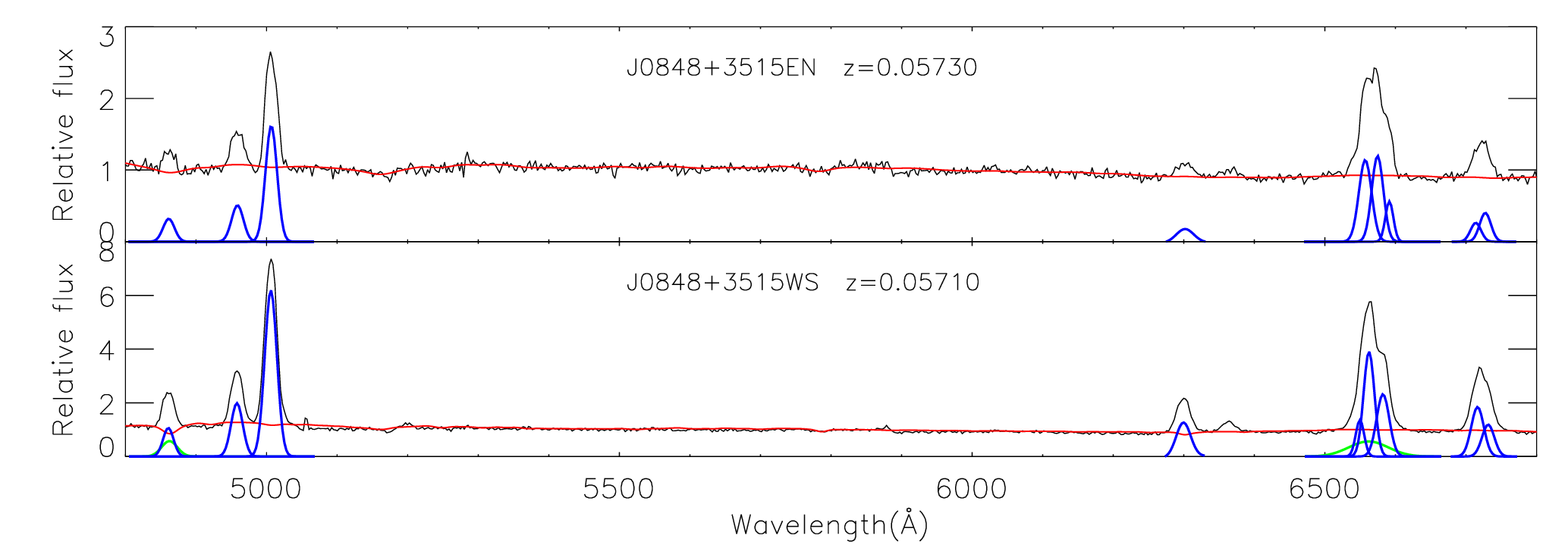}
\caption{Same as Fig.\,\ref{spectra_fitting:J0933+2114} but for J0848+3515. The spectra from LJT.}
\label{spectra_fitting:J0848+3515}
\end{figure*}

\begin{figure*}[ht]
  \centering
  \includegraphics[scale=0.58]{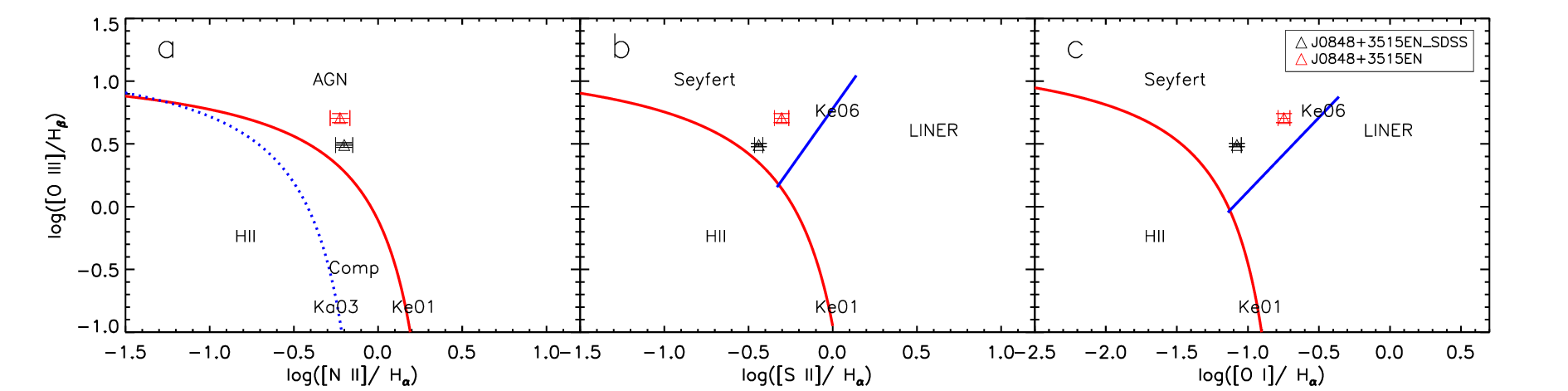}
\caption{Same as Fig.\,\ref{BPT_DAGN} but for J0848+3515.}

\label{BPT diagrams of J0848+3515}
\end{figure*}

\subsection*{\rm Dual AGN: J084809.69+351532.12}
\label{DAGN of J0848+3515}

%
%

Two sets of AGN spectra are spatially resolved as shown in Fig.\,\ref{slitimages of J0848+3515}, so the two cores, i.e. J0848+3515EN and J0848+3515WS can be identified separately. 

The fitting of extracted 1D spectra of the two cores are shown in Fig.\,\ref{spectra_fitting:J0848+3515}. The redshifts, FWHMs of emission lines and emission line flux ratios of the two cores, measured from the 1D spectra, are presented in Tables\,\ref{BPT classify} and \ref{finally DAGN}.
The spectrum of J0848+3515EN only has narrow line component.
We use the BPT diagram to distinguish this AGN shown in Fig.\,\ref{BPT diagrams of J0848+3515} and it is classified as Seyfert (AGN).
The spectrum of J0848+3515WS has broad line (Fig.\,\ref{spectra_fitting:J0848+3515}; FWHM\,$> 2000$\,km\,s$^{-1}$ as measured from H$\alpha$ broad line component) and thus it is a Type I AGN.

The object J084809.69+351532.12 has been revealed as a dual AGN composed of Seyfert (J0848+3515EN) and Type I AGN (J0848+3515WS). This dual AGN has a separation of 6.2 kpc and a velocity offset of $60 \pm 40 $ km $\mathrm{s}^{-1}$.

\clearpage

\begin{figure*}[ht]
  \centering
  \includegraphics[width=5.20cm,height=4.6cm]{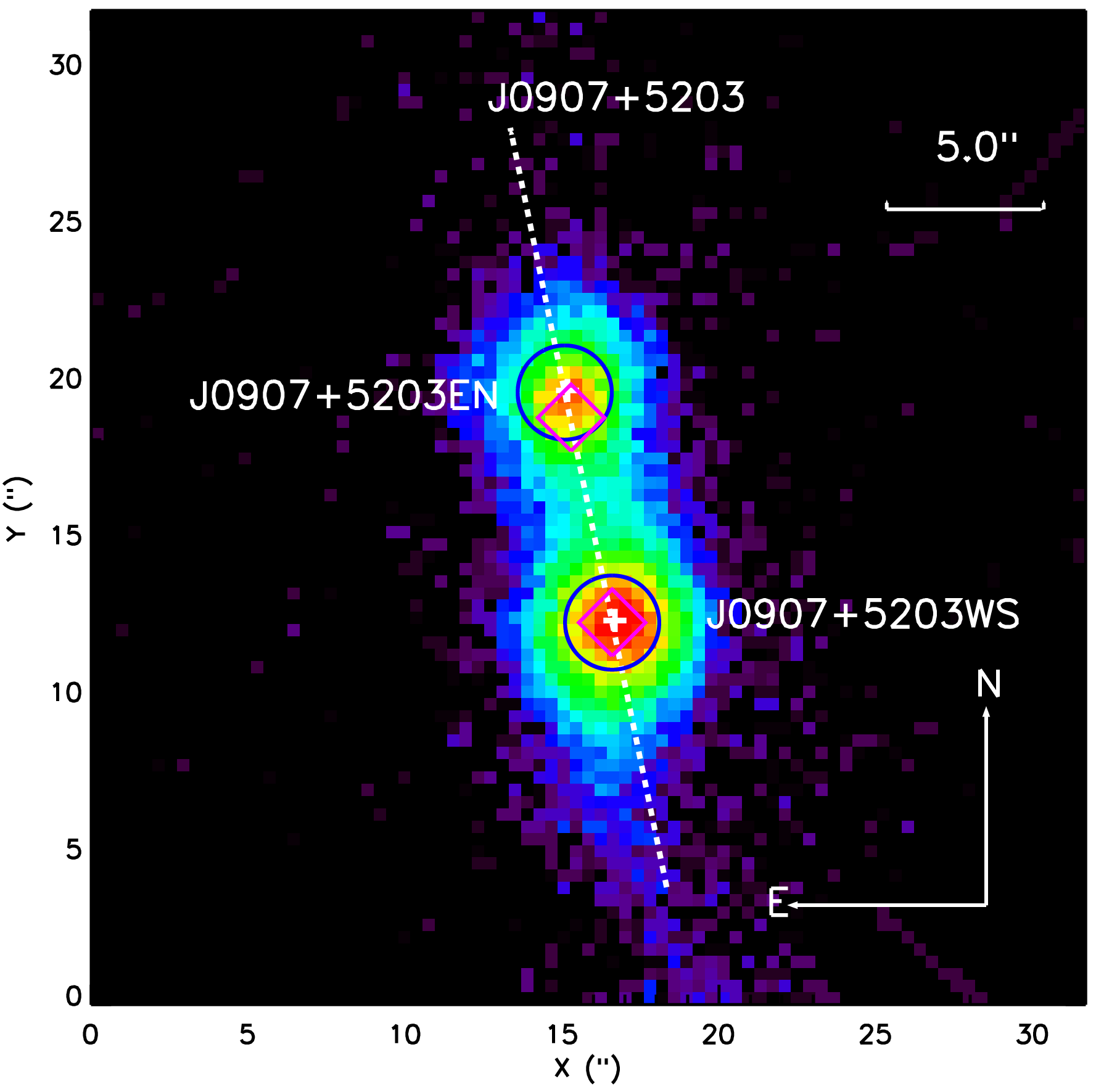}
  \includegraphics[width=12.2cm,height=4.6cm]{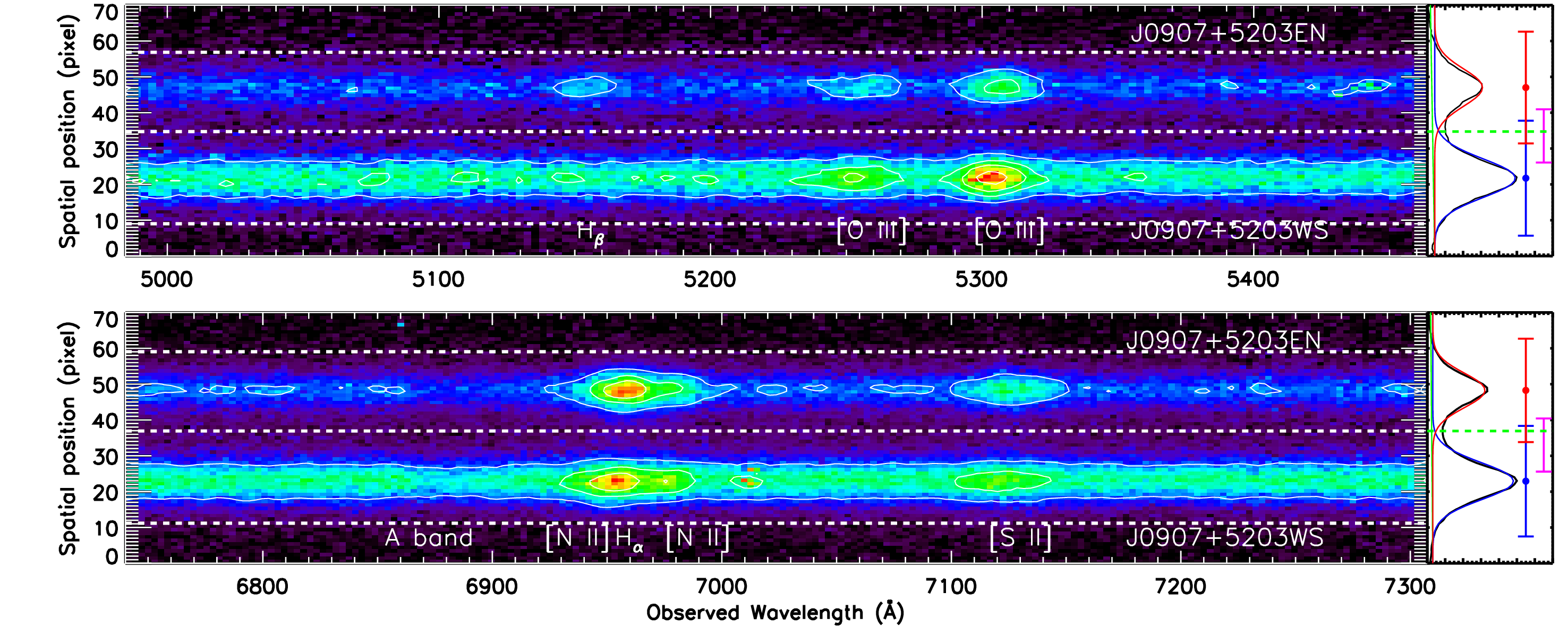}
\caption{Same as Fig.\,\ref{slitimages of J0933+2114} but for J0907+5203. }
\label{slitimages of J0907+5203}
\end{figure*}

\begin{figure*}[ht]
\centering
  \includegraphics[scale=0.51]{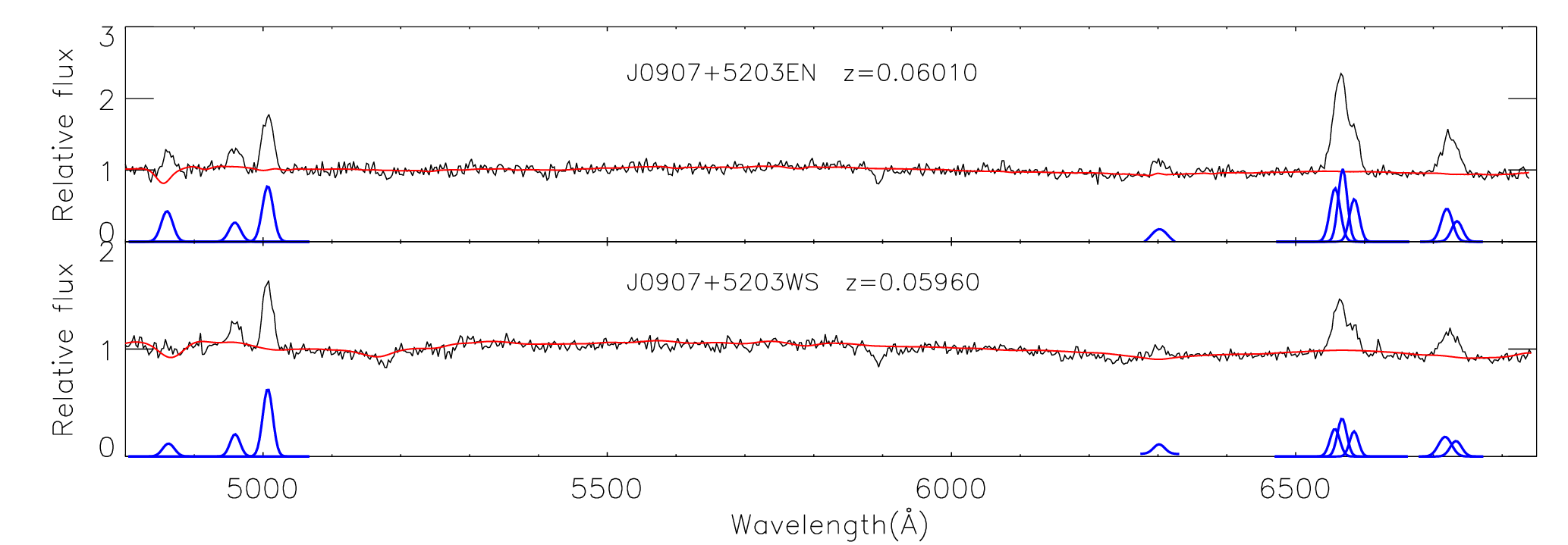}
\caption{Same as Fig.\,\ref{spectra_fitting:J0933+2114} but for J0907+5203. The spectra from LJT.}
\label{spectra_fitting:J0907+5203}
\end{figure*}

\begin{figure*}[ht]
  \centering
  \includegraphics[scale=0.58]{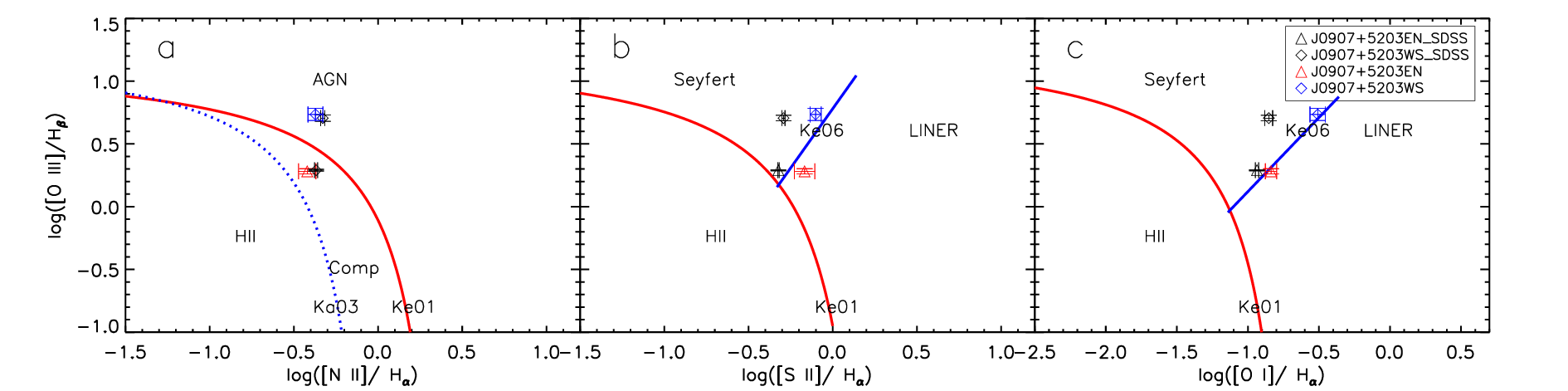}
\caption{Same as Fig.\,\ref{BPT_DAGN} but for J0907+5203.}

\label{BPT diagrams of J0907+5203}
\end{figure*}

\subsection*{\rm Dual AGN: J090714.61+520350.61}
\label{DAGN of J0907+5203}

%
%

Two sets of AGN spectra are spatially resolved as shown in Fig.\,\ref{slitimages of J0907+5203}, so the two cores, i.e. J0907+5203EN and J0907+5203WS can be identified separately.

The fitting of extracted 1D spectra of the two cores are shown in Fig.\,\ref{spectra_fitting:J0907+5203}. The redshifts, FWHMs of emission lines and emission line flux ratios of the two cores, measured from the 1D spectra, are presented in Tables\,\ref{BPT classify} and \ref{finally DAGN}.
For the two cores, no broad line components are detected, we therefore use BPT diagram to classsify their types (Fig.\,\ref{BPT diagrams of J0907+5203}). 
According to the diagnosis, J0907+5203EN is classified as ambiguous AGN and J0907+5203WS is classified as Seyfert (AGN).

The object J090714.61+520350.61 has been revealed as a dual AGN composed of ambiguous galaxy (J0907+5203EN) and Seyfert (J0907+5203WS). This dual AGN has a separation of 8.5 kpc and a velocity offset of $150 \pm 40 $ km $\mathrm{s}^{-1}$. 

\clearpage


\begin{figure*}[ht]
  \centering
  \includegraphics[width=5.20cm,height=5.0cm]{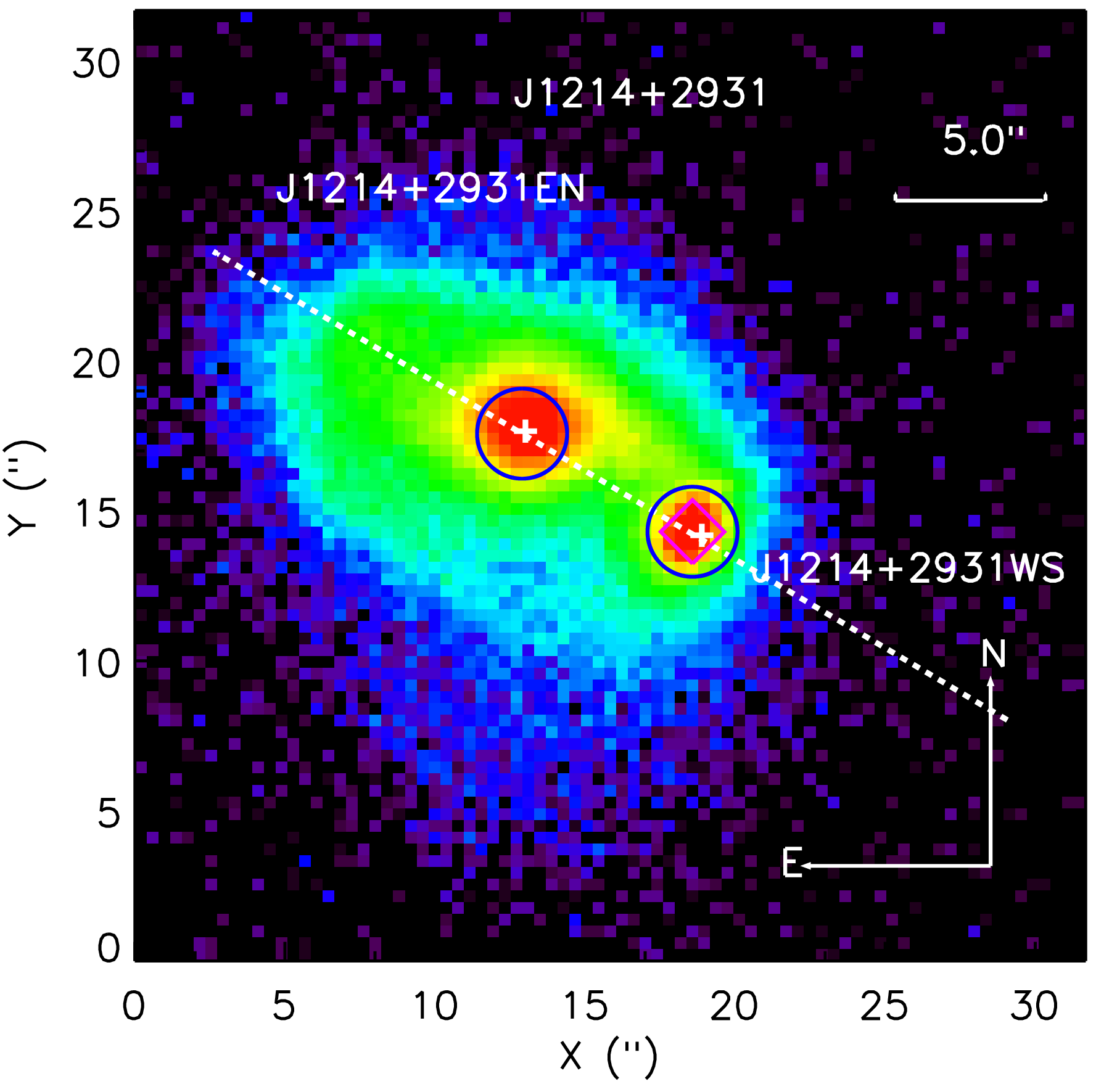}
  \includegraphics[width=12.2cm,height=5.0cm]{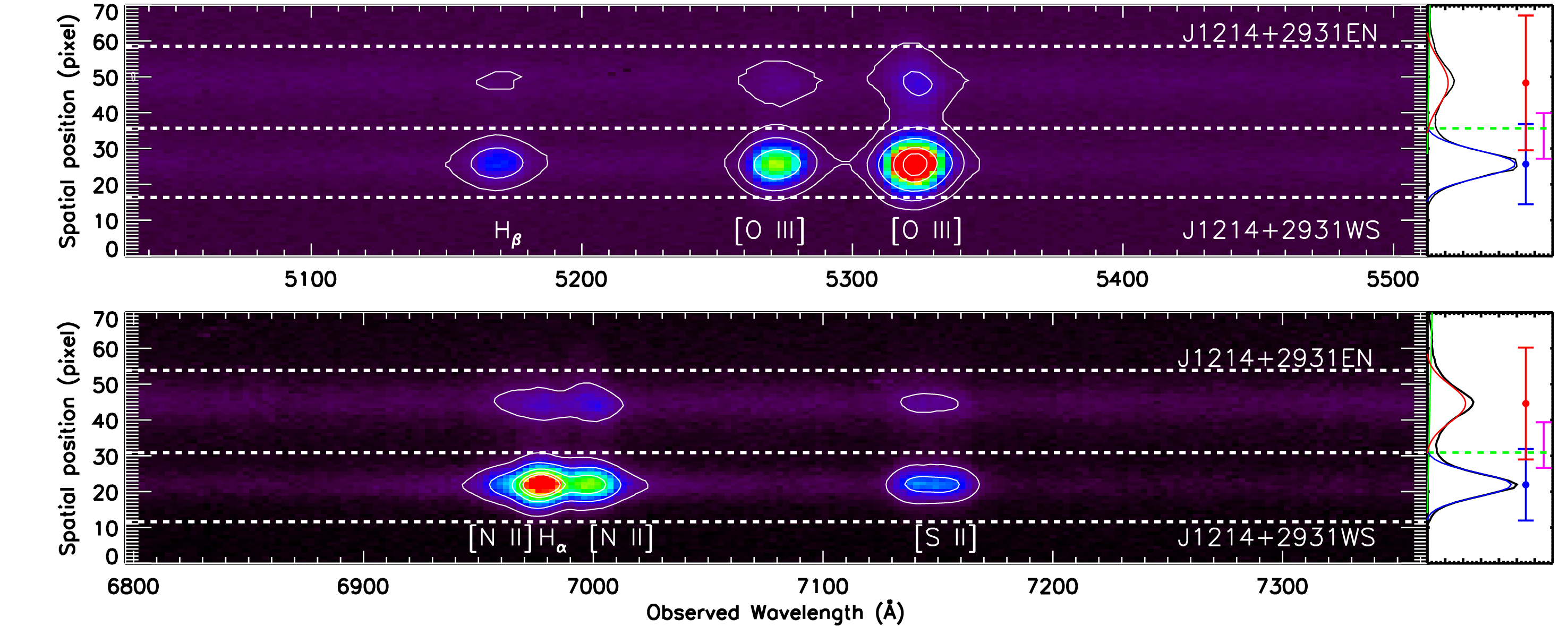}
\caption{Same as Fig.\,\ref{slitimages of J0933+2114} but for J1214+2931. }
\label{slitimages of J1214+2931}
\end{figure*}


\begin{figure*}[ht]
\centering
\includegraphics[scale=0.51]{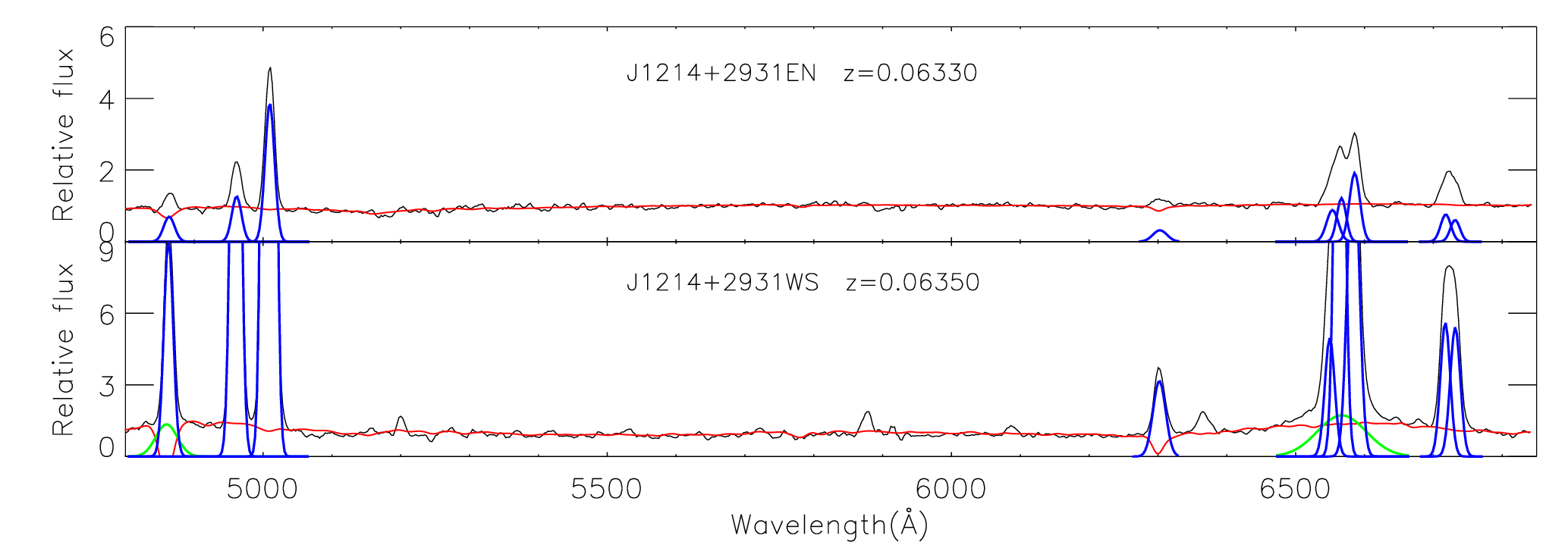}
\caption{Same as Fig.\,\ref{spectra_fitting:J0933+2114} but for J1214+2931. The spectra from LJT.}
\label{spectra_fitting:J1214+2931}
\end{figure*}

\begin{figure*}[ht]
  \centering
  \includegraphics[scale=0.58]{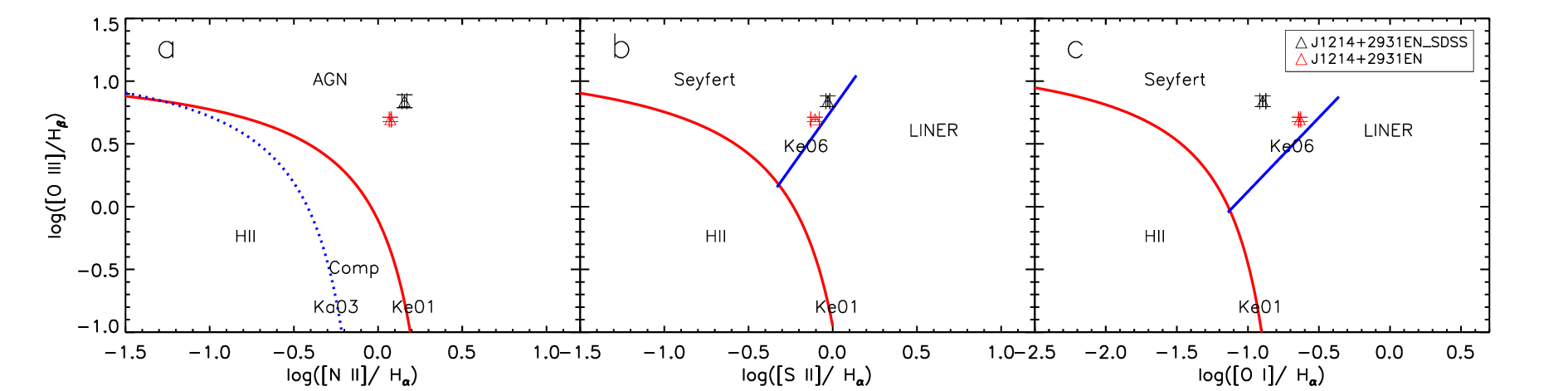}
\caption{Same as Fig.\,\ref{BPT_DAGN} but for J1214+2931.}

\label{BPT diagrams of J1214+2931}
\end{figure*}

\subsection*{\rm Dual AGN: J121418.25+293146.70}
\label{DAGN of J1214+2931}

%
%

Two sets of AGN spectra are spatially resolved as shown in Fig.\,\ref{slitimages of J1214+2931}, so the two cores, i.e. J1214+2931EN and J1214+2931WS can be identified separately. 

The fitting of extracted 1D spectra of the two cores are shown in Fig.\,\ref{spectra_fitting:J1214+2931}. The redshifts, FWHMs of emission lines and emission line flux ratios of the two cores, measured from the 1D spectra, are presented in Tables\,\ref{BPT classify} and \ref{finally DAGN}.
The spectrum of J1214+2931EN only has narrow line component.
We use the BPT diagram to distinguish this AGN shown in Fig.\,\ref{BPT diagrams of J1214+2931} and it is classified as Seyfert (AGN).
The spectrum of J1214+2931WS has broad line (Fig.\,\ref{spectra_fitting:J1214+2931}; FWHM\,$> 2000$\,km\,s$^{-1}$ as measured from H$\alpha$ broad line component) and thus it is a Type I AGN.

The object J121418.25+293146.70 has been revealed as a dual AGN composed of Seyfert (J1214+2931EN) and Type I AGN (J1214+2931WS). This dual AGN has a separation of 9.3 kpc and a velocity offset of $60 \pm 40 $ km $\mathrm{s}^{-1}$.

\clearpage


\begin{figure*}[ht]
  \centering
  \includegraphics[width=5.20cm,height=5.0cm]{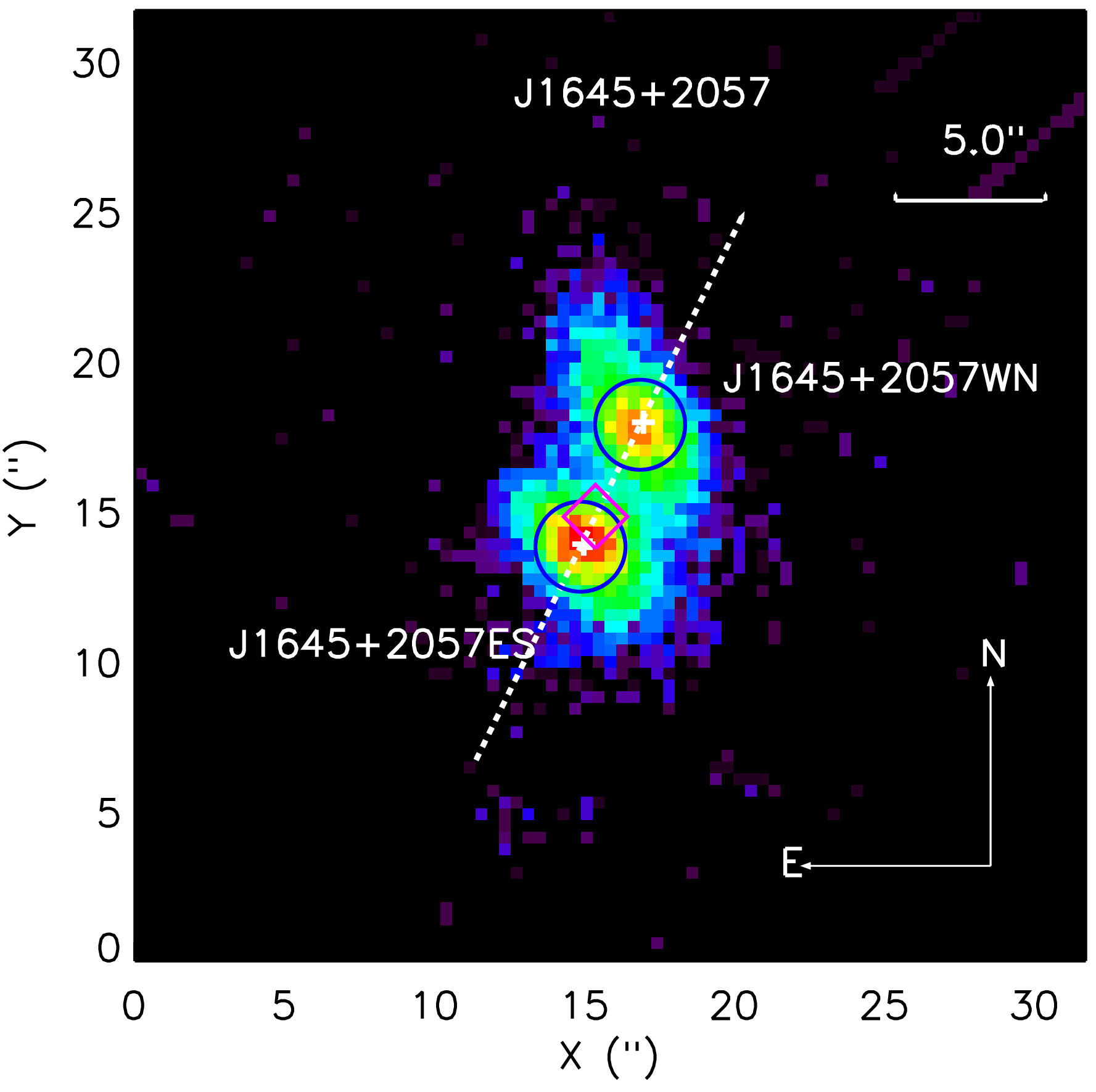}
  \includegraphics[width=12.2cm,height=5.0cm]{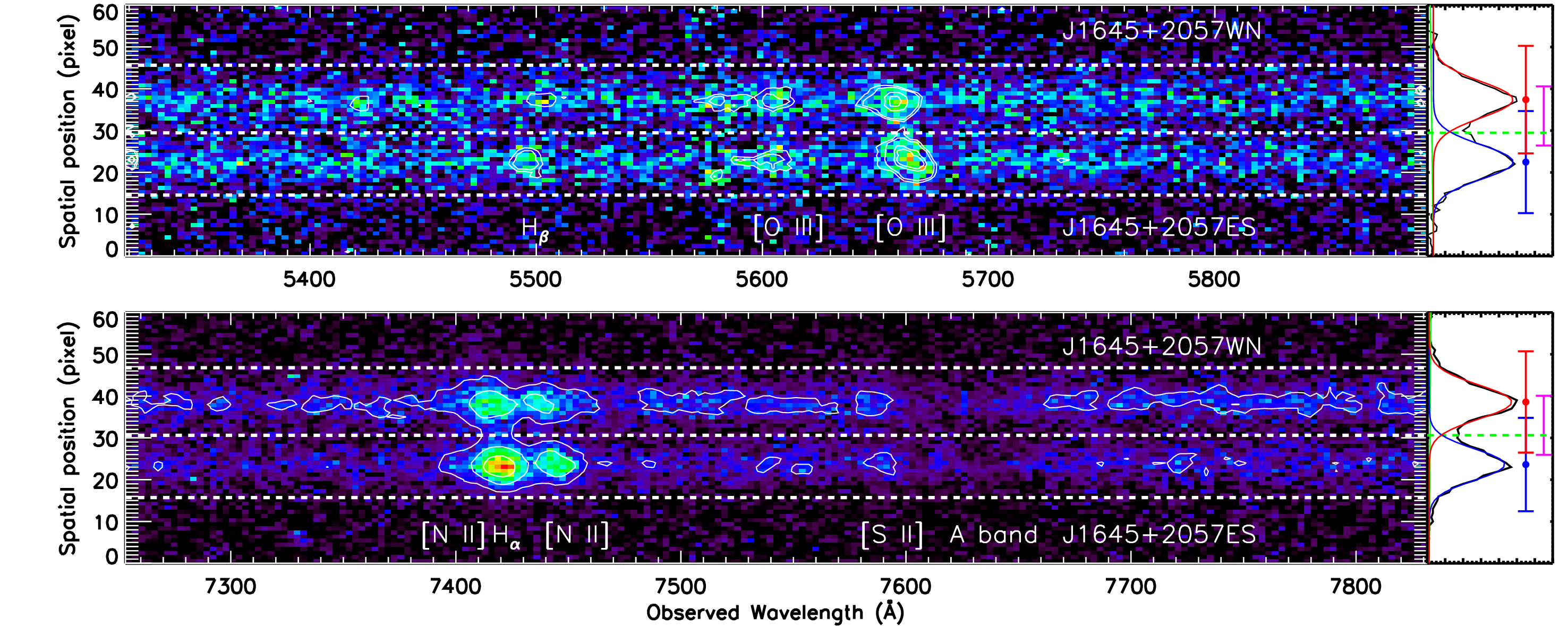}
\caption{Same as Fig.\,\ref{slitimages of J0933+2114} but for J1645+2057.}
\label{slitimages of J1645+2057}
\end{figure*}


\begin{figure*}[ht]
\centering
\includegraphics[scale=0.51]{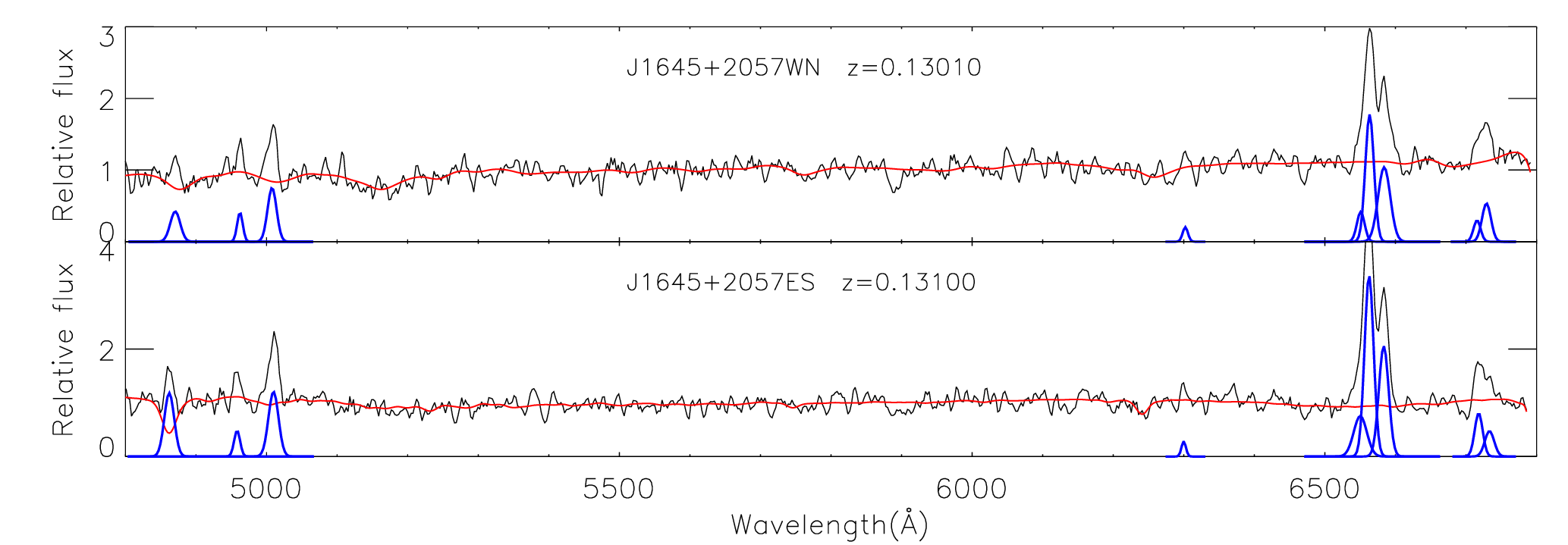}
\caption{Same as Fig.\,\ref{spectra_fitting:J0933+2114} but for J1645+2057. The spectra from LJT.}
\label{spectra_fitting:J1645+2057}
\end{figure*}

\begin{figure*}[ht]
  \centering
  \includegraphics[scale=0.58]{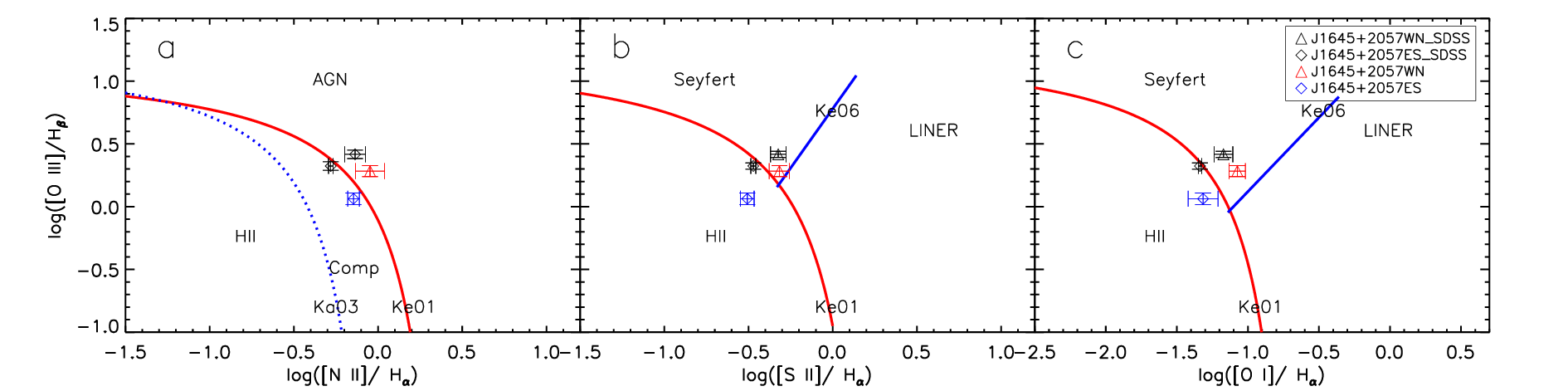}
\caption{Same as Fig.\,\ref{BPT_DAGN} but for J1645+2057.}

\label{BPT diagrams of J1645+2057}
\end{figure*}

\subsection*{\rm Dual AGN: J164507.91+205759.43}
\label{DAGN of J1645+2057}

%
%

Two sets of AGN spectra are spatially resolved as shown in Fig.\,\ref{slitimages of J1645+2057}, so the two cores, i.e. J1645+2057WN and J1645+2057ES can be identified respectively. 

The fitting of extracted 1D spectra of the two cores are shown in Fig.\,\ref{spectra_fitting:J1645+2057}. The redshifts, FWHMs of emission lines and emission line flux ratios of the two cores, measured from the 1D spectra, are presented in Tables\,\ref{BPT classify} and \ref{finally DAGN}.
For the two cores, no broad line components are detected, we therefore use BPT diagram to classsify their types (Fig.\,\ref{BPT diagrams of J1645+2057}). 
According to the diagnosis, J1645+2057WN is classified as Seyfert and J1645+2057ES is classified as Comp (AGN). 

The object J164507.91+205759.43 has been revealed as a dual AGN composed of Seyfert (J1645+2057WN) and Comp (J1645+2057ES). This dual AGN has a separation of 9.8 kpc and a velocity offset of $270 \pm 40 $ km $\mathrm{s}^{-1}$.

\clearpage

\begin{figure*}[ht]
  \centering
  \includegraphics[width=5.20cm,height=5.0cm]{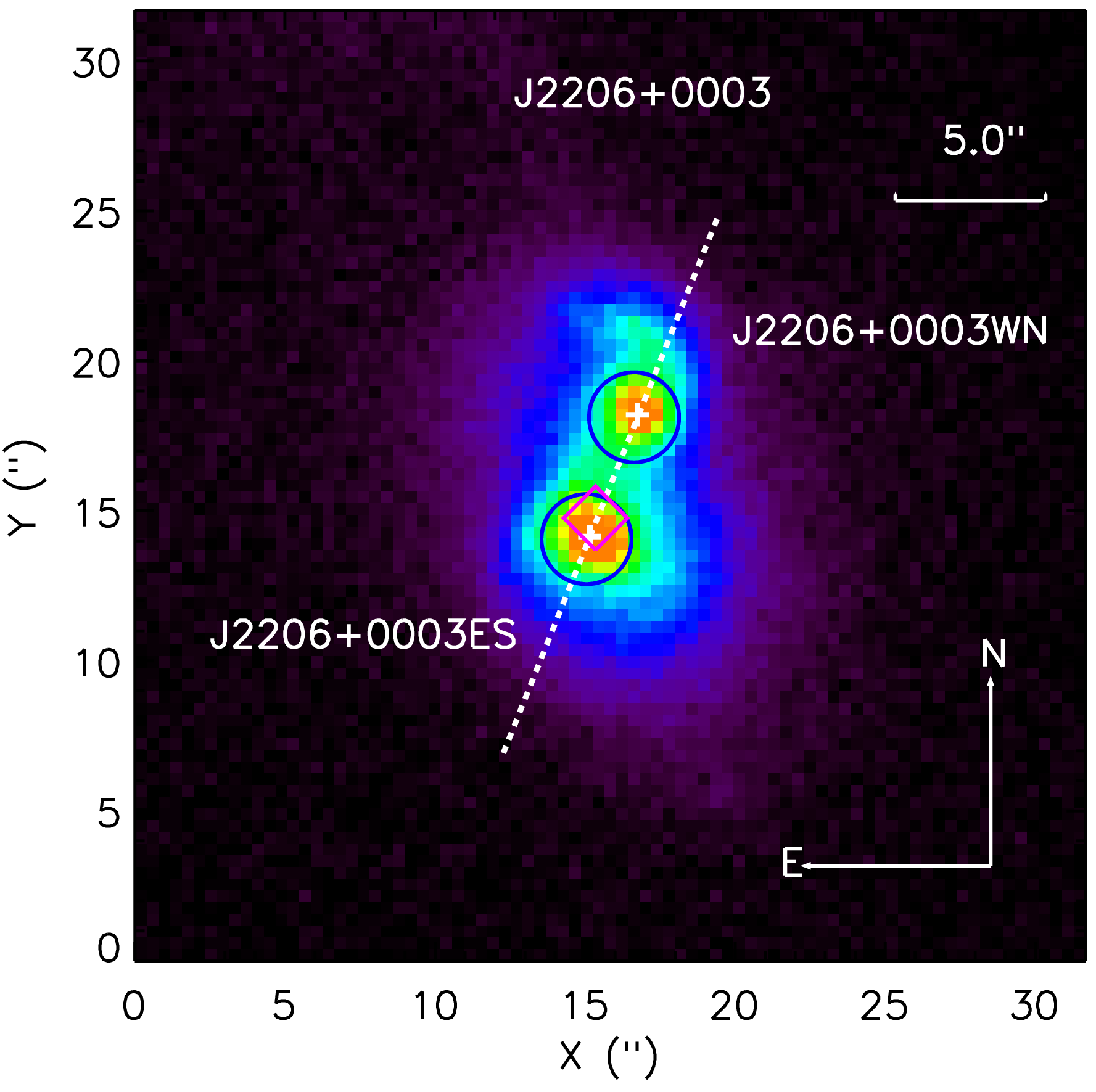}
  \includegraphics[width=12.2cm,height=5.0cm]{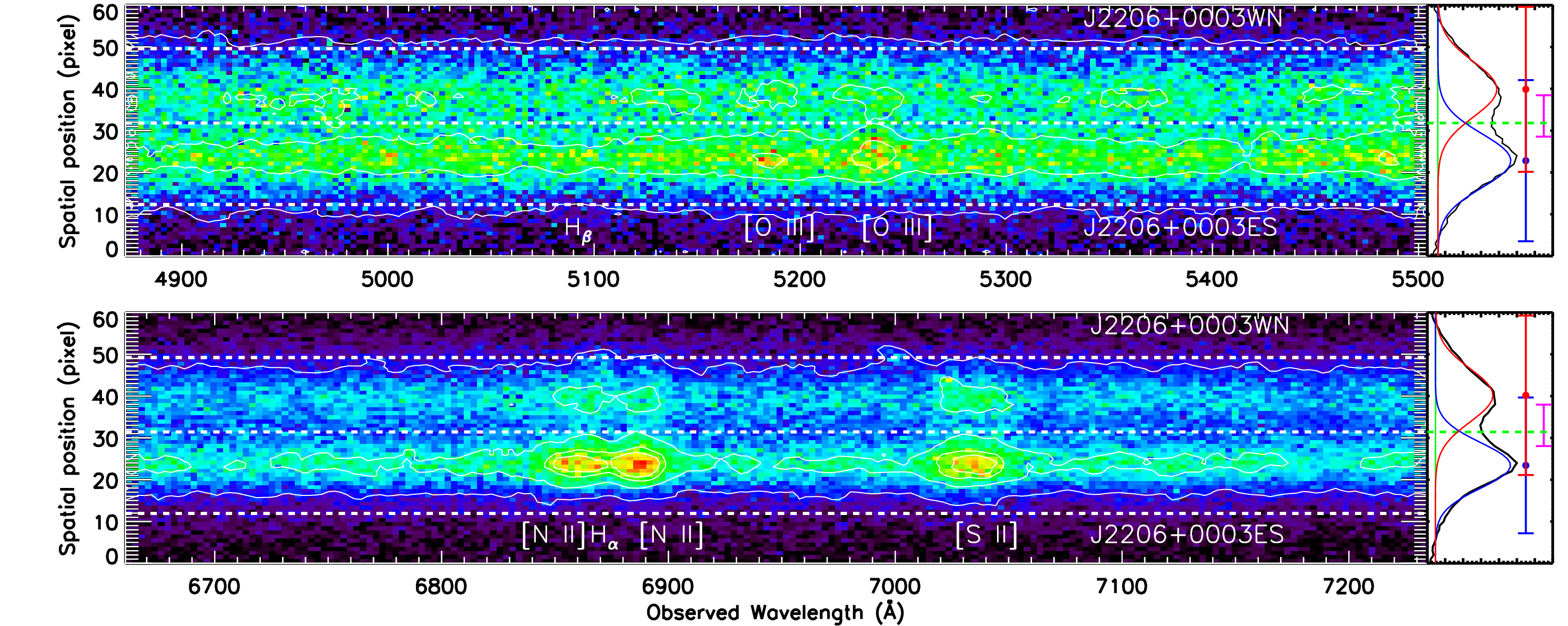}
\caption{Same as Fig.\,\ref{slitimages of J0933+2114} but for J2206+0003.}
\label{slitimages of J2206+0003}
\end{figure*}


\begin{figure*}[ht]
\centering
\includegraphics[scale=0.51]{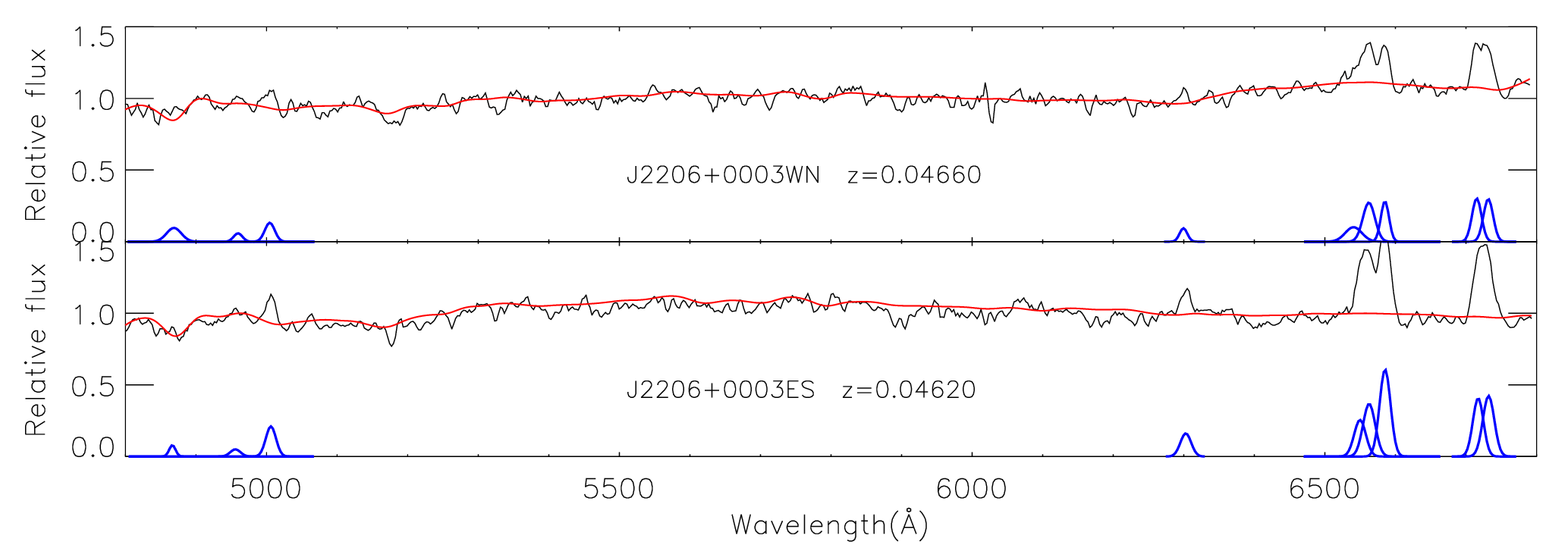}
\caption{Same as Fig.\,\ref{spectra_fitting:J0933+2114} but for J2206+0003. The spectra from LJT.}
\label{spectra_fitting:J2206+0003}
\end{figure*}

\begin{figure*}[ht]
  \centering
  \includegraphics[scale=0.58]{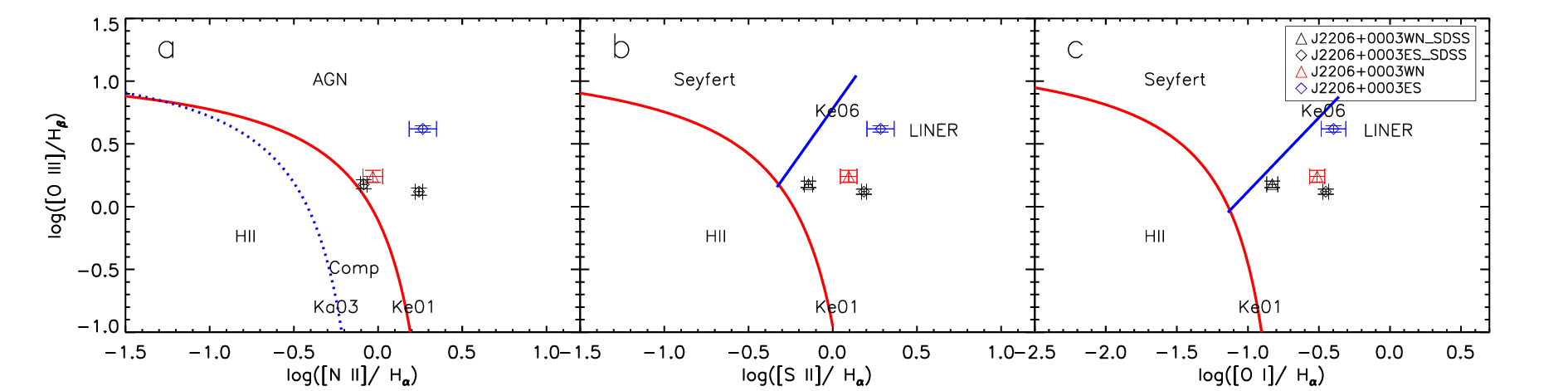}
\caption{Same as Fig.\,\ref{BPT_DAGN} but for J2206+0003.}

\label{BPT diagrams of J2206+0003}
\end{figure*}

\subsection*{\rm Dual AGN: J220634.97+000327.57}
\label{DAGN of J2206+0003}

%
%

Two sets of AGN spectra are spatially resolved as shown in Fig.\,\ref{slitimages of J2206+0003}, so the two cores, i.e. J2206+0003WN and J2206+0003ES can be identified separately. 

The fitting of extracted 1D spectra of the two cores are shown in Fig.\,\ref{spectra_fitting:J2206+0003}. The redshifts, FWHMs of emission lines and emission line flux ratios of the two cores, measured from the 1D spectra, are presented in Tables\,\ref{BPT classify} and \ref{finally DAGN}.
For the two cores, no broad line components are detected, we therefore use BPT diagram to classsify their types (Fig.\,\ref{BPT diagrams of J2206+0003}). 
According to the diagnosis, both cores are classified as LINERs (AGN). 

The object J220634.97+000327.57 has been revealed as a dual AGN composed of LINER (J2206+0003WN) and LINER (J2206+0003ES). This dual AGN has a separation of 4.3 kpc and a velocity offset of $120 \pm 40 $ km $\mathrm{s}^{-1}$.

\clearpage

\begin{figure*}[ht]
  \centering
  \includegraphics[width=5.20cm,height=4.5cm]{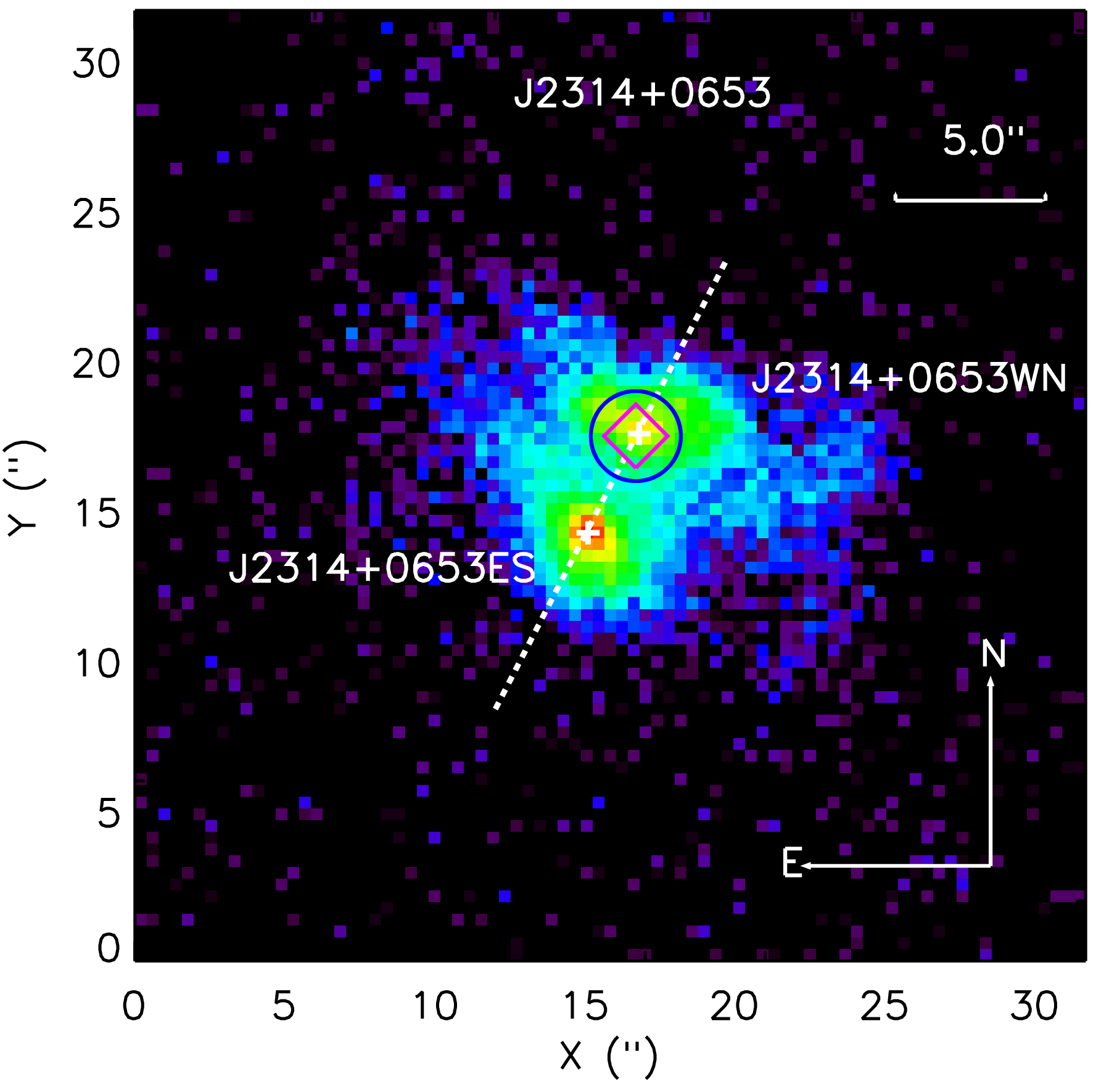}
  \includegraphics[width=12.2cm,height=4.5cm]{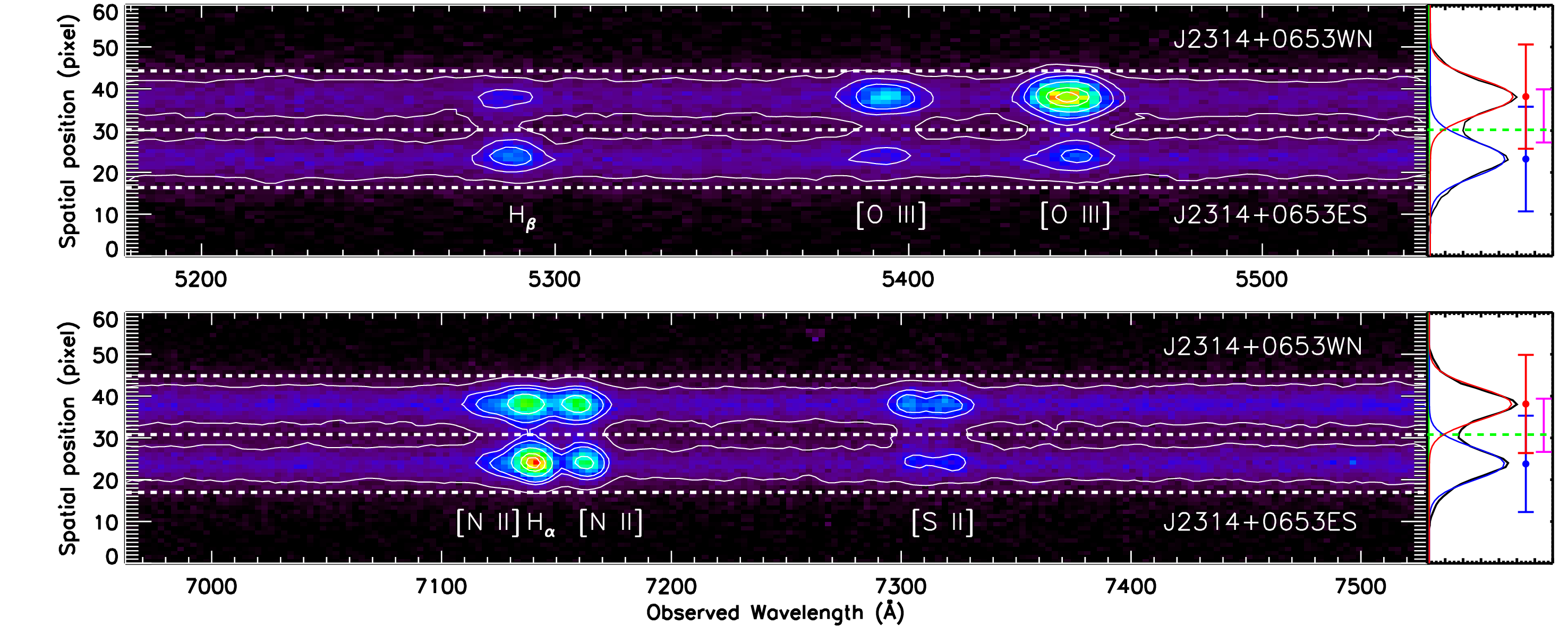}
\caption{Same as Fig.\,\ref{slitimages of J0933+2114} but for J2314+0653.}
\label{slitimages of J2314+0653}
\end{figure*}


\begin{figure*}[ht]
\centering
\includegraphics[scale=0.51]{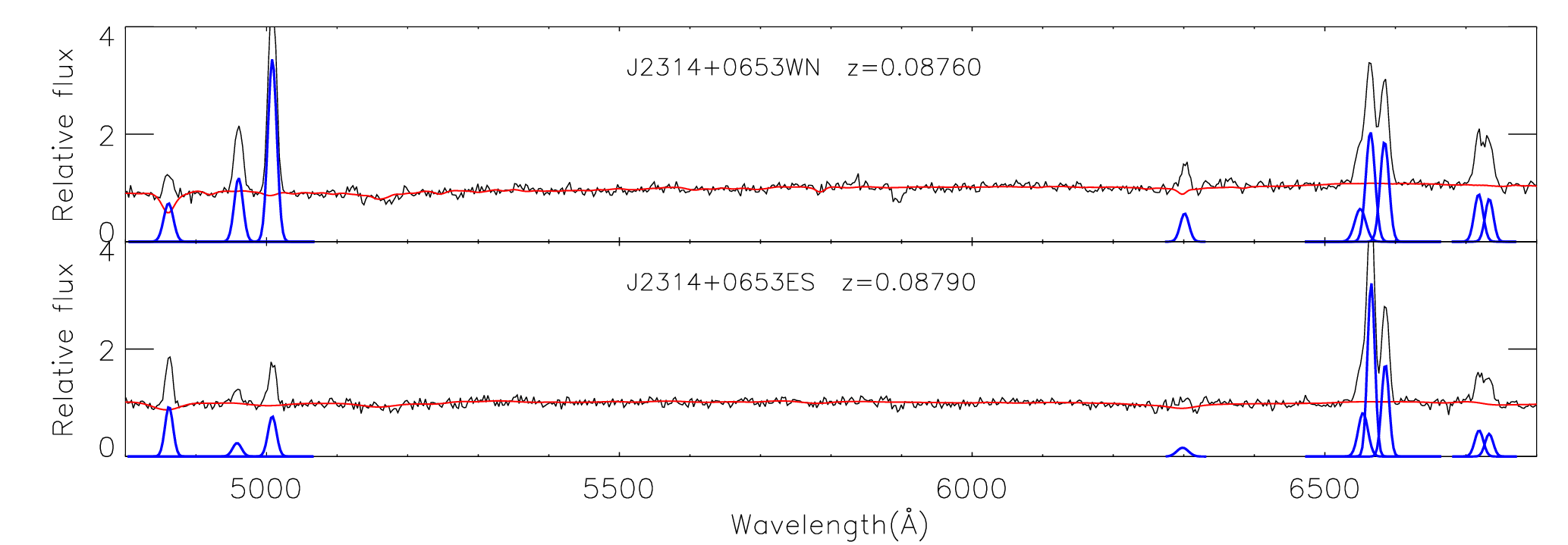}
\caption{Same as Fig.\,\ref{spectra_fitting:J0933+2114} but for J2314+0653. The spectra from LJT.}
\label{spectra_fitting:J2314+0653}
\end{figure*}

\begin{figure*}[ht]
  \centering
  \includegraphics[scale=0.58]{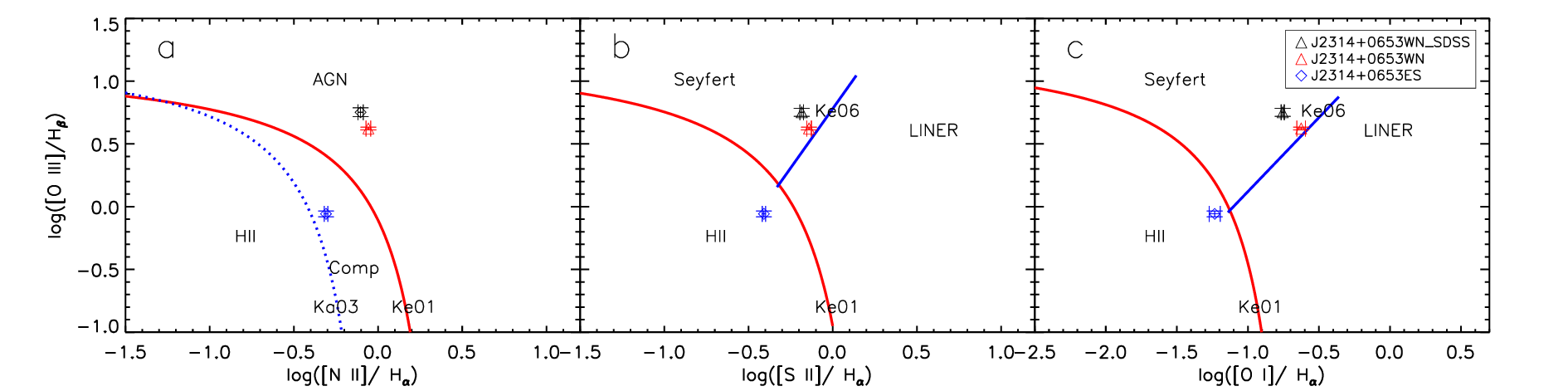}

\caption{Same as Fig.\,\ref{BPT_DAGN} but for J2314+0653.}
 
\label{BPT diagrams of J2314+0653}
\end{figure*}

\subsection*{\rm Dual AGN: J231439.21+065312.97}
\label{DAGN of J2314+0653}

%
%

Two sets of AGN spectra are spatially resolved as shown in Fig.\,\ref{slitimages of J2314+0653}, so the two cores, i.e. J2314+0653WN and J2314+0653ES can be identified separately.

The fitting of extracted 1D spectra of the two cores are shown in Fig.\,\ref{spectra_fitting:J2314+0653}. The redshifts, FWHMs of emission lines and emission line flux ratios of the two cores, measured from the 1D spectra, are presented in Tables\,\ref{BPT classify} and \ref{finally DAGN}.
For the two cores, no broad line components are detected, we therefore use BPT diagram to classsify their types (Fig.\,\ref{BPT diagrams of J2314+0653}). 
According to the diagnosis, J2314+0653WN is classified as Seyfert (AGN) and J2314+0653ES is classified as Comp (AGN). 

The object J231439.21+065312.97 has been revealed as a dual AGN composed of Seyfert (J2314+0653WN) and LINER (J2314+0653ES). This dual AGN has a separation of 6.7 kpc and a velocity offset of $90 \pm 40 $ km $\mathrm{s}^{-1}$.

\end{appendices} 
\label{lastpage}
\end{document}